\documentclass[11pt,aps,pra,eqsecnum,subeqn,graphicx]{article}
\usepackage{amssymb,amsmath,color,times,epsfig}
\newcommand{\be}{\begin{equation}}
\newcommand{\ee}{\end{equation}}
\newcommand{\ba}{\begin{array}}
\newcommand{\ea}{\end{array}}
\newcommand{\beqa}{\begin{eqnarray}}
\newcommand{\eeqa}{\end{eqnarray}}
\newcommand{\beqas}{\begin{eqnarray*}}
\newcommand{\eeqas}{\end{eqnarray*}}
\newcommand{\beqal}{\begin{lefteqnarray}}
\newcommand{\eeqal}{\end{lefteqnarray}}

\begin{document}

\title{Determining fundamental parameters from the chargino sector in Left-Right Supersymmetric models}

\author{Nibaldo Alvarez-Moraga${}^{\dag}$\thanks{{\it Email address:}
nibaldo.alvarez.m@exa.pucv.cl}
\\
{\small  ${}^\dag$  Autonomous Center of Theoretical Physics and
Applied Mathematics,} \\ {\small 3561 Hutchison $\#$ 3, Montr\'eal
(Qu\'ebec) H2X 2G9, Canada} }
 \maketitle
\baselineskip 0.6cm

\begin{abstract}
Analytical expressions relating the fundamental parameters
describing the chargino sector in the context of the Left-Right
Supersymmetric  model are constructed. A general complex extension
of the real non-symmetric chargino mass matrix including all
possible CP-violating phases is considered. The method used for such
a effects is the projector formalism based on the explicit knowledge
of two unitary matrices diagonalizing the chargino mass matrix. Some
possible scenarios allowing us to extract analytical and numerical
values for the unknown parameters are considered. Moreover, an
algorithm allowing us  to disentangle the fundamental parameters of
the chargino sector, based on possible measurements of some class of
cross-section observables related to the chargino pair production in
$e^+ e^-$ annihilation processes, is described. Some comparisons
with the corresponding results in the context of the Minimal
Supersymmetric Standard Model are given.
\end{abstract}


 \vspace{5mm}

\setcounter{equation}{0}
\newpage
\section{INTRODUCTION}
In the Left-Right Supersymmetric (L-R SUSY) model \cite{key1,key2},
which is based on the gauge group $SU(3)_C \times SU(2)_L \times
SU(2)_R \times U(1)_{B-L}$ \cite{key3}, the masses and mixing
matrices of the neutralinos and charginos are determined by $M_L$ ,
$M_R,$ the left-right gaugino mass parameters associated with the
gauge group $SU (2)_L$ and $SU(2)_R$ respectively, $M_V$ the gaugino
mass parameter associated with the gauge group $ U(1)_{B-L},$
$\mu_j, j=1,2,3,$ the Higgsino mass parameters, the ratio
$\tan\theta_k= {k_{u}/ k_{d}},$ where $k_{u}$ and $k_d$ are the
vacuum expectation values ($VEV's$) of the Higgs fields which couple
to $d$-type and $u$-type quarks respectively, $v_{\Delta_R}$  the
VEV of the Higgs triplet field $\Delta_R$ which together the triplet
Higgs field $\Delta_L$ are associated to the spontaneous symmetry
breaking of the group $S(2)_R \times U(1)_{B-L}$ to the hypercharge
symmetry group $U(1)_Y,$ and $v_{\delta_R},$ the VEV of the Higgs
triplet field $\delta_R$ which together the triplet Higgs field
$\delta_L$ are introduced  to insure cancellation of the anomalies
in the fermionic sector \cite{key4}-\cite{key10}.

In Ref. \cite{key11} several relations connecting the parameters
describing the neutralino sector in the context of the L-R SUSY
model were described in terms of projectors, pseudoprojectors,
reduced projectors and eigenphases\cite{key12,key13}. There,
analytic expressions for the neutralino masses were obtained by
diagonalizing the associated complex symmetric neutralino mass
matrix. Then, based on the explicit knowledge of the corresponding
diagonalizing unitary matrix novel formulae for the fundamental
reduced projectors were constructed. Moreover, several CP-conserving
and CP-violating possible scenarios were considered in the study of
the so-called inverse problem consisting to express the fundamental
parameters in terms of the neutralino physical masses, the
eigenphases and the reduced projectors.

The purpose of this work is extending this procedure to study the
exiting relations among the fundamental parameters describing the
chargino sector of the L-R SUSY model. More precisely, we consider
the terms in the total Lagrangian density which are relevant to
describe the lightest chargino masses. The principal difference with
the previous work is that in this case the chargino mass matrix is
not symmetric and therefore one requires two different unitary
matrices to diagonalize it. This implies the construction of two
types of fundamental reduced projectors, one for each diagonalizing
matrix, and the generalization of the method used in
\cite{key11,key13} to treat the inverse problem. The study is carry
out in a general context by introducing  arbitrary CP violating
phases in the mixing chargino matrix. Thus, we investigate, for
instance, the effects of these phases on the chargino mass spectrum
considering several CP-violating scenarios. Also, concerning the
inverse problem, we study scenarios conditioning the values of these
phases as well as the unknown fundamental parameters.

The paper is organized as follows, in Section \ref{sec-LRSUSY}, we
write the Lagrangian density describing the light chargino sector in
terms of the charged gaugino and Higgsino fields, the coupling
constants associated to the gauge groups of the L-R SUSY model and
the above mentioned fundamental parameters.  According to our
purpose, we write it in terms of the mixing chargino mass matrix.
Also, in this section, in order to compare some of our results with
the corresponding ones obtained in the context of the Minimal
Supersymmetric Standard Model (MSSM), we describe briefly some
aspects of this last model (for reviews see \cite{keyMSSM}).

In Section \ref{sec-numerical-results}, we analyze numerically the
chargino mass spectrum based on the analytical expressions for the
chargino masses obtained by diagonalizing the above mentioned
chargino mass matrix. The details of these calculus are given in
Appendix \ref{sec-chargino-MASS}.  We plot the masses versus the
Higgsino mass parameter in some possible CP-conserving and
CP-violating scenarios (effects of the CP violating phases on the
chargino mass spectrum) and we compare the results with those
obtained in \cite{key14} and also with those obtained using the
corresponding formulas in the context of the MSSM \cite{key19}. In
Section \ref{sec-unitary-U-V}, we  compute the analytical
expressions for the corresponding diagonalizing unitary matrices.

In Section \ref{sec-projectors}, the projector formalism
\cite{key16} for this model is implemented in a original way. As in
Ref. \cite{key11}, we define the reduced projectors and eigenphases
suitably and we express the entries of the diagonalizing chargino
matrices in terms of them. Then, using an appropriate connection
between the eigenvectors associated to these diagonalizing matrices
we obtain a system of equations involving the fundamental
parameters, the chargino masses, the eigenphases and the reduced
projectors. We use these equations to  express all the fundamental
L-R SUSY parameters of the chargino sector in terms of the reduced
projectors and eigenphases.

In Section \ref{sec-ML-disentangling}, the disentangle of some
fundamental parameters is considered. Using explicit expressions for
the reduced projectors we disentangle the complex parameter $M_L $
from the chargino physical masses, the eigenphases and the rest of
the parameters. Moreover, using the Jarlskog's formulas
\cite{key16},  we  get an alternative formula to compute the norm of
$M_L$ in terms of the last mentioned quantities, but with a given
combination of the CP-violating phases playing  the role of the
eigenphase arguments. In this section, we also propose an
alternative parametrization based on some redefined reduced
projectors and eigenphases allowing  to disentangle the real
parameter $M_R$  from the rest of the parameters, we compare with
the corresponding case in the context of the MSSM. In Section
\ref{sec-determinig-parameters} analytical and numerical
reconstruction of the fundamental parameters is performed
considering various possible CP-conserving and CP-violating
scenarios.

In Section \ref{sec-total-disentangle}, based on some possible
experimental measurements of cross-section-type observables related
to the chargino pair production in $e^+ e^-$ annihilation processes
a general algorithm for disentangle the fundamental parameters in
the chargino sector is proposed.  The general problem may be reduced
to compute twelve independent parameters which can be chosen to be
eight reduced projector norms (they can be parameterized by four
pairs of spherical angles) and either four reduced projectors phases
or four eigenphases. When the two lightest chargino masses are known
the amount of independent parameters is reduced from twelve to ten.
The expression for the parameter $\tan\theta_k,$ in terms of the
independent parameters and two lightest chargino masses is displayed
in Appendix \ref{sec-tanthetak}.

Finally, in Appendix \ref{sec-tablas}, in order to get a better
overview of the complex interplay between the whole parameter set in
that model, we construct some tables resuming the principal results
of this paper.

\section{Chargino sector in  Left-Right Supersymmetric model}
\label{sec-LRSUSY} The particle content of the L-R SUSY model is
different from that of the MSSM in the gauge sector, the Higgs
sector and also having a right-handed neutrino in a weak isosinglet
neutrino superfield. In this model, the gauge sector has an extra
neutral $Z^0_R$ and two charged $W_R^\pm$ gauge bosons corresponding
to the gauge group $SU(2)_R.$ The Higgs sector contains two
bi-doublet fields (in order to give masses to all the quarks)
\begin{eqnarray} \label{eq:bi-doublet}\phi_{u,\,d}=\left(\begin{array}{cc}
\phi^0_{1}&\phi^+_{1}\\
\phi^-_{2}&\phi^0_{2}
\end{array}\right)_{u,\,d}
\eeqa and four triplet fields, \beqa
\Delta_{L,\,R}=\left(\begin{array}{cc}
\frac{1}{\sqrt{2}}\,\Delta^{+}&\Delta^{++}\\
\Delta^{0}&-\frac{1}{\sqrt{2}}\,\Delta^{+}
\end{array}\right)_{L,\,R},\eeqa
and \beqa \delta_{L,\,R}=\left(\begin{array}{cc}
\frac{1}{\sqrt{2}}\,\delta^{+}&\delta^{++}\\
\delta^{0}&-\frac{1}{\sqrt{2}}\,\delta^{+}
\end{array}\right)_{L,\,R}.\eeqa
The Higgs $\phi_{u,d}$ both transform as $(1/2,1/2,0),$ and  the
Higgs $\Delta_{L,\,R}$ transform as $(1,0,2)$ and $(0,1,2),$
respectively. The triplet Higgs $\delta_{L,\,R}$ which transform as
$(1,0,-2)$ and $(0,1,-2),$ respectively, are introduced to cancel
anomalies in the fermionic sector that would otherwise occurs.

The full Lagrangian of the  L-R SUSY model is given by \cite{key1}
\be {\cal L} = {\cal L}_{\rm gauge} +  {\cal L}_{\rm matter} + {\cal
L}_{\rm Y} - {\cal V} + {\cal L}_{\rm soft}, \ee where ${\cal
L}_{\rm gauge}$ contains the kinetic and self-interactions terms for
the bosons vector fields $(W^\pm,W^0)_{L,R}$ and $ V^0,$ and the
Dirac Lagrangian of their corresponding superpartners, i.e., the
gaugino fields $(\lambda^\pm,\lambda^0)_{L,R}$ and $\lambda^0_V;$
${\cal L}_{\rm matter}$ contains the kinetic terms for the fermionic
and bosonic matter fields, the Higgs fields and interaction of the
gauge and matter multiplets; ${\cal L}_{\rm Y}$ (Yukawa Lagrangian)
contains the self-interaction terms of the matter multiplets,  as
well as the Higgs multiplets, e.g., it contains the self-interaction
terms of the matter multiplets involving the fundamental Higgsino
mass parameters $\mu_1 \equiv \mu, \mu_2$ and $\mu_3:$ $ {\rm Tr} [
\mu_1 (\tau_1 {\tilde \phi}_u \tau_1)^T {\tilde \phi}_d ], $ $ {\rm
Tr} [\mu_2 (\tau \cdot {\tilde \Delta}_L) (\tau \cdot {\tilde
\delta}_L)] $ and $ {\rm Tr} [\mu_3 (\tau \cdot {\tilde \Delta}_R)
(\tau \cdot {\tilde \delta}_R)],$ where $\tau_j, \, j=1,2,3$ are the
usual Pauli matrices,  ${\tilde \phi}_u, $ ${\tilde \phi}_d, $
${\tilde \Delta}_{L,R}$ and ${\tilde \delta}_{L,R}$ are the
superpartners of the bi-doublet field $\phi_u,$ $\phi_d$  and  the
four triplet fields $\Delta_{L,R}$ and $\delta_{L,R},$ respectively;
$\cal V$ is a scalar potential; ${\cal L}_{\rm soft}$ is the
soft-breaking Lagrangian, involving the fundamental gaugino mass
parameters $M_L, M_R$ and $M_V,$ which gives Majorana mass to the
gauginos:  \beqa \nonumber {\cal L}_{\rm soft} &=& M_L
(\lambda^a_L \lambda^a_L + {\bar \lambda}^a_L {\bar \lambda}^a_L) \\
\nonumber &+& M_R (\lambda^a_R \lambda^a_R + {\bar \lambda}^a_R
{\bar \lambda}^a_R)
\\ &+& M_V (\lambda^0_V \lambda^0_V + {\bar \lambda}^0_V {\bar \lambda}^0_V).
\eeqa

In order to generate mass for the gauge bosons we can choose the
VEV's of the Higgs fields in the form \cite{key1} \beqa
\label{vevdelta}\langle \Delta_{L} \rangle = \langle
\delta_{L}\rangle = 0, \qquad \langle \Delta_{R} \rangle &=&
\begin{pmatrix}
0 & 0 \cr \\ v_{\Delta_R} & 0 \cr \end{pmatrix}, \qquad \langle
\delta_{R} \rangle =
\begin{pmatrix}
0 & 0 \cr \\ v_{\delta_R} & 0 \cr \end{pmatrix},
\\ \label{vevkud} \langle \phi_{u} \rangle  &=& \begin{pmatrix}
k_{u}&0 \cr \\
0&0 \cr \end{pmatrix}, \qquad \langle \phi_{d}
\rangle=\begin{pmatrix}
0&0 \cr \\
0&k_{d} \end{pmatrix}. \label{eq:vacumm-expectation-values}\eeqa
Hence, the generation of mass for the gauge bosons can be analyzed
in two separate stages of symmetry breaking. In the first stage, the
break of $SU(2)_{R}\times U(1)_{B-L}$ into $U(1)_Y,$ according to
the VEV's of the $\Delta_R , \delta_R$ fields  given in Eq.
\eqref{vevdelta}, generates masses for $W^{\pm}_R, W^0_R$ and $V^0.$
The two neutral states $W^0_R$ and $V^0 $ mix yielding the physical
field $Z_R$ and the massless field $B.$ The vacuum expectation value
$v_{\Delta_R}$ of the triplet Higgs $\Delta_{R}$ must be chosen big
enough ($ >> $ 1 TeV ) to provide large masses to gauge bosons
$W^{\pm}_{R}$ and $Z_R$. In the second stage, taking place at a much
lower-energy scale, the spontaneous breaking of $SU(2)_{L}\times
U(1)_{Y}$ into $U(1)_{\rm em},$ according the VEV's for $\phi_{u,d}$
given in Eq. \eqref{vevdelta}, generates masses for the  weak bosons
$W^\pm_{L}$ and $W^0_L,$ as well as for  $B_\mu.$ Also in this
stage, the masses for four light neutralinos $m_{{\tilde
\chi}^{0}_{j}}, j=1,\ldots,4,$ and five charginos $m_{{\tilde
\chi}^{\pm}_{j}}, j=1,\ldots,5$ are generated.  Once again, the
neutral fields mix forming the massless photon $A_\mu$ and the
physical gauge field $Z_L.$

To find the chargino and neutralino masses, we consider the
interaction terms between the gauge bosons, the Higgs, and their
superpartners. In this way, for instance, the charged
gaugino-higgsino mixing interaction Lagrangian is given by
\cite{key14,key15} \beqa \nonumber \mathcal{L}_{\rm chargino} &=& i
v_{\Delta_R} \, g_R {\tilde \Delta}^+_R \lambda^-_R + i v_{\delta_R}
\, g_R {\tilde \delta}^-_R \lambda^+_R +  i g_R k_d
\tilde\phi^+_{d1} \lambda_R^- + i g_L k_d \bar\phi^+_{d1}
\lambda_L^- + i g_R k_u \tilde\phi^-_{u2} \lambda_R^+  \\
\nonumber &+& i g_L k_u {\tilde \phi}_{u 2}^- \lambda_L^+ + \mu_1
{\tilde \phi}^+_{u1} {\tilde \phi}^-_{d2} + \mu_1 {\tilde
\phi}^-_{u2} {\tilde \phi}^+_{d1} +  M_L \lambda_L^+ \lambda_L^-  +
M_R \lambda_R^+ \lambda_R^- \\ &+& \mu_3 {\tilde \Delta}^+_R {\tilde
\delta}^-_R + \mu_3 {\tilde \Delta}^{++}_R {\tilde \delta}^{--}_R +
H.C, \label{eq:lagrangiano-chargino}  \eeqa where where $g_{L} $ and
$g_{R} $ are the coupling constants of the gauge groups $SU(2)_{L},$
$SU(2)_{R},$ respectively; $\lambda^\pm_{L,R}$ are the $SU(2)_{L,R}$
gauginos fields,  ${\tilde \phi}^+_{1u},{\tilde \phi}^-_{2 u}$ and
${\tilde \phi}^+_{1 d},{\tilde \phi}^-_{2 d}$ the charged higgsino
fields associated with the $u$ and $d$-quarks, respectively, and
${\tilde \Delta}^{+}_R, {\tilde \Delta}^{++}_R, $ ${\tilde
\delta}^{\pm}_R, $ ${\tilde \delta}^{--}_R$ the charged right-handed
higgsino fields.

As we have  mentioned above, the VEVs $v_{\Delta_R}$ and
$v_{\delta_R}$ are responsible for giving masses to $W_R$ and $Z_R$
bosons, as well as implementing the seesaw mechanism \cite{key3}.
Thus, $v_{\Delta_R}$ is various orders greater than  $1$ TeV
\cite{key17}, i.e., we can consider, in a first approximation, the
$\Delta^+_R$ and $\delta^-_R$ field decoupled from the lightest
chargino states. The Lagrangian describing the lightest charginos
can be written in the form \be \mathcal{L}_{\rm light-char.} = - {1
\over 2}
\begin{pmatrix} {\psi^+}^T & {\psi^-}^T \cr
\end{pmatrix}
\begin{pmatrix} 0 & M^T \cr  M & 0\cr
\end{pmatrix}  \begin{pmatrix} \psi^+ \cr  \psi^- \cr
\end{pmatrix} + h.c,\ee where
\be \label{psimas}\psi^+ = (-i \lambda^+_L, -i \lambda^+_R, {\tilde
\phi}^+_{u1}, {\tilde \phi}^+_{d1} )^T,\ee \be \label{psimenos}
\psi^- = (-i \lambda^-_L, -i \lambda^-_R, {\tilde \phi}^-_{u2},
{\tilde \phi}^-_{d2} )^T\ee
 and  $M$ is the chargino
mass matrix ($\mu_1 \equiv \mu, \mu_3=0$) \beqa M =
\begin{pmatrix}
M_{L} & 0 & 0 & \sqrt{2} {\tilde M}_L \cos\theta_k \cr\\
0 & M_{R} & 0 & \sqrt{2} {\tilde M}_R \cos\theta_k \cr \\
\sqrt{2} {\tilde M}_L \sin\theta_k & \sqrt{2} {\tilde M}_R \sin\theta_k & 0 & -\mu \cr\\
0 & 0 & -\mu & 0 \cr \end{pmatrix},  \label{eq:Mmatrix} \eeqa where
we consider  $M_L = |M_L| e^{i \Phi_L},$  $\mu= |\mu| e^{i
\Phi_\mu}, $\be
 {\tilde M}_L =  M_{W_L} e^{i {\tilde \Phi}_L}  \ee
and \be
 {\tilde M}_R = {g_R \over g_L} M_{{W_L}}  e^{i {\tilde \Phi}_R}, \ee
 where $M_{W_L}$ denotes the mass of the left-handed gauge boson
  \be \label{eq:mass-WL} M_{W_L}= {1\over \sqrt{2}} g_L
(k_u^2 + k_d^2)^{1/2}.\ee Note that from Eq. \eqref{eq:mass-WL} and
the definition of $\theta_k,$  $k_u$ and $k_d$ can be expressed in
terms of $M_{W_L}, g_L$ and $\theta_k$ in
the form \beqa k_u &=& \sqrt{2} {M_{W_L} \over g_L} \sin\theta_k, \\
k_d&=& \sqrt{2} {M_{W_L}\over g_L} \cos\theta_k. \eeqa

In sum, in the CP-violating case, we assume  that the lightest
chargino mass matrix is parameterized by eight real parameters,
namely, $|M_L|,\Phi_L,|\mu|, \Phi_\mu, M_R ,{\tilde \Phi}_L, {\tilde
\Phi}_R$ and $\tan\theta_k.$ On the other hand, in the CP-conserving
case, when all the phases are equal to zero, we assume that the
chargino mass matrix is parameterized by four real parameters,
$M_L,\mu,M_R$ and $\tan\theta_k.$

\subsection{Some aspects about the  Minimal Supersymmetric Standard Model}

Along this work it will be  particulary interesting to compare the
results of some physical quantities obtained from the LR-SUSY model
with those obtained from the MSSM, which is based on the gauge group
$SU(3)_C \times SU(2)_L \times U(1)_Y$. The particle content of the
MSSM are the one of the Standard Model plus the corresponding
superpartners, but besides that, the MSSM includes two complex
scalar Higgs doublets field $H_u=(H_u^+,H_u^0)$ and $H_d=(H_d^0,
H_d^-)$ plus partners and a singlet partner field $\psi_x$ and
${\tilde \psi}_x.$ The  break of electroweak symmetry down to
electromagnetism $SU(2)_L \times U(1)_Y \mapsto U(1),$ is realized
by giving nonzero VEVs to the Higgs fields  $\langle
H_u^0\rangle=v_u$ and $\langle H_u^0 \rangle=v_d.$ These VEVs can be
connected to the known mass of the $Z^0$ boson and the electroweak
gauge couplings $g,g'$: \be v_u^2 + v_d^2 = {2 m_{Z}^2 \over (g^2 +
g'^2)} \approx (174 {\rm GeV})^2. \ee The ratio of the two VEVs is
written as $ \tan\beta=v_u /v_d.$ Among the effects of electroweak
symmetry breaking, the neutral higgsinos ${\tilde H}_u^0$ and
${\tilde H}_d^0$ and the neutral gauginos ${\tilde B}$ and ${\tilde
W}^0$ combine to form four neutral mass eigenstates, i.e., the
neutralinos. The charged higgssinos ${\tilde H}_u^+$ and ${\tilde
H}_d^-$ and winos ${\tilde W}^+$ and ${\tilde W}^-$ mix to form two
mass eigenstates with charge $\pm 1$ called charginos. Thus, the
mass content of the neutralino sector is described by a $4 \times 4$
symmetric matrix which, in the most general CP-violating
case\cite{key13}, is parameterized by two complex quantities,
$M_1=|M_1| e^{i \Phi_1}$ and  $\mu=e^{i \Phi_\mu},$ (the
supersymmetric higssino mass parameter) and two real quantities,
$M_2$ (the $SU(2)$ gaugino mass), and $\tan\beta=v_2 /v_1.$ Here,
the phase angles are defined such that $\pi \le \Phi_1\le \pi$ and
$\pi \le \Phi_\mu \le \pi.$ The chargino mass matrix is given by the
$2\times 2$ non-symmetric matrix \beqa {\cal M}_C &=&
\begin{pmatrix}
 M_{2} &  \sqrt{2} {m}_W \cos\beta \cr\\
 \sqrt{2} {m}_W \sin\beta & \mu \cr \\
\end{pmatrix},  \label{eq:MSSMatrix}\eeqa
where $m_W\approx 80,419$GeV is the mass of the vector boson gauge
field $W.$  Introducing a CP-violating phase in the higgsino
parameter $\mu=|\mu| e^{i \Phi_\mu},$ we can show that the chargino
mass spectrum for this model is given by \cite{key18} \be
m^2_{{\tilde \chi}_{1,2}^{\pm}}= {1 \over 2} \left(M_2^2 + |\mu|^2 +
2 m_W^2 \mp \Delta_C\right), \label{eq:mass-char-MSSM} \ee where \be
\Delta_C=\sqrt{(M_2^2 -|\mu|^2)^2 + 4 m_W^4 \cos^2(2 \beta) + 4
m_W^2 (M_2^2 + |\mu|^2) + 8 m_W^2 M_2 |\mu| \sin(2\beta)
\cos\Phi_\mu}. \label{eq:Deltac} \ee

The parameter determination problem in both cases, the CP-conserving
and CP-violating, has been studied in detail in the literature, see
for example \cite{key13,key18,key22,key23}.


\section{Chargino masses, numerical results}
\label{sec-numerical-results} In this section we investigate the
chargino mass spectrum of the LR-SUSY under some chosen
CP-conserving and CP-violating scenarios we compare the results with
those obtained in similar conditions in the context of the MSSM.

The physical chargino mass eigenstates are related to the states
given in Eqs. \eqref{psimas} and \eqref{psimenos} as \be
\psi^{+}_{i}= \sum_{j=1}^4  V_{ij} \,\chi^{+}_{j}, \qquad
i=1,\ldots,4,  \ee \be \, \psi^{-}_{i}= \sum_{j=1,4} U_{ij}\,
\chi^{-}_{j} \qquad i=1,\ldots,4, \ee respectively, where  the
unitary  $U$ and $V$ matrices satisfies \beqa
 M_{D} &=& U^T \,M \,V, \nonumber \\
&\equiv& \sum_{j=1}^{4}\, m_{{\tilde \chi}^{\pm}_{j}}\,E_{j},
\label{eq:VMV} \eeqa and
\beqa
\nonumber  M^{2}_{D} &=& V^{-1}\, M^{\dag} \,M\,V =  U^T \, M  \, M^\dag  \,U^\ast  \\
&\equiv& \sum_{j=1}^{4}\,m_{{\tilde \chi}^{\pm}_{j}}^{2}\,E_{j},
\label{eq:MD2} \eeqa
where $(E_{j})_{4\times4}$ are the basic matrices defined by \be
(E_{j})_{ik}= \delta_{ji} \, \delta_{jk}. \ee Here, we suppose that
the real eigenvalues of $M_{D}$ are ordered in the following way \be
\label{eq:orden-masas}  m_{{\tilde \chi}^{\pm}_{1}} \le   m_{{\tilde
\chi}^{\pm}_{2}}  \le  m_{{\tilde \chi}^{\pm}_{3}} \le  m_{{\tilde
\chi}^{\pm}_{4}} .\ee  The resulting diagonal matrix $M_D,$
containing the chargino masses, and the matrices $U$  and $V$ are
known only for the CP-conserving case under the assumptions that
$g_L \approx g_R$ and in limit of large $M_R,$ $M_L$ or $\mu,$ \be
|M_L \, \mu | \gg {\tilde M}^2_{L} \sin^2\theta_k, \quad  |M_R \,
\mu | \gg {\tilde M}^2_{R} \sin^2\theta_k,  \ee and similarly for
 $\sin^2\theta_k \leftrightarrow \cos^2\theta_k$ \cite{key14}. In
Appendix \ref{sec-chargino-MASS}, we put into practice a method
\cite{key11,key12,key13,key135,key140} giving analytic expressions
for these masses without limiting the values of the parameters. In
this way, in the most general case, we directly observe that to
determine the chargino masses in the context of  L-R SUSY model we
must to fix eight real parameters, namely
$|M_L|,\Phi_L,|\mu|,\Phi_\mu, M_R, {\tilde \Phi}_L, {\tilde
\Phi}_R,$ and $\tan\theta_k,$ whereas in the MSSM, the chargino
masses only depend on four real fundamental parameters, $M_2, |\mu|,
\Phi_\mu$ and $\tan\beta.$

\subsection{CP-conserving scenarios}
\begin{table}
\begin{center}
\begin{tabular}{c c c c c }\cline{1-5}\\
Scenario & $M_R \; $ & $ M_L \; $ & $ M_{W_L} $ & $ \tan\theta_k  $ \\
\\\cline{1-5} \\
$Scpc_1 $ & 300 & 50  &  50.271 &    1.6   \\ \\  \hline \\
$Scpc_2 $ & 200 & 150  &  80.456 &   3
\\ \\  \hline \\
\end{tabular}
\end{center}
\caption{Input parameters for scenarios $Scpc_1$ and $Scpc_2.$  All
mass quantities are given in  GeV.} \label{tab:tablauno}
\end{table}
Let us consider the CP-conserving scenarios $Scpc_1$ and $Scpc_2$
described in Tab. \ref{tab:tablauno}. These scenarios are similar to
the ones studied in Ref. \cite{key14} where they have been used to
compare the chargino masses predicted by the L-R SUSY model and the
MSSM.  For both scenarios, we consider the coupling constants values
$g_R \approx g_L =  0.65.$  To make suitable the general results
given in Eqs. \eqref{eq:EIGU} and \eqref{eq:EIGV} to these
situations we must take in Eqs. (\ref{eq:b-term}-\ref{eq:d-term})
the mixing phases $\Phi_L={\tilde \Phi}_L={\tilde \Phi}_R=0$ and
$\Phi_\mu=0,\pi.$

Figures \ref{fig:chargino1} and \ref{fig:chargino2}, show the
behaviour of the chargino physical masses $m_{{\tilde
\chi}^{\pm}_{i}}, \, i=1,\ldots,4,$ versus $\mu,$ computed from Eqs.
(\ref{eq:EIGU}-\ref{eq:EIGV}), for input parameters of scenarios
$Scpc_1$ and $Scpc_2,$ respectively.  We observe the correct size
ordering of the chargino masses, such as required by Eq.
\eqref{eq:orden-masas}. Also, in both  scenarios, we find that for $
|\mu| \sim 200 \, {\rm GeV}$ the chargino masse $m_{{\tilde
\chi}^{\pm}_{1}}$ is equal to $M_L,$ approximately,  and for $|\mu|
\sim 300 \, {\rm GeV}$ the the chargino masse $m_{{\tilde
\chi}^{\pm}_{3}}, $ is of the order of $ M_R.$ On the other hand,
considering the values of the chargino masse $m_{{\tilde
\chi}^{\pm}_{4}}, $ we observe that for all values of $|\mu|$ they
are always greater than $M_R.$

In Fig. \ref{fig:chargino1}, we observe that the mass $m_{{\tilde
\chi}^{\pm}_{1}}=|\mu|,$ when $-20 \, {\rm GeV} \lesssim \mu
\lesssim 60$ GeV and $m_{{\tilde \chi}^{\pm}_{2}}=|\mu|,$ in the
complementary region,  whereas in Fig. \ref{fig:chargino2} we
observe that the mass $m_{{\tilde \chi}^{\pm}_{2}}=|\mu|,$ when
$-175 \, {\rm GeV} \lesssim \mu - \lesssim 25$ GeV and $125 \, {\rm
GeV} \lesssim \mu \lesssim 185$ GeV, $m_{{\tilde
\chi}^{\pm}_{1}}=|\mu|,$ when $-25 \, {\rm GeV} \lesssim \mu
\lesssim 125$ GeV and $m_{{\tilde \chi}^{\pm}_{3}}=|\mu|,$ in the
complementary region. This is an important feature that we must
taking into account when we study the inverse problem. Indeed, we
can show that $|\mu|^2$ is an exact solution of the  characteristic
equation determining the chargino spectrum (in either CP-conserving
or CP-violating case) (see Appendix \ref{sec-chargino-MASS}), i.e.,
it is always possible to find a region in the fundamental parameter
space where one of the chargino masses exactly takes the value
$|\mu|.$

\begin{figure} \centering
\begin{picture}(31.5,21)
\put(1,2){\includegraphics[width=70mm]{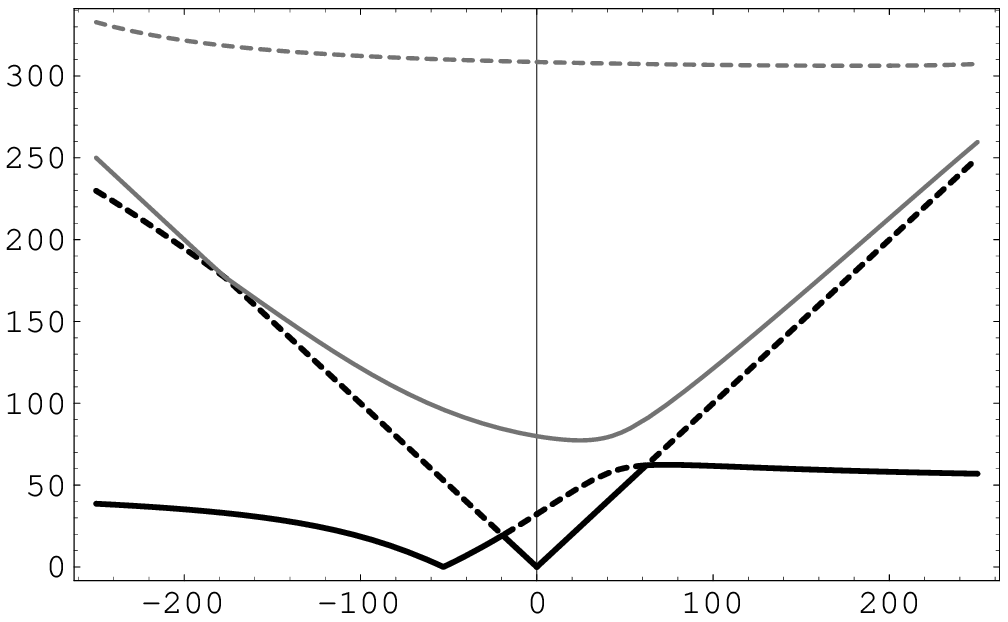}}
\put(15.75,0.7){\scriptsize{$\mu \, {\rm (GeV)}$}}
\put(1.05,19.25){\scriptsize{$m_{{\tilde \chi}^{\pm}_{i}} \, {\rm
(GeV)}$}}
\end{picture}
\caption{chargino masses $m_{{\tilde \chi}^{\pm}_{j}}, \,
i=1,\ldots,4,$ as functions of $\mu$ for scenario input parameters
of scenario $Scpc_1.$} \label{fig:chargino1}
\end{figure}
\begin{figure} \centering
\begin{picture}(31.5,21)
\put(1,2){\includegraphics[width=70mm]{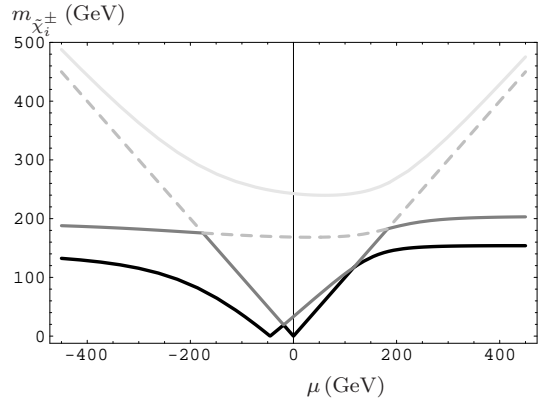}}
\put(15.75,0.7){\scriptsize{$\mu \, {\rm (GeV)}$}}
\put(1.05,19.25){\scriptsize{$m_{{\tilde \chi}^{\pm}_{i}}\, {\rm
(GeV)} $}}
\end{picture}
\caption{chargino masses $m_{{\tilde \chi}^{\pm}_{j}}, \,
j=1,\ldots,4,$ as functions of $\mu$ for input parameters of
scenario $Scpc_2.$ }\label{fig:chargino2}
\end{figure}

\subsection{CP-violating scenarios}
In this section we study the effects of the CP-violating phases on
the chargino mass spectrum. Also, we analyze  the singularities in
the curves  $m_{{\tilde \chi}^\pm_1}$ versus $\mu$ and we compare
with the corresponding ones obtained using the formulas for the
MSSM. Moreover, we study the behaviour of this mass when
$\tan\theta_k$ varies and again we compare with the results obtained
in the context of the MSSM.

\subsubsection{Chargino mass spectrum when varying \boldmath
$\mu$ and the phase angles}
\begin{table}[b]
\begin{center}
\begin{tabular}{c c c c c c }\cline{1-6}\\
Scenario & $  \; \Phi_L, {\tilde \Phi}_L, {\tilde \Phi}_R $ \; &  $M_R \; $ & $ |M_L| \; $ & $ k_u $ & $ \tan\theta_k  $ \\
\\\cline{1-6} \\
$Scpv_1$ & \parbox[c]{0.5cm}{0 \\ $\pi/8$ \\ $\pi/4$ \\  $\pi/3$} & 300 & 50  &  92.75 &   1.6  \\ \\  \hline \\
\end{tabular}
\end{center} \caption{Input parameters
for scenario $Scpv_1.$ All mass quantities are given in GeV.}
\label{tab:tablados}
\end{table}Let us now to study the effects produced
by the variation of the mixing phases on the curves describing the
behaviour of the chargino masses as a function of  $\mu.$  Let us
consider, for instance, the chargino masses $m_{{\tilde
\chi}^{\pm}_{i}}, i=1,2,$ given in Eq. \eqref{eq:EIGU} and the
CP-violating scenario $Scpv_1$ described in Tab. \ref{tab:tablados}.
Figure \ref{fig:chargino15} shows the effects produced by the
variation of the phase $\Phi_L$ on the curves describing the
chargino mass $m_{{\tilde \chi}^{\pm}_{1}}$ as a function of $\mu,$
for input parameters of scenario $Scpv_1$ with fixed phases ${\tilde
\Phi}_L = {\tilde \Phi}_R=0.$ We observe that for values of $ -20 \,
{\rm GeV} \lesssim \mu  \lesssim 60 \, {\rm GeV}, $ there is not
difference between the curves with $\Phi_L=0$ and those with $\Phi_L
=\pi/8,\pi/4,\pi/3.$ However, for values of $ \mu   \lesssim - 20 \,
{\rm GeV}$ or  $ \mu  \gtrsim 60 \, {\rm GeV},$ we observe for some
values of $\mu,$ differences attaining as far as $6$ Gev
approximately. Also, we observe that in these cases the differences
with respect to the variation of $\Phi_L$ are positives  when $ \mu
\lesssim - 20 \, {\rm GeV}$ and negatives when $ \mu  \gtrsim 60 \,
{\rm GeV}.$

The same arguments are valid when we  consider the effects produced
by the variation of the phase ${\tilde \Phi}_L$ on these curves,  as
we can see by observing Fig. \ref{fig:chargino16}. However, in this
last case, the differences between the curves for some values of
$\mu$ can attain as far as $16$ Gev and the lower limit of positif
values of $\mu$ where some differences are detected is $30$ Gev,
approximately. On the other hand, if we consider the effect on the
curves induced by the variation of the phase ${\tilde \Phi}_R,$  we
don't appreciate important differences among them. Indeed, for
positif values of $\mu$ there is none. Contrarily to the above
analysis, for the curves describing the chargino mass $m_{{\tilde
\chi}^{\pm}_{2}}$ as a function of $\mu,$ the intervals where we
appreciate differences among the curves which are consequences of
the variation of the phases $\Phi_L,$ ${\tilde \Phi}_L$ or ${\tilde
\Phi}_R$ are practically inverted.
\begin{figure} \centering
\begin{picture}(31.5,21)
\put(1,2){\includegraphics[width=70mm]{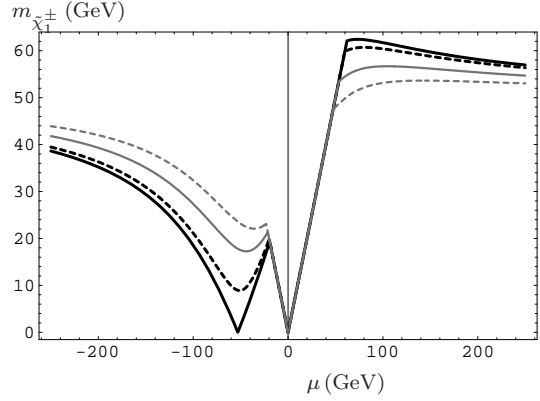}}
\put(15.75,0.7){\scriptsize{$\mu \, {\rm (GeV)}$}}
\put(1.05,19.25){\scriptsize{$m_{{\tilde \chi}^{\pm}_{1}} \, {\rm
(GeV)}$}}
\end{picture}
\caption{Chargino masse $m_{{\tilde \chi}^{\pm}_{1}} $ as a function
of $\mu$ for scenario $Scpv_1,$ assuming ${\tilde \Phi}_L= {\tilde
\Phi}_R=0.$ The curves are: $\Phi_L =0$ (heavy solid), $\Phi_L =\pi
/8$ (heavy dashed), $\Phi_L=\pi/4$ (light solid), $\Phi_L=\pi/3$
(light dashed).} \label{fig:chargino15}
\end{figure}
\begin{figure} \centering
\begin{picture}(31.5,21)
\put(1,2){\includegraphics[width=70mm]{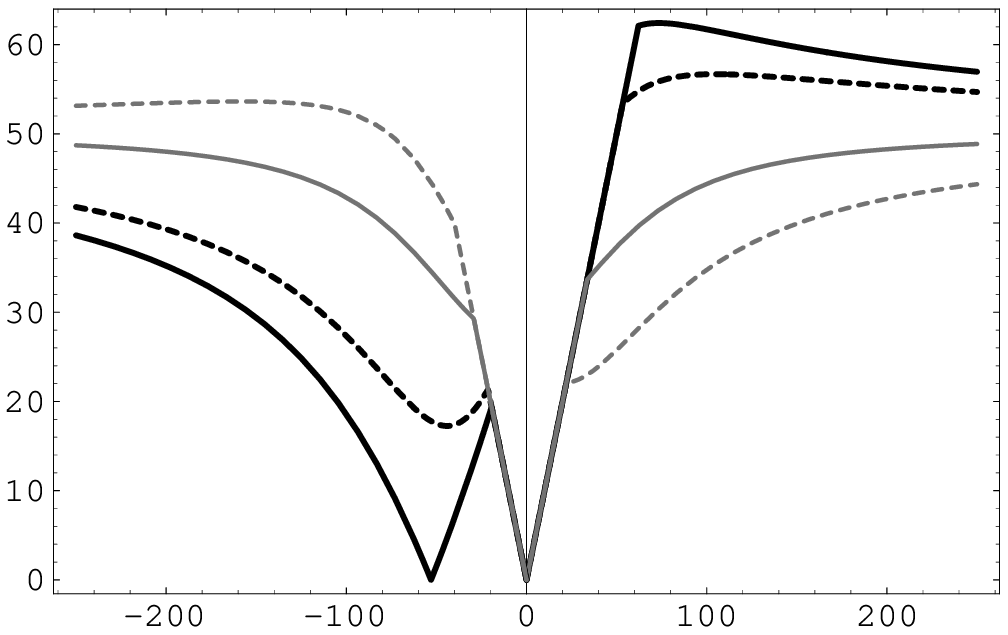}}
\put(15.75,0.7){\scriptsize{$\mu \, {\rm (GeV)}$}}
\put(1.05,19.25){\scriptsize{$m_{{\tilde \chi}^{\pm}_{1}} \, {\rm
(GeV)}$}}
\end{picture}
\caption{Chargino masse $m_{{\tilde \chi}^{\pm}_{1}} $ as a function
of $\mu$ for scenario $Scpv_1,$ assuming $\Phi_L= {\tilde
\Phi}_R=0.$ The curves are: ${\tilde \Phi}_L =0$ (heavy solid),
${\tilde \Phi}_L =\pi /8$ (heavy dashed), ${\tilde \Phi}_L=\pi/4$
(light solid), ${\tilde \Phi}_L=\pi/3$ (light dashed).}
\label{fig:chargino16}
\end{figure}
\begin{figure} \centering
\begin{picture}(31.5,21)
\put(1,2){\includegraphics[width=70mm]{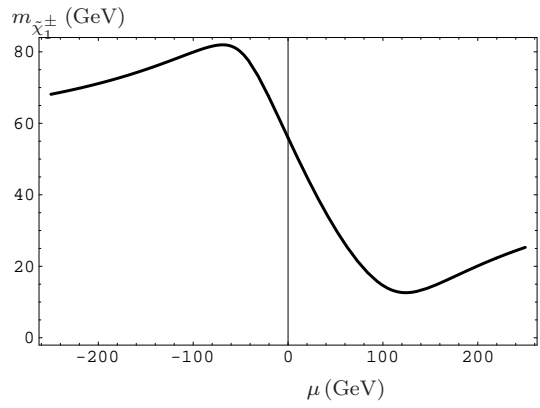}}
\put(15.75,0.7){\scriptsize{$\mu \, {\rm (GeV)}$}}
\put(1.05,19.25){\scriptsize{$m_{{\tilde \chi}^{\pm}_{1}} \, {\rm
(GeV)}$}}
\end{picture}
\caption{Chargino masse $m_{{\tilde \chi}^{\pm}_{1}} $ as a function
of $\mu,$ computed from Eq. \eqref{eq:mass-char-MSSM} in the context
of the MSSM, assuming $m_W=80.419$ GeV, $M_2=50$ GeV and
$\tan\beta=1.6.$} \label{fig:charMSSM-vs-mu-2006}
\end{figure}

Note that the irregularities in the plots of Figs.
\ref{fig:chargino15} and \ref{fig:chargino16}, produces because the
curve representing the variation with respect to $\mu$ of the factor
$\sqrt{\beta - {\bar w} - \lambda / 4 \alpha}$ in Eq.
\eqref{eq:EIGV}, take the forme of a ${\cal W},$ approximately,
whereas the curve representing the variation with respect to $\mu$
of the factor $ a/4 - \alpha /2$ in that equation take the forme of
a hyperbolic branch, approximately, both curves with the
ordinate-axis as its symmetry axis. The hyperbolic branch being
inscribed in the ${\cal W}$ curve approach asymptotically to this
one for positive values of $\mu$ and go away from this one, after
practically intersect it, for negative values of $\mu.$ Thus, in
this case, we can say that the change in the mixing character is
determined  by the factor $\sqrt{\beta - {\bar w} - \lambda / 4
\alpha}$ and tuned and displaced by the factor $ a/4 - \alpha /2.$
In the corresponding case of the MSSM, the tuning factor in Eq.
\eqref{eq:mass-char-MSSM} take the form of a parabola and the change
in the mixing character, determined in combination with the factor
$\Delta_C /2,$ is smooth, see Fig \ref{fig:charMSSM-vs-mu-2006}. We
don't observe an important displacement of the curve with respect to
the ordinate-axis as in the case of the LR-SUSY model. Also, for
large values of $|\mu|,$ we observe an inverted mixing behaviour
with respect to the one observed in the L-R SUSY model.

\subsubsection{Chargino  mass $m_{{\tilde \chi}^{\pm}_{1}}$ as a function  of \boldmath
$\tan\theta_k$}
\begin{table}[h]
\begin{center}
\begin{tabular}{c c c c c c c}\cline{1-7}\\
Scenario & $  \; \Phi_L= {\tilde \Phi}_L= {\tilde \Phi}_R $ \; &  $M_R \; $ & $ |M_L| \; $ & $|\mu| \;$ & $ M_{W_L}=m_{W} $ & $ M_2 $ \\
\\\cline{1-7} \\
$Scpv_2$ & \parbox[c]{0.5cm}{0} & 300 & 152  & 248 & 80.456 &   152  \\ \\  \hline \\
\end{tabular}
\end{center}
\caption{Input parameters for scenario $Scpv_2.$ All mass quantities
are given in GeV.} \label{tab:tabla-LR-MSSM}
\end{table}
In this section we investigate the behaviour of the lightest
chargino mass when the ratio $k_u / k_d = \tan\theta_k$ varies,  and
we compare with the behaviour of this mass in the context of the
MSSM when $\tan\beta$ is allowed to vary.

 Let us consider the scenario $Scpv_2,$
described in Tab. \ref{tab:tabla-LR-MSSM}, where in addition to the
the input parameters for the L-R SUSY model, we include the input
parameters representing similar conditions in the context of the
MSSM. Figure \ref{fig:charLRSUSY} shows the curves describing the
variation of the chargino mass $m_{{\tilde \chi}^{\pm}_{1}}$ (in the
context of the L-R SUSY model) with respect to $\tan\theta_k$ for
input parameters of scenario $Scpv_2$ when
$\Phi_\mu=0,\pi/4,\pi/2,3\pi/4,\pi$ (the curves are distingued by
its grey level, from the darkest one ($\Phi_\mu =0$) to the lightest
one ($\Phi_\mu =\pi$), respectively). We observe that, for values of
$0 \le \Phi_\mu < \pi /2, $ the chargino mass decrease when
$\tan\theta_k$ increase whereas for values of $\pi / 2 < \Phi_\mu
\le \pi, $ there is a change of the mixing character, i.e., the
chargino mass increase when $\tan\theta_k$ grows. For small values
of $\Phi_\mu$ and $\theta_k,$ the chargino mass $m_{{\tilde
\chi}^{\pm}_{1}}$ is of order of $|M_L|.$ For large  values of
$\tan\theta_k$ and for all $\Phi_\mu,$ the values of $m_{{\tilde
\chi}^{\pm}_{1}}$ approach to the common value $\approx 136$GeV,
i.e., the value corresponding to the phase angle $\Phi_\mu= \pi /2.$
In general, in this scenario the value of  $m_{{\tilde
\chi}^{\pm}_{1}}$ lies in the range $112$GeV-$152$GeV.

Figure \ref{fig:charMSSM} shows curves describing the variation of
the chargino mass $m_{{\tilde \chi}^{\pm}_{1}}$ (in the context of
the MSSM) with respect to $\tan\beta$ for input parameters of
scenario $Scpv_2,$ when $\Phi_\mu=0,\pi/4,\pi/2,3 \pi/4,\pi$ (
again, the curves are distinguished by its grey level, from the
darkest one ($\Phi_\mu =0$) to the lightest one ($\Phi_\mu =\pi$),
respectively). Comparing these curves with those of the LR-SUSY
model, we observe an opposite change rate of the chargino mass
$m_{{\tilde \chi}^{\pm}_{1}}$  with respect to $\tan\beta,$
approximately, into the same intervals of variation of $\Phi_\mu$
described in the previous paragraph. This last fact illustrates  the
way as the parameter $\Phi_\mu$ provides of  an inverted mixing
character to both models (a phase difference of $\pi,$ approximately
) .

In the MSSM the order of magnitude of the chargino mass $m_{{\tilde
\chi}^{\pm}_{1}}$ is dominated by the parameter $M_2,$ whereas in
the LR-SUSY model this magnitude is determined by both the $|M_L|$
and $M_R$ parameters. For instance, if we  take $M_R=100$GeV in
place of $300$GeV in Tab. \ref{tab:tabla-LR-MSSM}, then we find that
the values  of $m_{{\tilde \chi}^{\pm}_{1}}$ lies  in the range
$70$GeV-$110$GeV.

\begin{figure} \centering
\begin{picture}(31.5,21)
\put(1,2){\includegraphics[width=70mm]{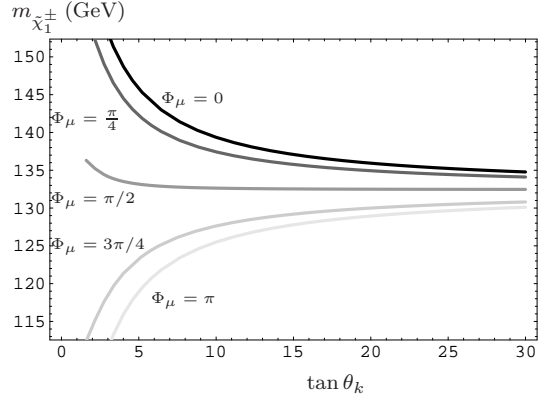}}
\put(15.75,0.7){\scriptsize{$\tan\theta_k$}}
\put(8.5,15){\tiny{$\Phi_\mu=0$}}
\put(3,14){\tiny{$\Phi_\mu={\pi\over 4}$}}
\put(3,10){\tiny{$\Phi_\mu=\pi/2$}} \put(3,7.7){\tiny{$\Phi_\mu=3
\pi/4$}} \put(8,5){\tiny{$\Phi_\mu=\pi$}}
\put(1.05,19.25){\scriptsize{$m_{{\tilde \chi}^{\pm}_{1}} \, {\rm
(GeV)}$}}
\end{picture}
\caption{Chargino mass $m_{{\tilde \chi}^{\pm}_{1}}$ as a function
of $\tan\theta_k$ for input parameters of  scenario $Scpv_2,$ in the
context of LR-SUSY model.} \label{fig:charLRSUSY}
\end{figure}

\begin{figure} \centering
\begin{picture}(31.5,21)
\put(1,2){\includegraphics[width=70mm]{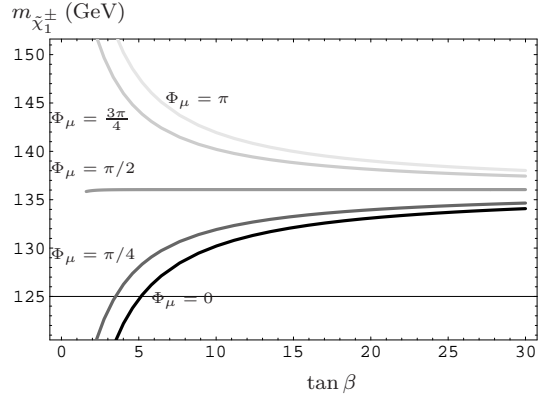}}
\put(8.7,15){\tiny{$\Phi_\mu=\pi$}} \put(3,14){\tiny{$\Phi_\mu={3
\pi\over 4}$}} \put(3,11.5){\tiny{$\Phi_\mu=\pi/2$}}
\put(3,7.2){\tiny{$\Phi_\mu= \pi/4$}} \put(8,5){\tiny{$\Phi_\mu=0$}}
\put(15.75,0.7){\scriptsize{$\tan\beta$}}
\put(1.05,19.25){\scriptsize{$m_{{\tilde \chi}^{\pm}_{1}} \, {\rm
(GeV)}$}}
\end{picture}
\caption{Chargino mass $m_{{\tilde \chi}^{\pm}_{1}}$ as a function
of $\tan\beta$ for input parameters of  scenario $Scpv_2,$ in the
context of MMSM.} \label{fig:charMSSM}
\end{figure}

\section{Construction of  the unitary $U$ and $V$ matrices}
\label{sec-unitary-U-V} The diagonalizing matrix $V$ can be obtained
by computing the eigenvectors $v_j$ of $H\equiv M^\dagger M,$
corresponding to the eigenvalues $m_{{\tilde \chi}^{\pm}_{j}} $
given in Eqs. \eqref{eq:EIGU} and \eqref{eq:EIGV}: \be \label{eq:vj}
v_j =\begin{pmatrix}  V_{1j} & V_{2j} & V_{3j} & V_{4j} & \cr
\end{pmatrix}^T, \quad j=1,\ldots,4.\ee
 Following the results in Ref.\cite{key11}, we get the $V_{ij}$ entries   \be
\label{eq:EVEVij} V_{ij} = {\Delta_{ij} \over \Delta_{1j}} \, {|
\Delta_{1j} | \, e^{i \theta_j}\over \sqrt{| \Delta_{1j} |^2 + |
\Delta_{2j} |^2 | + |\Delta_{3j} |^2 | + |\Delta_{4j} |^2}}, \ee
when i=1,\ldots,4. Here \be \label{eq:delta1j} \Delta_{1j}(H) =
\begin{vmatrix} H_{22} - m_{{\tilde \chi}^{\pm}_{j}}^2  & H_{23}&
H_{24}\cr H_{32} & H_{33} - m_{{\tilde \chi}^{\pm}_{j}}^2 &
H_{34}\cr H_{42} &H_{43} &H_{44} - m_{{\tilde \chi}^{\pm}_{j}}^2
\cr
\end{vmatrix}
\ee and $\Delta_{ij} (H), \, i=2,3,4,$ is formed from
$\Delta_{1j}(H)$ by substituting  the $(i-1)$th  column by $\bigl(
\begin{smallmatrix}- H_{21} \cr - H_{31} \cr  - H_{41} \cr \end{smallmatrix}\bigr).$

In the same way, the eigenvectors $u^\ast_j$ forming the $U^\ast$
matrix are given by  \be \label{eq:uj} u^\ast_j =
\begin{pmatrix}  U^\ast_{1j} & U^\ast_{2j} & U^\ast_{3j} & U^\ast_{4j} & \cr
\end{pmatrix}^T, \quad j=1,\ldots,4, \ee where the  $U^\ast_{ij}$ matrix's elements
are given by  \be \label{eq:EVEUij} U^\ast_{ij} = {{\tilde
\Delta}_{ij} \over {\tilde \Delta_{1j}}} \, {| {\tilde \Delta}_{1j}
| \, e^{- i {\tilde \theta}_j}\over \sqrt{| {\tilde \Delta}_{1j} |^2
+ | {\tilde \Delta}_{2j} |^2 | + |{\tilde \Delta_{3j}} |^2 | +
|{\tilde \Delta}_{4j} |^2}}, \ee when i=1,\ldots,4. Here, ${\tilde
\Delta}_{ij} ({\tilde H}) \equiv \Delta_{ij} (\tilde H ),
i,j=1,\ldots,4. $

At this point it is pertinent to clarify that considering both  the
particular values taken by the entries of the  $H$ matrix, namely
$H_{31}=H_{32}=H_{34}=0$ and $H_{33}=|\mu|^2,$ and the particular
value taken by one of the physical chargino masses $m_{{\tilde
\chi}^\pm_j}= |\mu|,$ when $j=1,2, 3$ or $4,$ (see Appendix
\ref{sec-chargino-MASS}), so for a given $j$ the corresponding
$\Delta_{ij}, i=1,2,\ldots,4, $ factors defined in Eq.
\eqref{eq:delta1j} become zero, i.e., the $V$ matrix becomes
singular.  This fact obliges  us to pay special attention at the
moment to compute the eigenvector associated to this particular
chargino mass value. Indeed, the usual theory of matrix eigenvalues
and eigenvectors provides us the tools we need to remedy this
situation. The same is valid  for the ${\tilde \Delta}_{ij}$
factors. Hereafter we suppose, without any loss of generality
(concerning the general method) that we work in a region of the
fundamental parameter space where $m_{{\tilde \chi}^\pm_3}=|\mu|.$
For other regions where a different chargino has mass equal to
$|\mu|,$ the subsequent analysis is modified only slightly to the
extent that role of the old $j=3$ index is played by the new $j$
index associated to the new fixed chargino mass. The general
formulation that we present here allows us to do it without any
problem.

Continuing with our arguments,  both the phases $ \theta_j $'s in
Eq. \eqref{eq:EVEVij} and the phases  ${\tilde \theta}_j$'s in Eq.
\eqref{eq:EVEUij} are arbitrary, they  will be fixed by the
requirement that $U$ and $V$ satisfy Eq. \eqref{eq:VMV}.  As we will
see in the next section, these phases are related to the so-called
CP eigenphases \cite{key11,key13}.

On the other hand,  as  $v_j$  is an eigenvector of  $H$  associated
to the eigenvalue $m_{{\tilde \chi}^{\pm}_{j}}, $ i.e.,  \be M^\dag
M v_j = m_{{\tilde \chi}^{\pm}_{j}} v_j, \ee then multiplying both
sides of this equation  by  $M,$ we get \be
 M M^\dag \, (M  v_j)  =  m_{{\tilde
\chi}^{\pm}_{j}} \, (M v_j), \ee i.e,  $M v_j$ is an eigenvector of
$\tilde H$ corresponding to the same eigenvalue $m_{{\tilde
\chi}^{\pm}_{j}}.$  Thus, according to Eq. \eqref{eq:MD2} and taking
in account the unitary character of $U^\ast,$ we can show that the
normalized eigenvectors $u^\ast_j$ forming this matrix  are given by
\be u^\ast_j = {1 \over m_{{\tilde \chi}^{\pm}_{j}} } M v_j,
\label{eq:ujvj}\ee when $j=1,\ldots,4.$ Thus, substituting Eq.
\eqref{eq:vj} in  Eq. \eqref{eq:ujvj} and comparing with  Eq.
\eqref{eq:uj}, we get  \be \label{eq:UIJ} U^\ast_{ij}= {1 \over
m_{{\tilde \chi}^{\pm}_{j}} } \sum_{k=1}^4 M_{i k} V_{kj}. \ee
Equations  \eqref{eq:EVEUij} and \eqref{eq:UIJ} represent two
equivalent ways to write the entries of the $U^\ast$ matrix.

On the other  hand, by the same arguments we get    \be
\label{eq:VIJ} V_{ij}= {1 \over m_{{\tilde \chi}^{\pm}_{j}} }
\sum_{k=1}^4 M^\dagger_{i k} U^\ast_{kj}. \ee Thus, equations
\eqref{eq:EVEVij} and \eqref{eq:VIJ} represent two equivalent ways
to write the entries of the $V$ matrix.

\section{The chargino projectors, reduced projectors, pseudoprojectors  and CP eigenphases}
\label{sec-projectors} In this section we show how to implement the
projector formalism \cite{key16} to describe the chargino
observables in the L-R SUSY model.

It has been demonstrated that the projector formalism is very
efficient to describe the neutralino observables in both the the
MSSM \cite{key13} and the L-R SUSY model \cite{key11}, where the
neutralino mass matrix was described by a real or complex $4\times
4$ symmetric matrix. So in both, the CP-conserving and CP-violating
cases, this formalism represents a systematic method to express the
fundamental parameters of the neutralino sector in terms of some
basic quantities, namely the reduced projectors and eigenphases.
Conversely, these last quantities can be expressed in terms of the
fundamental parameters and physical masses in a direct manner. Other
important characteristic of this formalism is that it can be easily
generalized to any neutralino number.

Concerning the chargino sector, in the MSSM we must deal with a  $2
\times 2$ non-symmetric matrix. The problem of determining the
fundamental chargino parameters including all possible CP-violating
phases into the chargino mixing matrix has been completely solved by
standard diagonalization matrix methods, see for instance
\cite{key17,key18} an references therein. In the case of the LR-SUSY
model, where we must deal with a $4 \times 4$ non-symmetric mass
matrix (or a $5 \times 5,$ if we consider the contribution of the
charged right-handed higgsino fields ${\tilde \Delta}_R$ and
${\tilde \delta}_R$), it is suitable to dispose of a method that
allows us to disentangle the parameters in a systematic manner.

 For a complete description of the chargino observables, in addition to
the projectors matrices, it is also necessary to compute the
so-called pseudoprojector matrices and CP eigenphases. In the
following we implement a method  based on the knowledge of the
general structure of the diagonalizing  matrices $U$ and $V$ to
obtain these quantities and demonstrate some of their properties.

Since two unitary matrices are needed to diagonalize $M,$ we must
define two classes of projector matrices $P^U$ and $P^V$ associated
to the diagonalizing matrices $U$ and $V,$ respectively, and a class
of pseudoprojector matrix ${\bar P}$.  They are defined
as\cite{key16}: \be \label{eq:PJU}
P^{U^\ast}_{j}={(P^{U^\ast}_j)}^{\dag}= U^\ast \, E_{j}\,U^T, \ee
\be \label{eq:PJV} P^V_{j}={(P^V_j)}^{\dag}= V \, E_{j}\,V^{-1}, \ee
\be \label{eq:PJUV} {\bar P}_{j}= U^\ast \, E_{j} \,V^{-1}, \ee so
that \be \label{eq:PjUUC} P^{U^\ast}_{j\alpha \beta} =
U^\ast_{\alpha j} \, U_{\beta j}, \ee \be \label{eq:PjVVC} P^V_{j
\alpha \beta} = V_{\alpha j} \, V^\ast_{\beta j}, \ee
\be\label{eq:PjUVC} {\bar P}_{j \alpha \beta} = U^\ast_{\alpha j} \,
V^\ast_{\beta j}. \ee
These projectors  and pseudoprojectors satisfy the relations
\be \label{eq:proj-U_properties}
P^{U^\ast}_{i}\,P^{U^\ast}_{j}=P^{U^\ast}_{j}\,\delta_{ij},\quad
Tr\,P^{U^\ast}_{j}=1,\quad\sum_{j=1}^{4}\,P^{U^\ast}_{j}=1, \ee \be
\label{eq:proj-V-properties}
P^V_{i}\,P^V_{j}=P^V_{j}\,\delta_{ij},\quad
Tr\,P^V_{j}=1,\quad\sum_{j=1}^{4}\,P^V_{j}=1, \ee

\be {\bar P}_j {\bar P}_j^\dag = P^{U^\ast}_j, \qquad {\bar
P}_j^\dag {\bar P}_j = P^V_j, \ee

\be  P^{U^\ast}_j {\bar P}_j = {\bar P}_j P^V_j= {\bar P}_j. \ee

Note that from  Eqs. \eqref{eq:MD2}, \eqref{eq:PJU} and
\eqref{eq:PJV} it is possible to write \be \label{eq:MMdU} M
\,M^\dag =\sum_{j=1}^{4}\,m_{{\tilde
\chi}^{\pm}_{j}}^{2}\,P^{U^\ast}_{j} \ee and \be \label{eq:MMdV}
M^{\dag}\,M=\sum_{j=1}^{4}\,m_{{\tilde
\chi}^{\pm}_{j}}^{2}\,P^V_{j}. \ee  Also, from  Eqs. \eqref{eq:VMV}
and \eqref{eq:PJUV}, we can write \be
 M = \sum_{j=1}^{4}\,m_{{\tilde \chi}^{\pm}_{j}}\,  U^\ast \,
 \,
E_{j} \,  V^{-1} = \sum_{j=1}^{4}\,m_{{\tilde \chi}^{\pm}_{j}}\,
{\bar P}_j. \label{eq:MVtVd} \ee

\subsection{Reduced projectors}
The projector and pseudoprojector matrices can be expressed in terms
of the most fundamental ones, the so-called reduced projectors
\cite{key13}. Indeed, by inserting \eqref{eq:EVEVij} into
\eqref{eq:PjVVC},
 we get \be \label{eq:CPjalpha-V} P^V_{j \alpha
\beta} ={ p^\ast_{j \alpha} p_{j \beta} \over |p_{j1}|^2 +
|p_{j2}|^2 + |p_{j3}|^2 + |p_{j4}|^2},\ee where we have  define the
type-$V$ reduced projectors \cite{key11} \be \label{eq:redu-proj-V}
p_{j \alpha} \equiv {\Delta^\ast_{\alpha j} (H) \over
\Delta^\ast_{1j} (H)}. \ee  Thus, from Eq. \eqref{eq:CPjalpha-V} we
deduce \be \label{eq:PVj11}
 P^V_{j11} ={ 1  \over |p_{j1}|^2 + |p_{j2}|^2 +
|p_{j3}|^2 + |p_{j4}|^2}. \ee Inserting this last result into
\eqref{eq:CPjalpha-V},  we  get  \be P^V_{j \alpha \beta} =
P^V_{j11} \, p^\ast_{j \alpha} \, p_{j \beta}.
\label{clave-Jarlskog-V}\ee

In the same way, substituting \eqref{eq:EVEUij} into
\eqref{eq:PjUUC}, we get \be \label{eq:CPjalpha-U} P^{U^\ast}_{j
\alpha \beta} ={ {\tilde p}^\ast_{j \alpha} {\tilde p}_{j \beta}
\over |{\tilde p}_{j1}|^2  + |{\tilde p}_{j2}|^2 + |{\tilde
p}_{j3}|^2 + |{\tilde p}_{j4}|^2},\ee where we have define the
type-$U^\ast$ reduced projectors \be \label{eq:redu-proj-U} {\tilde
p}_{j \alpha} \equiv {{\tilde \Delta}^\ast_{\alpha j} ({\tilde
H})\over {\tilde \Delta}^\ast_{1j}({\tilde H}) }. \ee  Thus, from
Eq. \eqref{eq:CPjalpha-U} we deduce \be \label{eq:PUj11}
 P^{U^\ast}_{j11} ={ 1  \over  |{\tilde p}_{j1}|^2 + |{\tilde p}_{j2}|^2 +
|{\tilde p}_{j3}|^2 + |{\tilde p}_{j4}|^2}. \ee Inserting this last
result into \eqref{eq:CPjalpha-U},  we  get  \be P^{U^\ast}_{j
\alpha \beta} = P^{U^\ast}_{j11} \,  {\tilde p}^\ast_{j \alpha} \,
{\tilde p}_{j \beta}. \label{clave-Jarlskog-U}\ee

On the other hand,  using  Eqs. \eqref{eq:redu-proj-V} and
\eqref{eq:PVj11}, we can express the entries of the diagonalizing
matrix $V$ given in Eq. \eqref{eq:EVEVij} in terms of the type-$V$
reduced projectors, that is \be \label{eq:Vaj} V_{\alpha j} =
\sqrt{P^V_{j11} \over \eta_j}  p^\ast_{j \alpha}, \ee where $\eta_j
\equiv  e^{- 2 i \theta_j }$ stands for the type-$V$ CP eigenphases.
Similarly, using Eqs. \eqref{eq:redu-proj-U} and \eqref{eq:PUj11},
we can express  the entries  of the diagonalizing matrix $U^\ast$
given in Eq. \eqref{eq:EVEUij} in terms of the type-$U^\ast$ reduced
projectors, that is \be \label{eq:Uaj} U^\ast_{\alpha j} =
\sqrt{P^{U^\ast}_{j11} \over {\tilde \eta}_j} {\tilde p}^\ast_{j
\alpha}, \ee where ${\tilde \eta}_j \equiv  e^{2 i {\tilde \theta}_j
}$ stands for the type-$U^\ast$ CP eigenphases.

Note that the reduced projector are not independent. As $V$ and
$U^\ast$ are unitary matrices, using
 Eqs. \eqref{eq:Vaj} and \eqref{eq:Uaj}, we get the  constraints
\be \sum_{k=1}^{4} p_{ik} p^\ast_{jk}=0 \label{eq:sysV} \ee and \be
\sum_{k=1}^{4} {\tilde p}_{ik} {\tilde p}^\ast_{jk}=0,
\label{eq:tsysU}\ee when $i,j=1,\ldots,4,    \diagup \, \, i>j.$
Eqs. \eqref{eq:sysV} and \eqref{eq:tsysU} represents each one a
system of six complex algebraic equations serving to reduce, in the
most general case, up to twelve the number of real independent
parameters on each set of reduced projectors.

Inserting Eqs. \eqref{eq:Vaj} and \eqref{eq:Uaj} into Eq.
\eqref{eq:PjUVC} we can express the pseudoprojector entries in terms
of the reduced projectors and eigenphases $\zeta_j = \sqrt{\eta_j
/{\tilde \eta}_j}:$  \be\label{eq:pseudoPab} {\bar P}_{j \alpha
\beta} = \zeta_j \, \sqrt{P^{U^\ast}_{j11} P^V_{j11}} \, {\tilde
p}^\ast_{j \alpha} \, p_{j \beta}. \ee Moreover, substituting Eq.
\eqref{eq:Vaj}  into Eq. \eqref{eq:UIJ}, we obtain \be
\label{eq:Uaj-Vk} U^\ast_{\alpha j} = {1 \over m_{{\tilde
\chi}^{\pm}_{j}}} \sqrt{P^V_{j11} \over \eta_j} \sum_{k=1}^4
M_{\alpha k} p^\ast_{j k}, \ee so from the equivalence between Eqs.
\eqref{eq:Uaj} and \eqref{eq:Uaj-Vk}, we deduce  ($\alpha
=1,\ldots,4$) \be \label{eq:gen-inversionU} m_{{\tilde
\chi}^{\pm}_{j}} \, \zeta_j \, \sqrt{P^{U^\ast}_{j11}\over
P^V_{j11}} \, {\tilde p}^\ast_{j \alpha} = \sum_{\beta=1}^4
M_{\alpha \beta} \, p^\ast_{j \beta}. \ee On the other hand,
following the same procedure, from the corresponding equivalence for
the entries of the $V$ matrix,  we get ($\alpha =1,\ldots,4$) \be
\label{eq:gen-inversionV} m_{{\tilde \chi}^{\pm}_{j}} \, \zeta_j \,
\sqrt{ P^V_{j11} \over P^{U^\ast}_{j11}} \, p_{j \alpha} =
\sum_{\beta=1}^4 M_{\beta \alpha} \, {\tilde p}_{j \beta}. \ee

Equations \eqref{eq:gen-inversionU} and \eqref{eq:gen-inversionV}
constitute a generalization of the corresponding formulas deduced in
the case of the neutralino \cite{key11,key13} which was based on a
complex symmetric mass matrix. These equations represent, for fixed
$j,$ a system of eight complex algebraic equations serving to
determine the fundamental parameters of the model in terms of the
reduced projectors, the chargino physical masses $m_{{\tilde
\chi}^\pm_j }$ and the eigenphases, or vice versa.

Note that, in Eqs. \eqref{eq:Vaj} and \eqref{eq:Uaj} as well as in
Eqs. \eqref{eq:gen-inversionU} and \eqref{eq:gen-inversionV},
without any loss of generality,  we could choose  the $U^\ast$-type
eigenphases either ${\tilde \eta_j}=1, j=1,\ldots,4,$ such that
$\zeta_j= \sqrt{\eta_j}$;  ${\tilde \eta}^\ast_j = \eta_j,
j=1,2,\ldots,4,$ such $\zeta_j=\eta_j,$ or any other suitable choice
allowing us to eliminate four superfluous parameters.

\subsection{Explicit form of the reduced projectors}
In general, according to Eqs. \eqref{eq:redu-proj-V} and
\eqref{eq:redu-proj-U}, to obtain the explicit form of the reduced
projectors of the $V$-type and $U^\ast$-type, in terms of the
fundamental parameters of the theory, we only need to know the
explicit form of quantities $ \Delta_{\alpha j}^\ast$ and  $ {\tilde
\Delta}_{\alpha j}^\ast,$ respectively.  For fixed $j=1,2,4$ the
quantities of the $V$-type are given by \beqa\nonumber
\Delta^\ast_{1 j} &=&  ( |\mu|^2 - m_{{\tilde \chi}^{\pm}_{j}}^2 )
\bigl\{( |\mu|^2 - m_{{\tilde
\chi}^{\pm}_{j}}^2 ) ( M_R^2 - m_{{\tilde \chi}^{\pm}_{j}}^2 ) \\
\nonumber  &-& 2 \cos^2\theta_k |{\tilde M}_L|^2  (m_{{\tilde
\chi}^{\pm}_{j}}^2 -
 M_R^2 - 2 |{\tilde M}_R|^2 \sin^2\theta_k)\\ \nonumber  &-&  2 |{\tilde M}_R|^2
\\ \nonumber &\times&\bigl[m_{{\tilde \chi}^{\pm}_{j}}^2 - M_R |\mu| \cos(2
{\tilde \Phi}_R - \Phi_\mu)\sin(2\theta_k)] \\ &+&  |{\tilde M}_R|^4
\sin^2(2\theta_k)\bigr\},
 \label{eq:cdelta1j-V}\eeqa
 \beqa \nonumber \Delta^\ast_{2 j} &=& \label{eq:cdelta2j-V}
2 |{\tilde M}_L|  |{\tilde M}_R| ( m_{{\tilde \chi}^{\pm}_{j}}^2 -
|\mu|^2 ) \\ \nonumber &\times& \bigl\{ \sin\theta_k \bigl[ |\mu|
M_R \cos\theta_k e^{i( \Phi_u - {\tilde \Phi}_L - {\tilde \Phi}_R)}
\\ \nonumber &+&  \sin\theta_k e^{i({\tilde \Phi}_R - {\tilde
\Phi}_L)} \\ \nonumber &\times& [ 2 (|{\tilde M}_L|^2  + |{\tilde
M}_R|^2)\cos^2\theta_k  - m_{{\tilde \chi}^{\pm}_{j}}^2 ] \bigr] \\
\nonumber &+& |M_L| \cos\theta_k \bigl[ |\mu| \sin\theta_k  e^{- i(
\Phi_u + \Phi_L - {\tilde \Phi}_L - {\tilde \Phi}_R)} \\  &-& M_R
\cos\theta_k e^{- i( \Phi_L - {\tilde \Phi}_L + {\tilde \Phi}_R)}
\bigr] \bigr\}, \eeqa \beqa  \Delta^\ast_{3 j} &=&0
\label{eq:cdelta3j-V} \eeqa and
 \beqa \nonumber \Delta^\ast_{4 j} &=& \sqrt{2} |{\tilde M}_L|( m_{{\tilde \chi}^{\pm}_{j}}^2 -
|\mu|^2 ) \\ \nonumber &\times& \bigl\{ |M_L| \cos\theta_k
e^{i({\tilde \Phi}_L - \Phi_L   )} \\ \nonumber &\times& [M_R^2 -
m_{{\tilde \chi}^{\pm}_{j}}^2 + 2 |{\tilde M}_R|^2 \sin^2\theta_k] +
\sin\theta_k \\ \nonumber &\times& \bigl[|\mu| ( m_{{\tilde
\chi}^{\pm}_{j}}^2 - |\mu|^2 ) e^{i( \Phi_\mu - {\tilde \Phi}_L )}
\\ &-& M_R |{\tilde M}_R|^2 \sin(2\theta_k) e^{i( 2 {\tilde \Phi}_R
-  {\tilde \Phi}_L )}\bigr] \bigr\} \label{eq:cdelta4j-V},\eeqa and
the ones of the $U^\ast$-type are given by \beqa\nonumber {\tilde
\Delta}^\ast_{1 j} &=&  ( |\mu|^2 - m_{{\tilde \chi}^{\pm}_{j}}^2 )
\bigl\{( |\mu|^2 - m_{{\tilde
\chi}^{\pm}_{j}}^2 ) ( M_R^2 - m_{{\tilde \chi}^{\pm}_{j}}^2 ) \\
\nonumber  &-& 2 \sin^2\theta_k |{\tilde M}_L|^2  (m_{{\tilde
\chi}^{\pm}_{j}}^2 -
 M_R^2 - 2 |{\tilde M}_R|^2 \cos^2\theta_k)\\ \nonumber  &-&  2 |{\tilde M}_R|^2
\\ \nonumber &\times&\bigl[m_{{\tilde \chi}^{\pm}_{j}}^2 - M_R |\mu| \cos(2
{\tilde \Phi}_R - \Phi_\mu)\sin(2\theta_k)] \\ &+&  |{\tilde M}_R|^4
\sin^2(2\theta_k)\bigr\},
 \label{eq:cdelta1j-U}\eeqa
 \beqa \nonumber {\tilde \Delta}^\ast_{2 j} &=& \label{eq:cdelta2j-U}
2 |{\tilde M}_L|  |{\tilde M}_R| ( m_{{\tilde \chi}^{\pm}_{j}}^2 -
|\mu|^2 ) \\ \nonumber &\times& \bigl\{ \cos\theta_k \bigl[ |\mu|
M_R \sin\theta_k e^{- i( \Phi_u - {\tilde \Phi}_L - {\tilde
\Phi}_R)}
\\ \nonumber &+&  \cos\theta_k e^{i({\tilde \Phi}_L - {\tilde
\Phi}_R)} \\ \nonumber &\times& [ 2 (|{\tilde M}_L|^2  + |{\tilde
M}_R|^2)\sin^2\theta_k  - m_{{\tilde \chi}^{\pm}_{j}}^2 ] \bigr] \\
\nonumber &+& |M_L| \sin\theta_k \bigl[ |\mu| \cos\theta_k  e^{i(
\Phi_u + \Phi_L - {\tilde \Phi}_L - {\tilde \Phi}_R)} \\  &-& M_R
\sin\theta_k e^{i( \Phi_L - {\tilde \Phi}_L + {\tilde \Phi}_R)}
\bigr] \bigr\}, \eeqa
 \beqa \nonumber {\tilde \Delta}^\ast_{3 j} &=& \sqrt{2} |{\tilde M}_L|( m_{{\tilde \chi}^{\pm}_{j}}^2 -
|\mu|^2 ) \\ \nonumber &\times& \bigl\{ |M_L| \sin\theta_k e^{-
i({\tilde \Phi}_L - \Phi_L   )} \\ \nonumber &\times& [M_R^2 -
m_{{\tilde \chi}^{\pm}_{j}}^2 + 2 |{\tilde M}_R|^2 \cos^2\theta_k] +
\cos\theta_k \\ \nonumber &\times& \bigl[|\mu| ( m_{{\tilde
\chi}^{\pm}_{j}}^2 - |\mu|^2 ) e^{- i( \Phi_\mu - {\tilde \Phi}_L )}
\\ &-& M_R |{\tilde M}_R|^2 \sin(2\theta_k) e^{- i( 2 {\tilde \Phi}_R
-  {\tilde \Phi}_L )}\bigr] \bigr\} \label{eq:cdelta4j-U},\eeqa and
\beqa  {\tilde \Delta}^\ast_{4 j} &=&0 \label{eq:cdelta4j-U}. \eeqa
We note that the quantities ${\tilde \Delta}^\ast_{i j}, i=1,2$ can
be obtained from $\Delta^\ast_{i j}, i=1,2$ by interchanging
$\sin\theta_k \leftrightarrow \cos\theta_k$ and then taking the
complex conjugate. The quantity ${\tilde \Delta}_{3 j},$ can be
obtained from $\Delta^\ast_{4 j}$ in the same way.

As the delta factors becomes  singular when $j=3,$ because the
particular values taken by the entries of the
 $H$ and ${\tilde H}$ matrices  and the fact that
 $m_{{\tilde \chi}^\pm_3} =|\mu|$, see Appendix \ref{sec-chargino-MASS}, the $p_{j\alpha}$'s and
 ${\tilde p}_{j,\alpha}$'s, when $j =3,$  must be computed separately taking into account the
unitary character of the $V$ and $U^\ast$ matrices. Using
Eqs.\eqref{eq:Vaj} and \eqref{eq:Uaj} , we get $p_{33}={\tilde
p}_{34}=1, p_{3 \alpha}=p_{\alpha 3}={\tilde p}_{\alpha 4}=0, \,
\alpha=1,2,4,$ and ${\tilde p}_{3\alpha}=0, \, \alpha=1,2,3.$

The quantities given in Eqs.
(\ref{eq:cdelta1j-V}-\ref{eq:cdelta4j-U})  allow us to express,
through the reduced projectors, all the essential quantities of the
model in terms of the original parameters.

\subsection{Fundamental parameters in terms of the reduced projectors}
Equations \eqref{eq:gen-inversionU} and \eqref{eq:gen-inversionV}
allows us to perform a change of parametrization from the original
fundamental parameters to the reduced projectors and eigenphases.
This change of parametrization is suitable when we essay to fix the
value of the fundamental parameters by measuring  physical
observables which explicitly involves the entries of the
diagonalizing $U^\ast$ and $V$ matrices such as the
cross-section-type observables associated to the chargino pair
production through unpolarized or polarized $e^+ e^-$ annihilation
processes, for instance, total cross sections,  asymmetries, T-odd
asymmetries and polarization vectors. Indeed, in such a case, the
relevant quantities depends directly on the entries of the
diagonalizing $V$ and $U^\ast$ matrices, i.e., they depends in a
relative simple forme on the reduced projectors and eigenphases.
Thus, experimental measurements of this class of observables provide
us of independent constraints serving to determinate, in principle,
these eigenphases and reduced projectors, i.e., inserting  these
results into the relations described immediately below allows us to
reconstruct all the fundamental L-R SUSY parameters.

Recalling that $M_R$ is a real parameter, from Eqs.
\eqref{eq:gen-inversionU} and \eqref{eq:gen-inversionV}, we get
(j=1,2,4)
 \be \label{eq:ML-pj-etaj-thetaj} M_L = m_{{\tilde
\chi}^{\pm}_{j}} \, \zeta_j \, \, { \sqrt{P^{U^\ast}_{j11}\over
P^V_{j11}} {\tilde p}_{j3} \tan\theta_k - \sqrt{P^V_{j11} \over
P^{U^\ast}_{j11}} p_{j4}^\ast
  \over {\tilde p}_{j3} \tan\theta_k - p_{j4}^\ast},\ee \be \label{eq:MtL-pj-etaj-thetaj} {\tilde M}_L = - { m_{{\tilde \chi}^{\pm}_{j}} \,
\zeta_j \over \sqrt{2}} \, \, {\sqrt{P^{U^\ast}_{j11}\over
P^V_{j11}} - \sqrt{P^V_{j11} \over P^{U^\ast}_{j11}}  \over {\tilde
p}_{j3} \sin\theta_k - p_{j4}^\ast \cos\theta_k },   \ee \be
\label{eq:MtR-pj-etaj-thetaj} {\tilde M}_R = { m_{{\tilde
\chi}^{\pm}_{j}} \,   \zeta_j \over \sqrt{2}} \, \,
{\sqrt{P^{U^\ast}_{j11}\over P^V_{j11}} |{\tilde p}_{j2}|^2 -
\sqrt{P^V_{j11} \over P^{U^\ast}_{j11}} |p_{j2}|^2 \over p^\ast_{j4}
{\tilde p}_{j2} \cos\theta_k - p^\ast_{j2} {\tilde p}_{j3}
\sin\theta_k }, \ee \beqa \mu &=& {m_{{\tilde \chi}^{\pm}_{j}} \,
\zeta_j \over ( p_{j4}^\ast  - {\tilde p}_{j3} \tan\theta_k ) (
{\tilde p}_{j3} p^\ast_{j2}
\tan\theta_k -  {\tilde p}_{j2} p_{j4}^\ast)} \nonumber  \\
&\times& \biggl\{ \sqrt{P^{U^\ast}_{j11}\over P^V_{j11}} \left[
{\tilde p}_{j2} {\tilde p}_{j3}^\ast p_{j4}^\ast  - \left(
(p^\ast_{j2} - {\tilde p}_{j2}) |{\tilde p}_{j2}|^2 + p^\ast_{j2}
|{\tilde p}_{j3}|^2\right) \tan\theta_k \right] \nonumber \\ &+&
\sqrt{P^{V}_{j11}\over P^{U^\ast}_{j11}} \tan\theta_k \left[
p_{j2}^\ast {\tilde p}_{j3} p_{j4} \tan\theta_k - \left( ({\tilde
p}_{j2} - p^\ast_{j2}) |p_{j2}|^2 + {\tilde p}_{j2}
|p_{j4}|^2\right)\right] \biggr\}, \label{eq:mu-pj-etaj-thetaj}
 \eeqa \be \label{eq:MR-pj-etaj-thetaj} M_R = {\rm sign}({\mathrm{Re\,}}Z_j) \,  m_{{\tilde \chi}^\pm_j } \,  |Z_j|, \quad  if \, \, {\rm Re}
Z_j \ne 0, \qquad  M_R = \mp \, m_{{\tilde \chi}^\pm_j } \,
{\mathrm{Im\,}}Z_j,  \quad if \, \, {\mathrm{Re\,}} Z_j =0,
 \ee and \be \zeta_j = {\rm sign}({\mathrm{Re\,}} Z_j)
\, {Z_j^\ast \over |Z_j|}, \quad if \, \, {\rm Re} Z_j \ne 0, \qquad
\zeta_j = \pm i, \quad  if \, \, {\mathrm{Re\,}} Z_j =0,
\label{eq:zetaj-ReZ-non-zero} \ee where \be Z_j ={
\sqrt{P^{U^\ast}_{j11}\over P^V_{j11}} {\tilde p}_{j2}^\ast {\tilde
p}_{j3} \tan\theta_k - \sqrt{P^V_{j11} \over P^{U^\ast}_{j11}}
p_{j2} p_{j4}^\ast
  \over p_{j2}^\ast {\tilde p}_{j3} \tan\theta_k - {\tilde p}_{j2} p_{j4}^\ast}.\ee

When $j=3,$   we only   have \be \zeta_3=  - {\mu \over |\mu|}.
\label{eq:zetaj3} \ee

Note that in the CP-conserving case ${\mathrm{Im\,}}Z_j=0,$ and
$\Phi_\mu =0, \pi$ i.e., from Eqs. \eqref{eq:zetaj-ReZ-non-zero} and
\eqref{eq:zetaj3}, we get $\zeta_j=\pm 1, j=1,\ldots, 4.$ Thus, in
the CP-conserving case, there is not an analogous to the equation
\eqref{eq:zetaj-ReZ-non-zero} that allows us to express
$\tan\theta_k$ in terms of the eigenphases and reduced projectors,
so in order to express the fundamental parameters in terms of them
using Eqs.(\ref{eq:ML-pj-etaj-thetaj}-\ref{eq:MR-pj-etaj-thetaj}),
we must known $\tan\theta_k$ from some other means. In the
CP-conserving case, the role of the eigenphases is to remedy the
sign ambiguity of the physical chargino masses represented by the
eigenvalues of an Hermitian matrix $H$ or ${\tilde H},$ which can be
either positive or negative \cite{key12,key13}. In addition to this,
in the CP-violating case, the eigenphases contain information on the
complex phases introduced in the chargino mass matrix .

On the other hand, we note that using Eqs. \eqref{clave-Jarlskog-V},
\eqref{clave-Jarlskog-U} and \eqref{eq:pseudoPab}, we can express
all the fundamental parameters completely  in terms of the
$U^\ast$-, $V$-type projectors, $P^{U^\ast}_j, P^V_j$ and
pseudoprojectors, ${\bar P}_j.$

\subsection{Equivalent formulas for the reduced projectors}
\label{sec-Jarlskog-equivalence}  Using the Jarlskog's formulas
\cite{key16}, we can show that
 the projectors of the  type $V$ and $U^\ast$ can be written
 in the form \cite{key11,key12,key13}($j=1,2,4$) \be P^V_j = {{\tilde
P}^V_j \over {\tilde \Delta}_j}, \label{eq:ptilde-grande-V} \ee \be
P^{U^\ast}_j = {{\tilde P}^{U^\ast}_j \over {\tilde \Delta}_j},
\label{eq:ptilde-grande-U} \ee respectively,   where \be {\tilde
\Delta}_j = - 3 m_{{\tilde \chi}^{\pm}_{j}}^8 + 2 a \, m_{{\tilde
\chi}^{\pm}_{j}}^6 - b \, m_{{\tilde \chi}^{\pm}_{j}}^4 + d. \ee
Thus entries  of these projector matrices can be written in the form
\cite{key11} \beqa  {\tilde P}^V_{j \alpha \beta} &=&- m_{{\tilde
\chi}^{\pm}_{j}}^6 \, H_{\alpha \beta} +
m_{{\tilde \chi}^{\pm}_{j}}^4 \, (a \, H_{\alpha \beta} - H^2_{\alpha \beta} ) \nonumber \\
\nonumber &+& m_{{\tilde \chi}^{\pm}_{j}}^2 \, (a \,  H^2_{\alpha
\beta} - b \, H_{\alpha \beta} - H^3_{\alpha \beta}) \\ &+&  d \,
\delta_{\alpha \beta} \label{eq:ptildej-V} \eeqa and \beqa  {\tilde
P}^{U^\ast}_{j \alpha \beta} &=&- m_{{\tilde \chi}^{\pm}_{j}}^6 \,
{\tilde H}_{\alpha \beta} + m_{{\tilde \chi}^{\pm}_{j}}^4 (a \,
{\tilde
H}_{\alpha \beta} - {\tilde H}^2_{\alpha \beta} ) \nonumber \\
\nonumber &+& m_{{\tilde \chi}^{\pm}_{j}}^2 \, (a \,  {\tilde
H}^2_{\alpha \beta}
- b \, {\tilde H}_{\alpha \beta} - {\tilde H}^3_{\alpha \beta}) \\
&+& d  \, \delta_{\alpha \beta}, \label{eq:ptildej-U} \eeqa

Now, combining  Eqs. \eqref{clave-Jarlskog-V} and
\eqref{eq:ptilde-grande-V} and  Eqs.  \eqref{clave-Jarlskog-U} and
\eqref{eq:ptilde-grande-U}, we  deduce
\be\label{eq:redu-proj-standar-V}
 p_{j \alpha} = { P^V_{j 1 \alpha} \over  P^V_{j11} } = { {\tilde P}^V_{j 1 \alpha} \over  {\tilde P}^V_{j11} }
 \ee and
\be\label{eq:redu-proj-standar-U}
 {\tilde p}_{j \alpha} = { P^{U^\ast}_{j 1 \alpha} \over  P^{U^\ast}_{j11} } = { {\tilde P}^{U^\ast}_{j 1 \alpha} \over  {\tilde P}^{U^\ast}_{j11} }. \ee

Equations \eqref{eq:redu-proj-V} and \eqref{eq:redu-proj-standar-V}
are equivalent expressions.  Thus, combining these equations and
comparing the expressions (\ref{eq:cdelta2j-V}-\ref{eq:cdelta4j-V})
with the corresponding ${\tilde P}_{j 1 \beta}, \,  \beta=2,3,4,$
computed from Eq. \eqref{eq:ptildej-V}, we can show that
\be\label{eq:ptilde-delta-V} {\tilde P}^V_{j 1 \alpha} = m_{{\tilde
\chi}^{\pm}_{j}}^2 \,
 \Delta^\ast_{\alpha j}, \quad \forall \, \alpha=1,\ldots,4,  \ee
with

\beqa \nonumber {\tilde P}^V_{j 1 1} &=& - {1 \over 2} (m_{{\tilde
\chi}^{\pm}_{j}}^2 - |\mu|^2) \bigl\{ 4 |{\tilde M}_L|^2
\sin^2\theta_k \bigl[m_{{\tilde \chi}^{\pm}_{j}}^2  \\ &\times &
\nonumber  ( m_{{\tilde \chi}^{\pm}_{j}}^2 - M_R^2 - 2
|{\tilde M}_R|^2 \cos^2\theta_k) - 2 |{\tilde M}_L|^2 \\
\nonumber &\times& (m_{{\tilde \chi}^{\pm}_{j}}^2 - M_R^2)
\cos^2\theta_k\bigr] + 4 |M_L| |{\tilde M}_L|^2 \sin(2\theta_k)
\\ \nonumber &\times& \bigl[ M_R |{\tilde M}_R|^2 \cos(\Phi_L- 2 {\tilde
\Phi}_L + 2 {\tilde \Phi}_R) \sin(2\theta_k) \\\nonumber &-& |\mu|
(m_{{\tilde \chi}^{\pm}_{j}}^2 - M_R^2)\cos(\Phi_L- 2 {\tilde
\Phi}_L + \Phi_\mu)  \bigr] \\ \nonumber &+& 2 |M_L|^2 \bigl[
(m_{{\tilde \chi}^{\pm}_{j}}^2 - M_R^2)(m_{{\tilde
\chi}^{\pm}_{j}}^2 - |\mu|^2) - 2 |{\tilde M}_R|^2 \\ \nonumber
&\times& (m_{{\tilde \chi}^{\pm}_{j}}^2 - |\mu| M_R \cos(2{\tilde
\Phi}_R - \Phi_\mu) \sin(2\theta_k) \\& +& |{\tilde M}_R|^4
\sin^2(2\theta_k) \bigr]\bigl\}. \label{eq:ptilde11-V}\eeqa

In the same way, combining Eqs. \eqref{eq:redu-proj-U} and
\eqref{eq:redu-proj-standar-U}  and comparing the expressions
(\ref{eq:cdelta2j-U}-\ref{eq:cdelta4j-U}),
 with the corresponding ${\tilde P}^{U^\ast}_{j 1 \beta}, \beta=2,3,4,$
computed from Eq. \eqref{eq:ptildej-U}, we can show that
\be\label{eq:ptilde-delta-U} {\tilde P}^{U^\ast}_{j 1 \alpha} =
m_{{\tilde \chi}^{\pm}_{j}}^2 \,
 {\tilde \Delta}^\ast_{\alpha j}, \quad \forall \, \alpha=1,\ldots,4,  \ee
with \beqa \nonumber {\tilde P}^{U^\ast}_{j 1 1} &=& - {1 \over 2}
(m_{{\tilde \chi}^{\pm}_{j}}^2 - |\mu|^2) \bigl\{ 4 |{\tilde M}_L|^2
\cos^2\theta_k \bigl[m_{{\tilde \chi}^{\pm}_{j}}^2  \\ &\times &
\nonumber  ( m_{{\tilde \chi}^{\pm}_{j}}^2 - M_R^2 - 2
|{\tilde M}_R|^2 \sin^2\theta_k) - 2 |{\tilde M}_L|^2 \\
\nonumber &\times& (m_{{\tilde \chi}^{\pm}_{j}}^2 - M_R^2)
\sin^2\theta_k\bigr] + 4 |M_L| |{\tilde M}_L|^2 \sin(2\theta_k)
\\ \nonumber &\times& \bigl[ M_R |{\tilde M}_R|^2 \cos(\Phi_L- 2 {\tilde
\Phi}_L + 2 {\tilde \Phi}_R) \sin(2\theta_k) \\\nonumber &-& |\mu|
(m_{{\tilde \chi}^{\pm}_{j}}^2 - M_R^2)\cos(\Phi_L- 2 {\tilde
\Phi}_L + \Phi_\mu)  \bigr] \\ \nonumber &+& 2 |M_L|^2 \bigl[
(m_{{\tilde \chi}^{\pm}_{j}}^2 - M_R^2)(m_{{\tilde
\chi}^{\pm}_{j}}^2 - |\mu|^2) - 2 |{\tilde M}_R|^2 \\ \nonumber
&\times& (m_{{\tilde \chi}^{\pm}_{j}}^2 - |\mu| M_R \cos(2{\tilde
\Phi}_R - \Phi_\mu) \sin(2\theta_k) \\& +& |{\tilde M}_R|^4
\sin^2(2\theta_k) \bigr]\bigl\}. \label{eq:ptilde11-U}\eeqa

Note that Eq. \eqref{eq:ptilde-delta-V}, for $\alpha=2,3,4$
constitute an identity whereas for $\alpha=1,$ it constitutes an
equivalence. The same argument is valid for  Eq.
\eqref{eq:ptilde-delta-U}. Also, note that  ${\tilde P}^{U^\ast}_{j
1 1}$ can be obtained from ${\tilde P}^{V}_{j 1 1}$ by interchanging
$\sin\theta_k \leftrightarrow \cos\theta_k.$

\section{Disentangling L-R SUSY parameters}
\label{sec-ML-disentangling} In this section we use the results of
previous section to disentangle $M_L$ from the rest of the
fundamental parameters. We present two equivalent ways to express
the norm of $M_L,$ first in terms of the eigenphases and second in
terms of the phase angle $\Phi_L.$ Also, we propose an alternative
change of parametrization to disentangle $M_R$ from the rest of
parameters.

\subsection{Expressing \boldmath  $M_{L}$ in terms of the eigenphases}
\label{sec-genral-ml-eigenphases}
 From Eq. \eqref{eq:gen-inversionU} for $\alpha=1,$ considering that ${\tilde p}_{j1}=1, j=1,2,4,$  and using
 \eqref{eq:Mmatrix}, we get
\be   m_{{\tilde \chi}^{\pm}_{j}} \,  \zeta_j \, \sqrt{{\tilde
P}^{U^\ast}_{j11} \over {\tilde P}^V_{j11}} = M_{L} + \sqrt{2}
{\tilde M}_L \cos\theta_k p^\ast_{j4} . \label{eq:xij-mj-pj} \ee
Inserting Eqs. \eqref{eq:ptilde-grande-V} and
\eqref{eq:ptilde-grande-U} into Eq. \eqref{eq:xij-mj-pj} and taking
into account the equivalences given in Eqs.
\eqref{eq:ptilde-delta-V} and \eqref{eq:ptilde-delta-U} as well as
the fact that $p^\ast_{j4}=\Delta_{4j} / \Delta_{1j},$ we obtain \be
 m_{{\tilde \chi}^{\pm}_{j}} \, \zeta_j \, \sqrt{{\tilde \Delta}_{1j}
\over \Delta_{1j}} = M_{L} + \sqrt{2} {\tilde M}_L \cos\theta_k
{\Delta_{4j} \over \Delta_{1j}} . \label{eq:xij-mj-deltaj} \ee  Now,
substituting into \eqref{eq:xij-mj-deltaj} the values of
$\Delta_{1j}$ and $\Delta_{4j}$  given in Eqs. \eqref{eq:cdelta3j-V}
and \eqref{eq:cdelta4j-V}, respectively,  and solving a linear
algebraic equation for $M_{L},$ we get \be  M_{L} = {\tilde A}_j \,
\zeta_j + {\tilde B}_j, \qquad j=1,2,4, \label{genMLformula}\ee
where \be {\tilde A}_j= - {\sqrt{{\tilde \Delta}_{1j} \,
\Delta_{1j}} \over (m_{{\tilde \chi}^{\pm}_{j}}^2 - |\mu|^2)}
{m_{{\tilde \chi}^{\pm}_{j}} \over {\tilde {\cal D}}_j},
\label{eq:tildeAJ}\ee and \beqa \nonumber {\tilde B}_j&=&{ {\tilde
M}_L^2 \sin(2\theta_k) \over {\tilde {\cal D}}_j} \bigl[(m_{{\tilde
\chi}^{\pm}_{j}}^2 - M_R^2)|\mu| e^{ -i \Phi_\mu}
\\ &-& M_R |{\tilde M}_R|^2 e^{ -2i  {\tilde \Phi}_R}
\sin(2\theta_k)],\label{eq:tildeBJ}\eeqa where \beqa \nonumber
{\tilde {\cal D}}_j&=& |{\tilde M}_R|^4 \sin^2(2\theta_k) +
(m_{{\tilde \chi}^{\pm}_{j}}^2 - M_R^2) (m_{{\tilde
\chi}^{\pm}_{j}}^2 - |\mu|^2)\\ &-&  \nonumber 2 |{\tilde M}_R|^2 \\
&\times& \bigl[m_{{\tilde \chi}^{\pm}_{j}}^2 -|\mu| M_R \cos(2
{\tilde \Phi}_R- \Phi_\mu) \sin(2\theta_k) \bigr].
\label{eq:tilde-cal-dej} \eeqa

Equation \eqref{genMLformula} allows us to determinate the behaviour
of $|M_L|$ and $\Phi_L$ in terms of the eigenphases $\zeta_j$ and
the physical masses $m_{{\tilde \chi}^{\pm}_{j}},$ when the rest of
fundamental parameters are known.  The method used to obtain it is
direct and is based essentially on the fact that $\Delta_{4j} $ is a
linear function of $M_L$ and both $\Delta_{1 j}$ and
$\tilde{\Delta}_{1j} $ are independent of $|M_L|$ and $\Phi_L.$

Note that the same  result for $M_L$ can be obtained  if first we
solve the system of equations
(\ref{eq:gen-inversionU}-\ref{eq:gen-inversionV}) for the reduced
projectors $p_{j \alpha}, {\tilde p}_{j,\alpha}, \alpha=2,3,4,$
without introducing any explicit dependence on the parameters
$|M_L|,\Phi_L,$  and then we insert these results into either Eq.
\eqref{eq:gen-inversionU} or \eqref{eq:gen-inversionV} when
$\alpha=1.$  From the results of previous sections and using Eqs.
\eqref{eq:gen-inversionU} and \eqref{eq:gen-inversionV}, we get
(j=1,2,4) \beqa p_{j2}&=& {2 \over {\tilde {\cal D}}_j} \biggl\{
{\tilde M}_L^\ast {\tilde M}_R^\ast \sin\theta_k \biggl[ e^{2i
{\tilde \Phi}_R} \sin\theta_k ( m^2_{{\tilde \chi}^\pm_j} - 2
|{\tilde M}_R|^2
\cos^2\theta_k ) - \mu M_R \cos\theta_k \biggr] \nonumber \\
&+& {\tilde M}_L {\tilde M}_R  \cos\theta_k \biggl[e^{- 2i {\tilde
\Phi}_R} M_R \cos\theta_k - \mu^\ast \sin\theta_k\biggr] m_{{\tilde
\chi}^\pm_j} \zeta_j^\ast \sqrt{{\tilde \Delta}_{1j} \over
\Delta_{1j}} \biggr\},\eeqa \beqa p_{j4}&=& {{\sqrt 2} \over {\tilde
{\cal D}}_j} \biggl\{ {\tilde M}_L^\ast \sin\theta_k \biggl[ \mu
(M_R^2 - m^2_{{\tilde
\chi}^\pm_j}) + M_R {\tilde M}_R^2 \sin(2 \theta_k)\biggr] \nonumber \\
&+& {\tilde M}_L \cos\theta_k \biggl[m^2_{{\tilde \chi}^\pm_j} -
M_R^2 - 2 |{\tilde M}_L|^2  \sin^2\theta_k \biggr] m_{{\tilde
\chi}^\pm_j} \zeta_j^\ast \sqrt{ {\tilde \Delta}_{1j} \over
\Delta_{1j} } \biggr\},\eeqa and ${\tilde p}_{j2}$ and ${\tilde
p}_{j3}$ are obtained from $p_{j2}^\ast$ and $p_{j4}^\ast$ by
interchanging $\sin\theta_k \leftrightarrow \cos\theta_k,$
respectively.

\subsection{Expressing \boldmath $|M_L|$ in terms of $\Phi_L$} \label{sec-FUND-PARAMETERS}
 When $\alpha=1,$ either Eq. \eqref{eq:ptilde-delta-V}  combined with  Eqs. \eqref{eq:cdelta1j-V} and
 \eqref{eq:ptilde11-V} or Eq. \eqref{eq:ptilde-delta-U}  combined with  Eqs. \eqref{eq:cdelta1j-U} and
 \eqref{eq:ptilde11-U}, allow us to express the norm of $M_L$ in terms
 of the physical masses $m_{{\tilde \chi}^{\pm}_{j}},$ $\Phi_L$  and the
rest of the fundamental parameters . In both cases we obtain
identical results. Thus, for instance, inserting
\eqref{eq:cdelta1j-V} and \eqref{eq:ptilde11-V} into
\eqref{eq:ptilde-delta-V} (with $\alpha=1$) we get \be
\label{eq:ML-second-degre} {{\tilde {\cal D}}}_j \; |M_L|^2 +
{\tilde {\cal B}}_j \; |M_L| +
 {\tilde {\cal C}}_j =0, \qquad j=1,2,4, \ee
where \beqa \nonumber {\tilde {\cal B}}_j &=&  2 |{\tilde M}_L|^2
\sin(2\theta_k) \\ \nonumber &\times& \bigl[ M_R |{\tilde M}_R|^2
\cos(\Phi_L - 2 {\tilde \Phi}_L + 2 {\tilde \Phi}_R) \sin(2\theta_k)
\\ &-& |\mu| (m_{{\tilde \chi}^{\pm}_{j}}^2 - M_R^2)  \cos(\Phi_L -
2 {\tilde \Phi}_L + \Phi_\mu) \bigr] ,\eeqa \beqa \nonumber {\tilde
{\cal C}}_j &=& -  |{\tilde M}_L|^4 ( m_{{\tilde \chi}^{ \pm}_{j}}^2
- M_R^2) \sin^2(2\theta_k) -  m_{{\tilde \chi}^{\pm}_{j}}^2  \bigl[
{\tilde {\cal D}}_j \\ &-& 2|{\tilde M}_L|^2 \bigl[ m_{{\tilde
\chi}^{\pm}_{j}}^2 - M_R^2 - |{\tilde M}_R|^2\sin^2(2\theta_k)
\bigr] \bigr]\eeqa and ${\tilde {\cal D}}_j$ is given by Eq.
\eqref{eq:tilde-cal-dej}. Thus, solving a quadratic algebraic
equation for $|M_L|,$ we get \be \label{eq:normeML} |M_L| ={ -
{\tilde {\cal B}}_j \pm \sqrt{{\tilde {\cal B}}_j^2 - 4 {{\tilde
{\cal D}}}_j \, {\tilde {\cal C}}_j} \over 2 {\tilde {\cal D}}_j},
\qquad j=1,2,4. \ee  From  Eq. \eqref{eq:normeML} we deduce the
constraints ${\tilde {\cal B}}_j^2 - 4 {{\tilde {\cal D}}}_j \,
{\tilde {\cal C}}_j \ge 0 $ and ${\tilde {\cal B}}_j / {\tilde {\cal
D}}_j < 0.$

Equation \eqref{genMLformula} for $|M_L|$  is equivalent to Eq.
\eqref{eq:normeML}.
 For instance, in the CP-conserving case, when
all the mixing phases except $\Phi_L$ are equal to zero, the choice
of the eigenphase values $\xi_j=\pm 1$  in Eq. \eqref{genMLformula}
correspond to the choice of the values $\Phi_L= \pm \pi,$ in Eq.
\eqref{eq:normeML}, respectively. Thus, in the CP-violating case,
when all the fundamental parameters except $M_L, \Phi_L$ and
${\tilde \Phi}_L$ are known,  Eq. \eqref{eq:normeML} allows the
mixing angle $\Phi_L - 2 {\tilde \Phi}_L$ to play the role of the
eigenphases.

\subsection{Disentangling  \boldmath $M_R$}
The explicit form of $M_R$ in terms of some redefined eigenphases
$\zeta^{(2)}_j$ and the rest of fundamental parameters  can be
deduced from Eqs. (\ref{genMLformula}-\ref{eq:tilde-cal-dej}) by
interchanging $M_L$ with $M_R$ and  ${\tilde M}_L$ with ${\tilde
M_R}.$  This last affirmation  is based on the fact that $M_L$ and
$M_R$  play a similar role in Eqs. (\ref{eq:Hij}-\ref{eq:THij})
provided ${\tilde M}_L$ and ${\tilde M}_R$ are interchanged.  Hence,
we get \be M_{R} = {\tilde A}^{(2)}_j \, \zeta^{(2)}_j + {\tilde
B}^{(2)}_j, \qquad j=1,2,4, \label{genMRformula}\ee where \be
{\tilde A}^{(2)}_j= - {\sqrt{{\tilde \Delta}^{(2)}_{2j} \,
\Delta^{(2)}_{2j}} \over (m_{{\tilde \chi}^{\pm}_{j}}^2 - |\mu|^2)}
{m_{{\tilde \chi}^{\pm}_{j}} \over {\tilde {\cal D}}^{(2)}_j},
\label{eq:tildeA2J}\ee with \beqa\nonumber \Delta^{(2)}_{2 j} &=&  (
|\mu|^2 - m_{{\tilde \chi}^{\pm}_{j}}^2 ) \bigl\{( |\mu|^2 -
m_{{\tilde
\chi}^{\pm}_{j}}^2 ) ( |M_L|^2 - m_{{\tilde \chi}^{\pm}_{j}}^2 ) \\
\nonumber  &-& 2 \cos^2\theta_k |{\tilde M}_R|^2  (m_{{\tilde
\chi}^{\pm}_{j}}^2 -
 |M_L|^2 - 2 |{\tilde M}_L|^2 \sin^2\theta_k)\\ \nonumber  &-&  2 |{\tilde M}_L|^2
\\ \nonumber &\times&\bigl[m_{{\tilde \chi}^{\pm}_{j}}^2 - |M_L| |\mu| \cos(2
{\tilde \Phi}_L - \Phi_\mu - \Phi_L)\sin(2\theta_k)] \\ &+& |{\tilde
M}_L|^4 \sin^2(2\theta_k)\bigr\},
 \label{eq:delta(2)2j-V} \eeqa \beqa\nonumber {\tilde \Delta}^{(2)}_{2 j} &=&  ( |\mu|^2 -
m_{{\tilde \chi}^{\pm}_{j}}^2 ) \bigl\{( |\mu|^2 - m_{{\tilde
\chi}^{\pm}_{j}}^2 ) ( |M_L|^2 - m_{{\tilde \chi}^{\pm}_{j}}^2 ) \\
\nonumber  &-& 2 \sin^2\theta_k |{\tilde M}_R|^2  (m_{{\tilde
\chi}^{\pm}_{j}}^2 -
 |M_L|^2 - 2 |{\tilde M}_L|^2 \cos^2\theta_k)\\ \nonumber  &-&  2 |{\tilde M}_L|^2
\\ \nonumber &\times&\bigl[m_{{\tilde \chi}^{\pm}_{j}}^2 - |M_L| |\mu| \cos(2
{\tilde \Phi}_L - \Phi_\mu - \Phi_L) \sin(2\theta_k)] \\ &+&
|{\tilde M}_L|^4 \sin^2(2\theta_k)\bigr\},
 \label{eq:delta(2)2j-U}\eeqa and \beqa \nonumber {\tilde B}^{(2)}_j&=&{ {\tilde M}_R^2
\sin(2\theta_k) \over {\tilde {\cal D}}^{(2)}_j} \bigl[(m_{{\tilde
\chi}^{\pm}_{j}}^2 - |M_L|^2)|\mu| e^{ -i \Phi_\mu}
\\ &-& M_L |{\tilde M}_L|^2 e^{ -2i  {\tilde \Phi}_L}
\sin(2\theta_k)],\label{eq:tildeB2J}\eeqa where \beqa \nonumber
{\tilde {\cal D}}^{(2)}_j&=& |{\tilde M}_L|^4 \sin^2(2\theta_k) +
(m_{{\tilde \chi}^{\pm}_{j}}^2 - |M_L|^2) (m_{{\tilde
\chi}^{\pm}_{j}}^2 - |\mu|^2)\\ &-&  \nonumber 2 |{\tilde M}_L|^2 \\
&\times& \bigl[m_{{\tilde \chi}^{\pm}_{j}}^2 -|\mu| |M_L|
\cos(\Phi_L - 2 {\tilde \Phi}_L + \Phi_\mu) \sin(2\theta_k) \bigr].
\label{eq:tilde-cal-de2j} \eeqa  Formulas
(\ref{genMRformula}-\ref{eq:tilde-cal-de2j}) expressing $M_R$  in
terms of the redefined eigenphases $\zeta^{(2)}_j$ and the rest of
fundamental parameters may  be obtained using a generalized
projector formalism method as well. This method will be explained
elsewhere.

Note that  $ {\tilde A}^{(2)}_j$ and ${\tilde {\cal D}}^{(2)}_j$ are
real quantities and,  as $M_R$ has been chosen from the start to be
a real parameter, then in the CP-violating  case  Eq.
 \eqref{genMRformula} implies a constraint for
$\tan\theta_k.$ Indeed, writing  $\zeta^{(2)}_j = e^{-i (\theta_{2j}
+ {\tilde \theta}_{2j})}= e^{i \Theta_{2j}},$ we get \be M_R =
{\tilde A}^{(2)}_j \, \cos\Theta_{2j} + {\mathrm{Im\,}}({\tilde
B}^{(2)}_j), \label{eq:MRTheta2j} \ee where \be \sin\Theta_{2j}= -
{{\mathrm{Im\,}}({\tilde B}^{(2)}_j) \over {\tilde A}^{(2)}_j },
\label{eq:Theta2j}\ee with  \beqa \nonumber {\mathrm{Re\,}} ({\tilde
B}^{(2)}_j) &=& { |{\tilde M}_R |^2 \sin(2\theta_k) \over {\tilde
{\cal D}}^{(2)}_j} \bigl[(m_{{\tilde \chi}^{\pm}_{j}}^2 -
|M_L|^2)|\mu| \cos(2 {\tilde \Phi}_R - \Phi_\mu)
\\ &-& |M_L| |{\tilde M}_L|^2 \cos(2{\tilde \Phi}_R-  2 {\tilde
\Phi}_L + \Phi_L) \sin(2\theta_k)]\label{eq:RetildeB2J}\eeqa and
\beqa \nonumber {\mathrm{Im\,}} ({\tilde B}^{(2)}_j) &=& { |{\tilde
M}_R |^2 \sin(2\theta_k) \over {\tilde {\cal D}}^{(2)}_j}
\bigl[(m_{{\tilde \chi}^{\pm}_{j}}^2 - |M_L|^2)|\mu| \sin(2 {\tilde
\Phi}_R - \Phi_\mu)
\\ &-& |M_L| |{\tilde M}_L|^2 \sin(2{\tilde \Phi}_R-  2 {\tilde
\Phi}_L + \Phi_L) \sin(2\theta_k)]\label{eq:ImtildeB2J}.\eeqa Note
that combining Eqs. \eqref{eq:MRTheta2j} and \eqref{eq:Theta2j} we
obtain a  quadratic equation determining $M_R,$ namely \be
\label{eq:quadraticMR} M_R^2 - 2 M_R \, {\mathrm{Re\,}}( {\tilde
B}^{(2)}_j )+ (|{\tilde B}^{(2)}_j|)^2 - ({\tilde A}^{(2)}_j)^2=0,
\ee where the solutions are given by \be M_R = \pm \,  \sqrt{\left(
{\tilde A}^{(2)}_j \right)^2 - { \left( {\mathrm{Im\,}}({\tilde
B}^{(2)}_j)\right)^2} }+ {\mathrm{Re\,}}({\tilde B}^{(2)}_j), \ee
with the constraint $\left( {\tilde A}^{(2)}_j \right)^2 -  \left(
{\mathrm{Im\,}}({\tilde B}^{(2)}_j)\right)^2 \ge0.$

Equation \eqref{eq:quadraticMR} can be compared  with a similar one
found for $M_2$ in the context of the MSSM \cite{key19,key13}.
Indeed, using  the present method  in the context of the MSSM, we
get $j=1,2$ \be M_2^2 - 2 M_2 {\mathrm{Re\,}}( {\tilde b}^{(1)}_j )
+ (|{\tilde b}^{(1)}_j|)^2 - ({\tilde a}^{(1)}_j)^2=0,
\label{eq:M2MSSM} \ee where \be {\tilde a}^{(1)}_j = {\sqrt{ {\tilde
\delta}^{(1)}_{1j} \delta^{(1)}_{1j}} \, m_{{\tilde \chi}^\pm_j}
\over |\mu|^2 - m^2_{ {\tilde \chi}^\pm_j }} \ee with \be
\delta^{(1)}_j = |\mu|^2 - m^2_{ {\tilde \chi}^\pm_j } + 2 m^2_W
\cos^2\beta \ee and
 \be {\tilde \delta}^{(1)}_j = |\mu|^2 -
m^2_{ {\tilde \chi}^\pm_j } + 2 m^2_W \sin^2\beta, \ee and \be
{\mathrm{Re\,}}( {\tilde b}^{(1)}_j ) = {m^2_W |\mu| \sin(2\beta)
\cos\Phi_\mu \over |\mu|^2 - m^2_{ {\tilde \chi}^\pm_j }}, \ee \be
{\mathrm{Im\,}}( {\tilde b}^{(1)}_j ) = - {m^2_W |\mu| \sin(2\beta)
\sin\Phi_\mu \over  |\mu|^2 - m^2_{ {\tilde \chi}^\pm_j }}. \ee
Therefore, as before, from Eq. \eqref{eq:M2MSSM}, the solutions for
$M_2$ are given by  \be M_2 = \pm \, \sqrt{\left( {\tilde a}^{(1)}_j
\right)^2 - { \left( {\mathrm{Im\,}}({\tilde b}^{(1)}_j)\right)^2}
}+ {\mathrm{Re\,}}({\tilde b}^{(1)}_j), \ee with the constraint
$\left( {\tilde a}^{(1)}_j \right)^2 -  \left(
{\mathrm{Im\,}}({\tilde b}^{(1)}_j)\right)^2 \ge0.$

We note that in the L-R SUSY model, $M_R$ becomes  singular at the
points where either $(|\mu|^2 - m^2_{{\tilde \chi}^\pm_j})$ or $
{\tilde {\cal D}}^{(2)}_j $ vanishes whereas in the MSSM the
singularities of $M_2$ are only present at the point where $(|\mu|^2
- m^2_{{\tilde \chi}^\pm_j})$  vanishes.  Apart from that, the role
played by $M_R,$ $\mu$ and $\tan\theta_k$ in the chargino sector in
the context of the L-R SUSY model, when the remaining parameters are
fixed, is similar to the role played by $M_2,$ $\mu$ and $\tan\beta$
in the chargino sector in the context of the MSSM. This suggests us
to make use of  some useful technics implemented in the literature
\cite{key19,key18,key22,key23,key24} in the context of the MSSM,
concerning to the measurement of some independent cross section
physical observables related to the chargino and neutralino pair
production in $e^+ e^-$ annihilation processes, that could be
applied in the context of the L-R SUSY model to fix the chargino and
neutralino parameters.

\section{Determining L-R SUSY parameters}
\label{sec-determinig-parameters} In this section we give some
examples on the analytical and numerical reconstruction of the
fundamental L-R SUSY parameters of the ino sector by considering
some possible CP-conserving and CP-violating scenarios.

\subsection{Scenario $Scpv_3$}
Let us first to consider the scenario $Scpv_3$ given in  Tab.
\ref{tab:tablacuatro}, where the L-R SUSY parameters $M_R,
\tan\theta_k$ and  $|\mu|$ are known and where the phases
$\Phi_\mu={\tilde \Phi}_L={\tilde \Phi}_R=0.$  As before,  we assume
that  $g_L=g_R=0.65$, and $k_u=92.75$.

Figures \ref{fig:chargino21} and \ref{fig:chargino22} show the
values of $(|M_L|-m_{{\tilde \chi}^{\pm}_{j}})/2$ predicted by Eq.
\eqref{genMLformula} as a function of the chargino physical masses
$m_{{\tilde \chi}^{\pm}_{j}}$ for different values of the eigenphase
argument Arg$(\xi_j)=0,\pi/4,\pi/2,3 \pi /4,\pi,$ under the
conditions described in  scenario $Scpv_3$ with $\tan\theta_k=4$ and
$\tan\theta_k=30,$ respectively. Comparing these figures we observe
that for a given  value of the chargino mass, the variation of
$|M|_L - m_{{\tilde \chi}^{\pm}_{j }},$ with respect to the argument
of the eigenphases $\xi_j$ when the ratio $\tan\theta_k=k_u /k_d$ is
small is greater than the variation of this quantity when this ratio
is large. In the case where $\tan\theta_k =4,$ we observe that for
values of  $m_{{\tilde \chi}^{\pm}_{j}}\approx 470\, {\rm GeV},$ the
values of the parameter  $|M_L|$ are approximately the same ( $\sim
470\, {\rm GeV}$),  whereas for values of $m_{{\tilde
\chi}^{\pm}_{j}} \approx 160 \, {\rm GeV},$  the values of this
parameter lie in the range of $160 \, {\rm GeV}$-$ 167 \, {\rm
GeV},$ approximately. The singularities in the plot of  Fig.
\ref{fig:chargino21} occurs when the factor $(m^2_{{\tilde
\chi}^{\pm}_{j}}- |\mu|^2)$ in the denominator of Eq.
\eqref{eq:tildeAJ}  vanishes, i.e., when $m_{{\tilde
\chi}^{\pm}_{j}}=248$ GeV and also when the factor ${\tilde {\cal
D}}_j=0$ in the denominator of Eqs. \eqref{eq:tildeAJ} and
\eqref{eq:tildeBJ} vanishes, i.e., when $m_{{\tilde
\chi}^{\pm}_{j}}\approx 247.7$ GeV and $m_{{\tilde
\chi}^{\pm}_{j}}\approx 498.6$ GeV. By the same reasons, the
singularities in the plot of  Fig. \ref{fig:chargino22} occurs at
the points $m_{{\tilde \chi}^{\pm}_{j}}=248$ GeV, $m_{{\tilde
\chi}^{\pm}_{j}}=246$ GeV and $m_{{\tilde \chi}^{\pm}_{j}}=498.4$
GeV, respectively.  On the other hand, contrary to the case of the
neutralino \cite{key11}, in the case where $\tan\theta_k=30,$ we
don't observe a distinguishable value of the chargino mass where the
different curves intersect, i.e., the corresponding  graph don't
allows us  to determine an optimal value of $|M_L|.$
\begin{table}
\begin{center}
\begin{tabular}{c c c c c c }\cline{1-6}\\
Scenario & $|\mu| $\; &  $M_R \; $ & $\Phi_\mu= {\tilde \Phi}_L= {\tilde \Phi}_R$ &  $ \tan\theta_k  $ \; &  Arg$(\xi_j) $   \\
\\\cline{1-6} \\
$Spvc_3 $ &  248 & 500 & 0  &   \parbox{0.5cm}{4 \\ 30} & \parbox[c]{0.5cm}{0 \\ $\pi/4$  \\ $\pi/2$ \\ $3\pi/4$ \\ $\pi$ } \\ \\  \hline \\
\end{tabular}
\end{center}
\caption{Input parameters for scenario $Scpv_3.$  All mass
quantities are in GeV.} \label{tab:tablacuatro}
\end{table}
\begin{figure} \centering
\begin{picture}(31.5,21)
\put(1,2){\includegraphics[width=70mm]{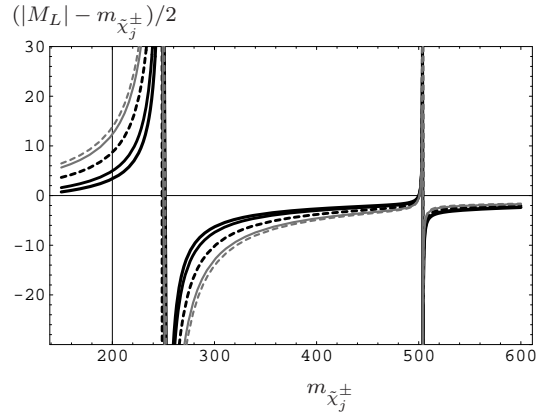}}
\put(15.75,0.7){\scriptsize{$m_{{\tilde \chi}^{\pm}_{j}}$}}
\put(1.05,19.25){\scriptsize{$(|M_L| - m_{{\tilde
\chi}^{\pm}_{j}})/2$}}
\end{picture}
\caption{Reconstruction of  $|M_L|$ as a function of the physical
chargino masses $m_{{\tilde \chi}^{\pm}_{j}},$ using the general
formula \eqref{genMLformula}, for inputs of scenario $Scpv_3$ with
$\tan\theta_k=4.$ The curves are:  Arg$(\xi_j)=0$ (heavy solid),
Arg$(\xi_j)=\pi/4$ (solid),
 Arg$(\xi_j)=\pi/2$ (dashed),  Arg$(\xi_j)=3\pi/4$ (light solid),
 Arg$(\xi_j)=\pi$ (light dashed). } \label{fig:chargino21}
\end{figure}
\begin{figure} \centering
\begin{picture}(31.5,21)
\put(1,2){\includegraphics[width=70mm]{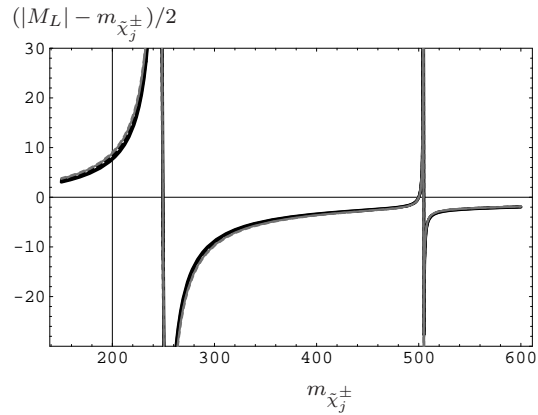}}
\put(15.75,0.7){\scriptsize{$m_{{\tilde \chi}^{\pm}_{j}}$}}
\put(1.05,19.25){\scriptsize{$(|M_L| - m_{{\tilde
\chi}^{\pm}_{j}})/2$}}
\end{picture}
\caption{The same as in Fig. \ref{fig:chargino21} but considering
$tan\theta_k=30.$} \label{fig:chargino22}
\end{figure}

\subsection{Scenario $Scpv_4$}
Let us now to consider a scenario where the physical mass of two
lightest charginos, $m_{{\tilde \chi}_1^\pm}$ and  $m_{{\tilde
\chi}_2^\pm}$ are known.
 Then, from Eq. \eqref{genMLformula}, the parameters  $|M_L|$ and $\Phi_L$
and the eigenphases  $\zeta_1$ and $ \zeta_2,$ can be expressed
analytically, up to a twofold discrete ambiguity, in terms of
parameters $|\mu|, \Phi_\mu, M_R, \tan\theta_k, {\tilde \Phi}_R$ and
${\tilde \Phi}_L.$ Indeed, following  the same treatment as in the
case of the neutralino in the MSSM\cite{key13}, if the two lightest
chargino masses are known, then  from Eq. \eqref{genMLformula}, we
deduce the following expression for the chargino eigenphases \be
\label{eq:zeta12} \zeta_{1,2}= {{\tilde A}^2_1 - {\tilde A}^2_2 \pm
|\Delta {\tilde B}|^2 + i\epsilon \sqrt{{\tilde \Delta}_{ch}} \over
2 {\tilde A}_{1,2} \Delta {\tilde B}^\ast}, \ee where \be \Delta
{\tilde B}= {\tilde B}_2 - {\tilde B}_1, \ee  \be {\tilde
\Delta}_{ch}= 4 {\tilde A}^2_1 {\tilde A}^2_2 - ( {\tilde A}^2_1 +
{\tilde A}^2_2 - |\Delta {\tilde B}|^2)^2 \ge 0\ee and $\epsilon=\pm
1.$   Then, inserting Eq. \eqref{eq:zeta12} into
\eqref{genMLformula} we get \be \label{eq:ML12} M_L= {{\tilde A}_1^2
- {\tilde A}_2^2 \pm ({\tilde B}_1 + {\tilde B}_2) \Delta {\tilde B}
+ i \epsilon \sqrt{{\tilde \Delta}_{ch}} \over 2 \Delta {\tilde
B}^\ast}.\ee

On the other hand, under similar conditions, we can give an
equivalent description to the one given above. Indeed, writing
${\tilde {\cal B}}_j, \, j=1,2, $ in the form \be
\label{eq:caltildeBjangles} {\tilde {\cal B}}_j = {{\tilde {\cal
P}}_j + {\tilde {\cal Q}}_j \tan(\Phi_L - 2 {\tilde \Phi}_L ) \over
\sqrt{1+ \tan^2 (\Phi_L - 2 {\tilde \Phi}_L )}},\ee with \beqa \
\nonumber
{\tilde {\cal P}}_j &=& 2 |{\tilde M}_L|^2 \sin(2\theta_k) \\
\nonumber &\times& \bigl[M_R |{\tilde M}_R|^2 \cos(2 {\tilde
\Phi}_R) \sin(2\theta_k)
\\ &-& |\mu| (m_{{\tilde \chi}^\pm_j}^2- M_R^2) \cos\Phi_\mu
\bigr]\label{eq:calPj}\eeqa and \beqa \nonumber {\tilde {\cal Q}}_j
&=& 2 |{\tilde M}_L|^2 \sin(2\theta_k) \\ \nonumber &\times& \bigl[-
M_R |{\tilde M}_R|^2 \sin(2 {\tilde \Phi}_R) \sin(2\theta_k)
\\ &+& |\mu| (m_{{\tilde \chi}^\pm_j}^2- M_R^2) \sin\Phi_\mu
\bigr],\label{eq:calQj}\eeqa and inserting it into Eq.
\eqref{eq:normeML}, after some algebraic manipulations we get \be
\label{eq:FILmenos2TFIL} \tan (\Phi_L - 2 {\tilde \Phi}_L)= {\tilde
{\mathbb R}}\equiv  { - {\tilde {\mathbb B}} - {\tilde \epsilon}
\sqrt{{\tilde {\mathbb B}}^2 - 4 {\tilde {\mathbb A}} {\tilde
{\mathbb C}}} \over 2 {\tilde {\mathbb A}} },\ee where ${\tilde
{\mathbb B}}^2 - 4 {\tilde {\mathbb A}} {\tilde {\mathbb C}} \ge 0,$
\be \label{eq:tildematA} {\tilde {\mathbb A}} = {1 \over 2}
F({\tilde {\cal Q}}_1,{\tilde {\cal Q}}_2,{\tilde {\cal
Q}}_1,{\tilde {\cal Q}}_2) - ({\tilde {\cal D}}_1 {\tilde {\cal
C}}_2 - {\tilde {\cal D}}_2 {\tilde {\cal C}}_1)^2,\ee \be
\label{eq:tildematB} {\tilde {\mathbb B}} = F({\tilde {\cal
P}}_1,{\tilde {\cal P}}_2,{\tilde {\cal Q}}_1,{\tilde {\cal Q}}_2)
\ee and \be \label{eq:tildematC} {\tilde {\mathbb C}} = {1 \over 2}
F({\tilde {\cal P}}_1,{\tilde {\cal P}}_2,{\tilde {\cal
P}}_1,{\tilde {\cal P}}_2) - ({\tilde {\cal D}}_1 {\tilde {\cal
C}}_2 - {\tilde {\cal D}}_2 {\tilde {\cal C}}_1)^2,\ee with \beqa
\nonumber  F({\tilde {\cal P}}_1,{\tilde {\cal P}}_2,{\tilde {\cal
Q}}_1,{\tilde {\cal Q}}_2) &=&  ({\tilde {\cal D}}_1 {\tilde {\cal
C}}_2 + {\tilde {\cal D}}_2 {\tilde {\cal C}}_1) ({\tilde {\cal
P}}_1 {\tilde {\cal Q}}_2 + {\tilde {\cal P}}_2 {\tilde {\cal Q}}_1)
\\\nonumber &-& 2 ( {\tilde {\cal D}}_1 {\tilde {\cal C}}_1 {\tilde
{\cal P}}_2 {\tilde {\cal Q}}_2  + {\tilde {\cal D}}_2
{\tilde {\cal C}}_2 {\tilde {\cal P}}_1 {\tilde {\cal Q}}_1),\label{eq:FP1P2Q1Q2} \\
\eeqa where $\tilde \epsilon=\pm 1.$

Moreover, combining  Eq. \eqref{eq:ML-second-degre} for $j=1$ with
Eq. \eqref{eq:ML-second-degre} for  $j=2,$ and then using Eq.
\eqref{eq:caltildeBjangles} for $j=1,2,$ we get \be
\label{eq:ML2charginos}|M_L|={ ({\tilde {\cal D}}_1 {\tilde {\cal
C}}_2 - {\tilde {\cal D}}_2 {\tilde {\cal C}}_1)\, \sqrt{1 + {\tilde
{\mathbb R}}^2} \over ({\tilde {\cal D}}_2 {\tilde {\cal P}}_1 -
{\tilde {\cal D}}_1 {\tilde {\cal P}}_2 )+ ({\tilde {\cal D}}_2
{\tilde {\cal Q}}_1 - {\tilde {\cal D}}_1 {\tilde {\cal Q}}_2)
{\tilde {\mathbb R}}}.\ee Equations \eqref{eq:FILmenos2TFIL} and
\eqref{eq:ML2charginos} allows us to determine the phase difference
$\Phi_L - 2 {\tilde \Phi}_L$ and the norm $|M_L|,$ respectively, up
to a twofold discrete ambiguity, in terms of the two lightest
chargino masses and the fundamental parameters $|\mu|, \Phi_\mu,
M_R, \tan\theta_k$ and ${\tilde \Phi}_R.$

\begin{table}
\begin{center}
\begin{tabular}{c c c c c c c}\cline{1-7}\\
Scenario & $|\mu| $\; &  $M_R \; $ & $m_{{\tilde \chi}_1^\pm} \; $ &  $m_{{\tilde \chi}_2^\pm} \; $ & $ \tan\theta_k  $ \; &  $ {\tilde \Phi}_R $   \\
\\\cline{1-6} \\
$Scpv_4 $ &  248 & 300 & 158.5  &   \parbox{0.5cm}{247.30\\ 247.38} &   30  & 0 \\
\hline
\end{tabular}
\end{center}
\caption{Input parameters for scenario $Scpv_4.$ All mass quantities
are in GeV.} \label{tab:Snc4char}
\end{table}

For instance, let us consider the CP-violating scenario $Scpv_4$
given in Tab. \ref{tab:Snc4char}. Figure \ref{fig:chargino30}, show
the behaviour of $\Phi_L - 2 {\tilde \Phi}_L$ as a function of
$\Phi_\mu$ when this last parameter is allowed to vary in the
interval $ [-0.95({\rm rad}),0.95({\rm rad})].$ This interval
correspond approximately to the rang where the discriminant under
the root in  Eq. \eqref{eq:FILmenos2TFIL}, is greater or equal to
zero. The two different curves observed in this figure demonstrate
the twofold ambiguity corresponding  to the two possible values of
${\tilde \epsilon}=\pm 1.$ Figure \ref{fig:chargino31}, show the
behaviour of $|M_L|,$ computed from Eq.\eqref{eq:ML2charginos},  as
a function of $\Phi_\mu,$ in the same interval as in Figure
\ref{fig:chargino30}. The values of $|M_L|$ fluctuate in the range
of $166.7\, {\rm GeV}$-$167.3\, {\rm GeV}$ approximately. The two
fold ambiguity induced by the two different signs of ${\tilde
\epsilon}$ is practically not  observed in this case.

The values of the phase angles and $|M_L|$ are very sensible to the
change of the physical masses of the charginos.  For instance,
according to the input parameters of scenario $Scpv_4,$ if the mass
of the second chargino is chosen to be $m_{{\tilde
\chi}_2^\pm}=247.38$ GeV, then the values of $\Phi_\mu$ where the
discriminant under the root in  Eq. \eqref{eq:FILmenos2TFIL} is
greater or equal to zero are restricted to the interval  $[-0.85 \,
{\rm rad},0.85 \, {\rm rad}].$  Also, the values of the phase
difference $\Phi_L - 2 {\tilde \Phi}_L $ and $|M_L|,$ for a given
value of $\Phi_\mu$ vary  with respect to precedent case, now the
values of $|M_L|$ fluctuate in the range of $166.8\, {\rm
GeV}$-$167.5\, {\rm GeV}$ approximately, as we can see from Figures
\ref{fig:chargino30b} and \ref{fig:chargino31b}, respectively.

\begin{figure} \centering
\begin{picture}(31.5,21)
\put(1,2){\includegraphics[width=70mm]{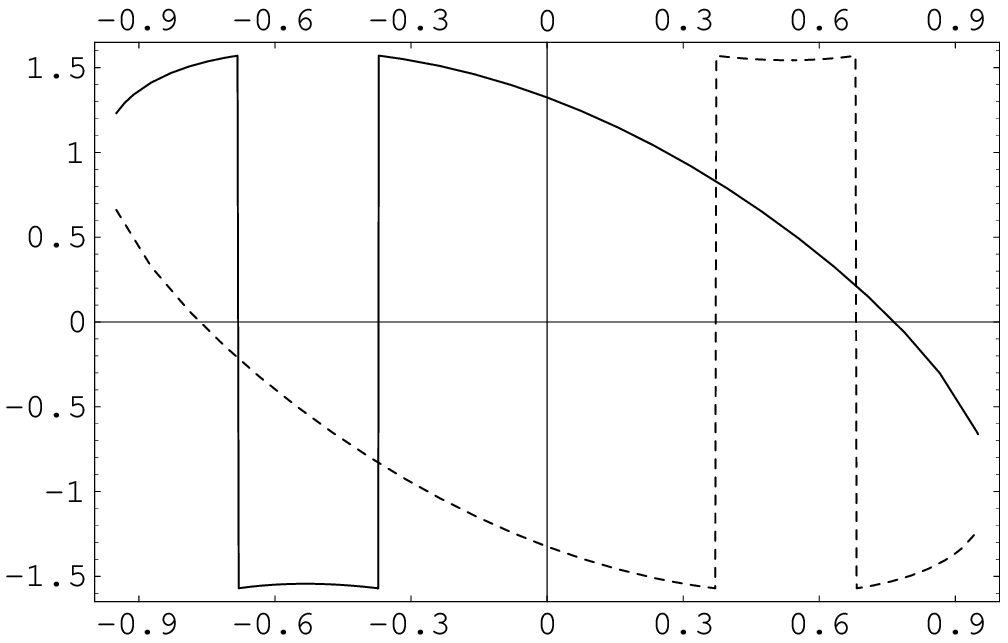}}
\put(15.75,0.7){\scriptsize{$\Phi_\mu \, {\rm (rad)}$}}
\put(1.05,19.25){\scriptsize{$(\Phi_L - 2 {\tilde \Phi}_L)\, {(\rm
rad)}$}}
\end{picture}
\caption{behaviour of  $\Phi_L - 2 {\tilde \Phi}_L$  as a function
of $\Phi_\mu,$ computed from Eq. \eqref{eq:FILmenos2TFIL}, for input
parameters of scenario $Scpv_4,$ with $m_{{\tilde
\chi}_2^\pm}=247.30$ GeV.  The curves are: ${\tilde \epsilon}=1$
(solid), ${\tilde \epsilon}=-1$ (dashed).} \label{fig:chargino30}
\end{figure}

\begin{figure} \centering
\begin{picture}(31.5,21)
\put(1,2){\includegraphics[width=70mm]{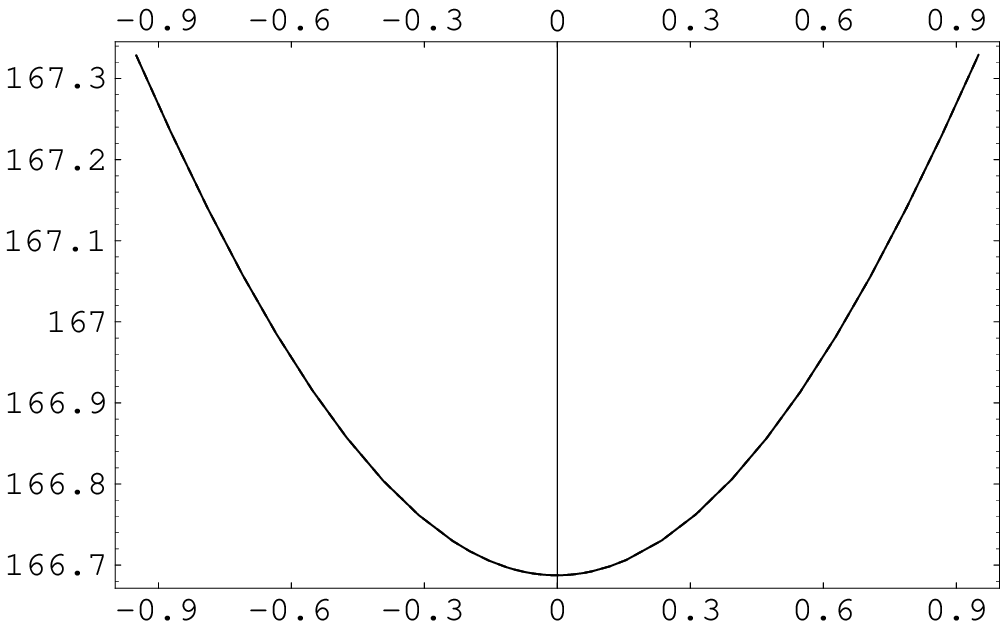}}
\put(15.75,0.7){\scriptsize{$\Phi_\mu \, {\rm (rad)}$}}
\put(1.05,19.25){\scriptsize{$|M_L|\, ({\rm GeV)}$}}
\end{picture}
\caption{Norm  of  $M_L$  as a function of $\Phi_\mu,$ computed from
Eq. \eqref{eq:ML2charginos}, for input parameters of scenario
$Scpv_4,$ with $m_{{\tilde \chi}_2^\pm}=247.30$ GeV. The curve are
practically the same for ${\tilde \epsilon}=1$ as for ${\tilde
\epsilon}=-1.$} \label{fig:chargino31}
\end{figure}

\begin{figure} \centering
\begin{picture}(31.5,21)
\put(1,2){\includegraphics[width=70mm]{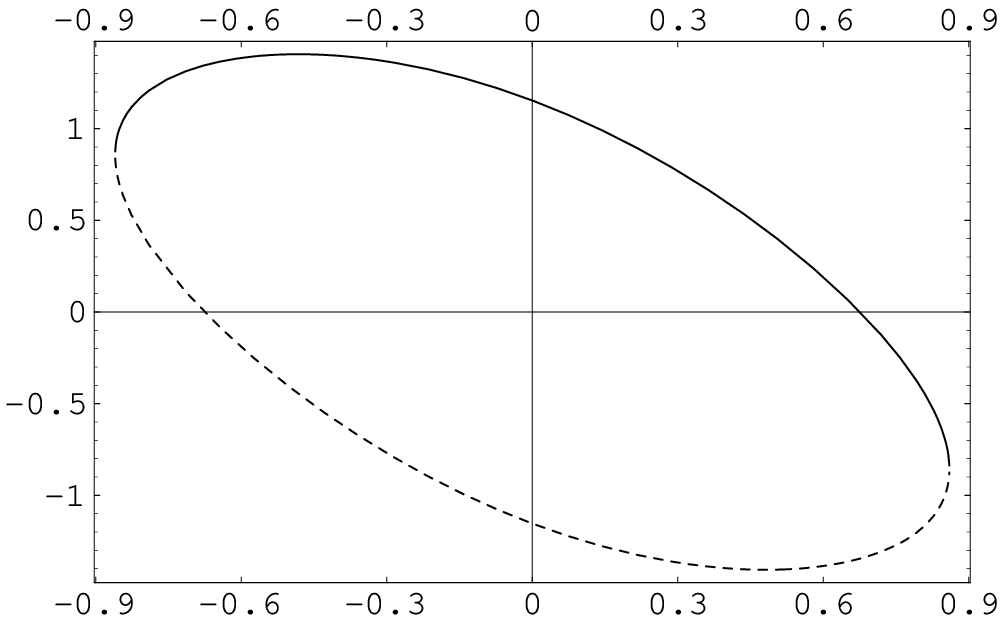}}
\put(15.75,0.7){\scriptsize{$\Phi_\mu \, {\rm (rad)}$}}
\put(1.05,19.25){\scriptsize{$(\Phi_L - 2 {\tilde \Phi}_L)\, {(\rm
rad)}$}}
\end{picture}
\caption{behaviour of  $\Phi_L - 2 {\tilde \Phi}_L$  as a function
of $\Phi_\mu,$ computed from Eq. \eqref{eq:FILmenos2TFIL}, for input
parameters of scenario $Scpv_4,$ with $m_{{\tilde
\chi}_2^\pm}=247.38$ GeV. The curves are: ${\tilde \epsilon}=1$
(solid), ${\tilde \epsilon}=-1$ (dashed).} \label{fig:chargino30b}
\end{figure}

\begin{figure} \centering
\begin{picture}(31.5,21)
\put(1,2){\includegraphics[width=70mm]{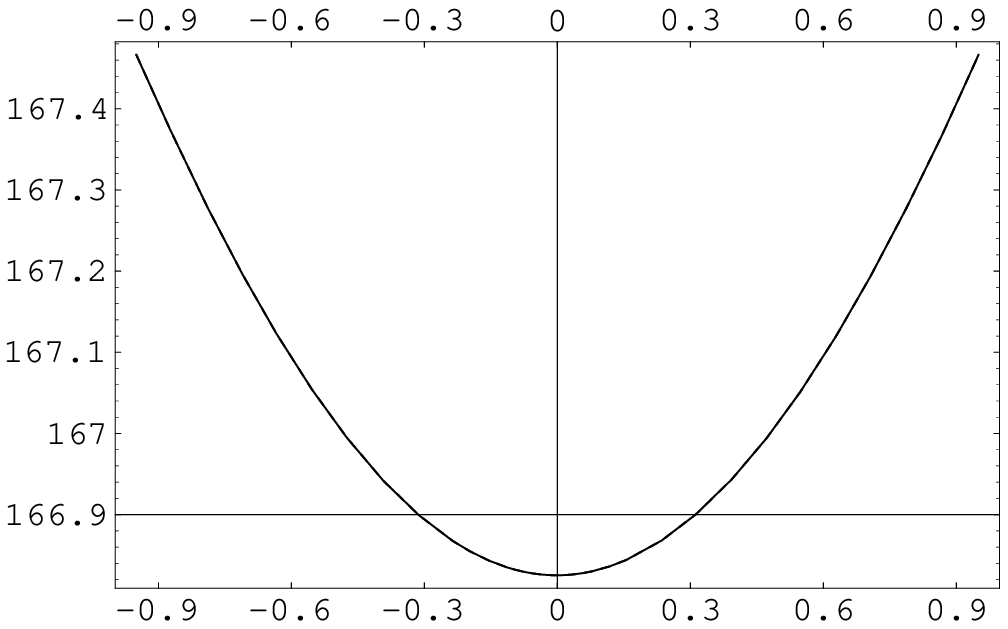}}
\put(15.75,0.7){\scriptsize{$\Phi_\mu \, {\rm (rad)}$}}
\put(1.05,19.25){\scriptsize{$|M_L|\, ({\rm GeV)}$}}
\end{picture}
\caption{Norm  of  $M_L$  as a function of $\Phi_\mu,$ computed from
Eq. \eqref{eq:ML2charginos}, for input parameters of scenario
$Scpv_4,$ with $m_{{\tilde \chi}_2^\pm}=247.30$ GeV. The curve are
practically the same for ${\tilde \epsilon}=1$ as for ${\tilde
\epsilon}=-1.$} \label{fig:chargino31b}
\end{figure}

\subsection{Scenario $Scpv_5$}
Let us assume now that in addition to the two lightest chargino
masses $m_{{\tilde \chi}_1^\pm}$ and $m_{{\tilde \chi}_2^\pm},$ we
also known  the physical masses of the two lightest neutralinos
$m_{{\tilde \chi}^{0}_{1}}$ and $m_{{\tilde \chi}^{0}_{2}}.$

In the same way as in Section \eqref{sec-FUND-PARAMETERS}, from the
neutralino sector of the L-R SUSY model, we can express the norm of
$M_L$ in terms of the neutralino physical masses $m_{{\tilde
\chi}^{0}_{j}}, \, j=1,\ldots,4$ and the fundamental parameters
$\Phi_L, M_R, M_V, |\mu|, \Phi_\mu$ and $\tan\theta_k,$  by solving
the algebraic equation \cite{key11} \be {\cal D}_j \, |M_L|^2 +
{\cal B}_j \,|M_L| + {\cal C}_j =0, \ee where \beqa \nonumber {\cal
B}_j &=& - 4 (M_{W_L})^2 \bigl\{ |\mu|(m_{{\tilde \chi}^{0}_{j}}^2 -
M_{RV}^2 )(m_{{\tilde \chi}^{0}_{j}}^2 - 4 |\mu|^2) \\ \nonumber
&\times&
\cos(\Phi_L + \Phi_\mu) \sin(2 \theta_k) + 2\kappa^2 (M_{W_L})^2 M_{RV} \\
&\times& [ m_{{\tilde \chi}^{0}_{j}}^2 - 4 |\mu|^2
\sin^2(2\theta_k)] \cos\Phi_L \bigr\} ,\eeqa \beqa \nonumber {\cal
C}_j &=&  \,m_{{\tilde \chi}^{0}_{j}}^2 \, \bigl(m_{{\tilde
\chi}^{0}_{j}}^2 - {M_{RV}}^2 \bigr) \,
   {\bigl(m_{{\tilde \chi}^{0}_{j}}^2 - 4\,{|\mu|}^2 \bigr)}^2
   \\ \nonumber &-&
  2\,m_{{\tilde \chi}^{0}_{j}}^2\,(M_{W_L})^2  \bigl(m_{{\tilde \chi}^{0}_{j}}^2 - 4\,{|\mu|}^2 \bigr) \,
   \bigl[ m_{{\tilde \chi}^{0}_{j}}^2\,\bigl(1 + 4\,{\kappa }^2 \bigr)  \\ \nonumber  &-& {M_{RV}}^2  +
     8\,  |\mu|\, M_{RV}\,{\kappa }^2\,\cos (\Phi_\mu)\,
     \sin (2\,\theta_k) \bigr]  \\\nonumber &+&
  (M_{W_L})^4\,\bigl[ m_{{\tilde \chi}^{0}_{j}}^2\,
     {\bigl(1 + 4\,{\kappa }^2 \bigr) }^2  -{M_{RV}}^2 \bigr] \\ &\times&
   \bigl[m_{{\tilde \chi}^{0}_{j}}^2 - 4\,{|\mu|}^2\,
      {\sin (2\,\theta_k)}^2 \bigr]\eeqa and \beqa \nonumber {\cal D}_j &=& - \bigl\{ (m_{{\tilde
\chi}^{0}_{j}}^2 - M_{RV}^2 )(m_{{\tilde \chi}^{0}_{j}}^2 - 4
|\mu|^2)^2 - 8 \kappa^2 (M_{W_L})^2 \\ \nonumber &\times&(m_{{\tilde
\chi}^{0}_{j}}^2 - 4 |\mu|^2) [m_{{\tilde \chi}^{0}_{j}}^2 + 2 |\mu|
M_{RV} \cos\Phi_\mu \sin (2 \theta_k) ] \\ &+& 16 \kappa^4
(M_{W_L})^4  [ m_{{\tilde \chi}^{0}_{j}}^2 - 4 |\mu|^2
\sin^2(2\theta_k)] \bigr\}, \label{eq:cal-dej} \eeqa where $M_{W_L}$
is  the mass of the left-handed gauge bosons given in Eq.
\eqref{eq:mass-WL}, $M_{RV}= (4 g_V^2 M_R + g_R^2 M_V) / g_1^2$ and
$ \kappa= {g_R g_V \over g_1 g_L}, $ where $g_1= (g_R^2 + 4
g_V^2)^{1/2}$ and $g_V$ is the coupling constant associated to the
gauge group $U(1)_{B-L}.$

Thus, writing ${\cal B}_j, \, j=1,2,$ in the form \be
\label{eq:calBjangles} {\cal B}_j = { {\cal P}_j + {\cal Q}_j
\tan\Phi_L  \over \sqrt{1+ \tan^2\Phi_L }},\ee with \beqa \nonumber
{\cal P}_j &=& - 4 (M_{W_L})^2 \bigl\{ |\mu|(m_{{\tilde
\chi}^{0}_{j}}^2 - M_{RV}^2 )(m_{{\tilde \chi}^{0}_{j}}^2 - 4
|\mu|^2) \\ \nonumber &\times&
\cos\Phi_\mu \, \sin(2 \theta_k) + 2 \kappa^2 \,  (M_{W_L})^2 \,  M_{RV} \\
&\times& [ m_{{\tilde \chi}^{0}_{j}}^2 - 4 |\mu|^2
\sin^2(2\theta_k)] \bigr\}\eeqa and \beqa \nonumber {\cal Q}_j &=& 4
|\mu| \,  (M_{W_L})^2 \,   (m_{{\tilde \chi}^{0}_{j}}^2 - M_{RV}^2 )
\, (m_{{\tilde \chi}^{0}_{j}}^2 - 4 |\mu|^2) \\ \nonumber &\times&
\sin\Phi_\mu \, \sin(2 \theta_k),\eeqa and proceeding as in the
previous  subsection, we get \be \label{eq:tanFIL}\tan\Phi_L =
{\mathbb R}\equiv {- {\mathbb B} - \epsilon \sqrt{{\mathbb B}^2 - 4
{\mathbb A} \, {\mathbb C}} \over 2 {\mathbb A},}  \ee where
 $\epsilon=\pm 1,$  $ {\mathbb B}^2 - 4 {\mathbb A} \, {\mathbb C} \ge 0, $ and
${\mathbb A},{\mathbb B}$ and ${\mathbb C}$ are obtained from Eqs.
\eqref{eq:tildematA}, \eqref{eq:tildematB} and \eqref{eq:tildematC},
respectively, by removing the tilde on the variables.

Moreover, as in the precedent subsection, we also  get \be
\label{eq:ML2neutralinos}|M_L|={ ( {\cal D}_1 {\cal C}_2 - {\cal
D}_2 {\cal C}_1)\, \sqrt{1 + {\mathbb R}^2} \over ({\cal D}_2 {\cal
P}_1 - {\cal D}_1 {\cal P}_2 )+ ({\cal D}_2 {\cal Q}_1 - {\cal D}_1
{\cal Q}_2 ) {\mathbb R} }.\ee

Equations \eqref{eq:tanFIL} and \eqref{eq:ML2neutralinos} express
the phase $\Phi_L$ and norm  $|M_L|,$  in terms of the two lightest
neutralino masses and the parameters $|\mu|,\Phi_\mu,M_R, M_V,$
$\tan\theta_k,$ respectively.

Now, combining Eqs. \eqref{eq:FILmenos2TFIL} and \eqref{eq:tanFIL},
we obtain an expression for the  chargino mixing angle ${\tilde
\Phi}_L,$ in terms of the two lightest neutralino physical masses,
the two lightest chargino physical masses and the set of parameters
described previously in addition to the mixing angle ${\tilde
\Phi}_R,$ namely, \be \label{eq:2TFIL} \tan (2 {\tilde \Phi}_L) =
{{\mathbb R} - {\tilde {\mathbb R}} \over 1 +  {\mathbb R} \,
{\tilde {\mathbb R}}}, \ee where $ {\tilde {\mathbb R}} (m_{{\tilde
\chi}^{\pm}_{1}}, m_{{\tilde \chi}^{\pm}_{2}}, |\mu|, \Phi_\mu,
M_R,{\tilde \Phi}_R,\tan\theta_k)$ and $ {\mathbb R} (m_{{\tilde
\chi}^{0}_{1}}, m_{{\tilde \chi}^{0}_{2}}, |\mu|, \Phi_\mu, M_R,
M_V, \tan\theta_k)$ are given by the right side member of Eqs.
\eqref{eq:FILmenos2TFIL} and \eqref{eq:tanFIL}, respectively.
Moreover, equating Eqs. \eqref{eq:ML2charginos} and
\eqref{eq:ML2neutralinos}, we obtain an equation serving to
determine, at least numerically, one of the remaining parameters.

\subsubsection{Case \boldmath ${\tilde \Phi}_L= 0$}
In the case ${\tilde \Phi_L}=0,$ from Eq. \eqref{eq:2TFIL} we get
${\tilde {\mathbb R}} = \mathbb R$ and,  taking in account this
result, combining \eqref{eq:ML2charginos} and
\eqref{eq:ML2neutralinos}, after some manipulations we obtain \be
{\tilde {\mathbb R}} =  {\mathbb R} = - {g[{\cal P}_1, {\cal P}_2,
{\tilde {\cal P}}_1,{\tilde {\cal P}}_2 ] \over  g[{\cal Q}_1, {\cal
Q}_2, {\tilde {\cal Q}}_1,{\tilde {\cal Q}}_2 ]},
\label{eq:equality-R-TildeR}\ee where \be
 g[{\cal P}_1, {\cal P}_2, {\tilde {\cal P}}_1,{\tilde
{\cal P}}_2 ]=  ({\cal D}_1 {\cal C}_2 - {\cal D}_2 {\cal C}_1)
({\tilde {\cal D}}_1 {\tilde {\cal P}}_2 - {\tilde {\cal D}}_2
{\tilde {\cal P}}_1) - ({\tilde {\cal D}}_1 {\tilde {\cal C}}_2 -
{\tilde {\cal D}}_2 {\tilde {\cal C}}_1) ({\cal D}_1 {\cal P}_2 -
{\cal D}_2 {\cal P}_1). \ee

The equality ${\tilde {\mathbb R}} =  {\mathbb R}$ in Eq.
\eqref{eq:equality-R-TildeR} produces an algebraic equation relating
the parameters $M_R,$$ M_V,$ $|\mu|,$ $\Phi_\mu,$ ${\tilde \Phi}_R$
and $\tan\theta_k,$ namely \be ({\mathbb A} {\tilde {\mathbb C}} -
{\mathbb C} {\tilde {\mathbb A}} )^2 - ({\mathbb B} {\tilde {\mathbb
C}} -  {\mathbb C} {\tilde {\mathbb B}})({\mathbb A} {\tilde
{\mathbb B}} - {\mathbb B} {\tilde {\mathbb A}} )=0,
\label{eq:MRorder32} \ee whereas the equality of ${\mathbb R}$  with
the third member in Eq. \eqref{eq:equality-R-TildeR} leads to the
following algebraic equation for these parameters: \be {\mathbb A}
\, \, g[{\cal P}_1, {\cal P}_2, {\tilde {\cal P}}_1,{\tilde {\cal
P}}_2 ]^2 - {\mathbb B} \, \, g[{\cal P}_1, {\cal P}_2, {\tilde
{\cal P}}_1,{\tilde {\cal P}}_2 ] \, g[{\cal Q}_1, {\cal Q}_2,
{\tilde {\cal Q}}_1,{\tilde {\cal Q}}_2 ] + {\mathbb C} \, \,
g[{\cal Q}_1, {\cal Q}_2, {\tilde {\cal Q}}_1,{\tilde {\cal Q}}_2
]^2=0. \label{eq:MRorder16} \ee For fixed $M_V,$ $|\mu|,$ $\Phi_\mu$
and ${\tilde \Phi}_R,$ Equations \eqref{eq:MRorder32} and
\eqref{eq:MRorder16} represent a system of equations serving to
determine, at least numerically,  $M_R$ and $\tan\theta_k.$ Indeed,
\eqref{eq:MRorder32} corresponds to an algebraic equation of order
$32$ in the variables $M_R$ and $\sin(2 \theta_k)$ whereas
\eqref{eq:MRorder16}, which can be factorized by $({\cal D}_1 {\cal
C}_2 - {\cal D}_2 {\cal C}_1)^2,$  corresponds to an algebraic
equation of order $16$ in these variables. Clearly, the analytical
treatment to obtain the solutions of these equations is hard, the
order of them is too large, however numerical extraction of
solutions shouldn't  represent, in principle, any problem. Apart
from the fact that the solutions have to verify Eqs.
(\ref{eq:MRorder32}-\ref{eq:MRorder32}), they also must verify the
constraints described above, namely ${\tilde {\mathbb B}}^2 - 4
{\tilde {\mathbb A}} {\tilde {\mathbb C}} \ge 0,$ $ {\mathbb B}^2 -
4 {\mathbb A} {\mathbb C} \ge 0,$  ${\tilde {\cal B}}_j^2 - 4
{{\tilde {\cal D}}}_j \, {\tilde {\cal C}}_j \ge 0, $  ${\tilde
{\cal B}}_j / {\tilde {\cal D}}_j < 0,$ ${\cal B}_j^2 - 4 {\cal D}_j
\, {\cal C}_j \ge 0, $  and $ {\cal B}_j / {\cal D}_j < 0,$  this
fact reduce considerably the number of possible solutions.

\section{Disentangle of the chargino sector L-R SUSY parameters
 based on cross-section physical observables}
 \label{sec-total-disentangle}
Further independent relations serving  determine  the fundamental
parameters can be obtained by computing some physical observables
such as the total cross section of the chargino pair production in
$e^+ e^-$ annihilation,  Left-Right asymmetries and polarization
vectors, similarly as it has been in the case of the MSSM
\cite{key22}.  In this section, we give an outline of the general
procedure that we could follow to fix the fundamental L-R SUSY based
on experimental measurements of cross-section-type observables.

The production of the chargino pairs at $e^+ e^-$ colliders in the
context of the L-R SUSY model at the tree level arise from $e^+ e^-
\rightarrow \gamma, Z_L , Z_R \rightarrow {\tilde \chi}^+_i {\tilde
\chi}^-_j $ in the s-channel, and $e^+ e^- \rightarrow {\tilde
\nu}_{L,R} \rightarrow {\tilde \chi}^+_i {\tilde \chi}^-_j $in the
t-channel. The Lagrangian corresponding to these interactions is
given by \cite{key14,key15} \beqa {\cal L} &=& - e A_\mu
\overline{{\tilde \chi}^+_i} \gamma^\mu {\tilde \chi}^+_j  + {g_L
\over {\cos\theta_W}} Z^\mu_L \overline{{\tilde \chi}^+_i}
\gamma^\mu \bigl(O^{'L}_{ij} \gamma_L + O^{'R}_{ij} \gamma_R \bigr)
{\tilde \chi}^+_j \nonumber
\\  &+& {g_R \sqrt{\cos(2 \theta_W)} \over {\cos\theta_W}} Z^\mu_R
\overline{{\tilde \chi}^+_i} \gamma^\mu \bigl(O^{L}_{ij} \gamma_L +
O^{R}_{ij} \gamma_R \bigr) {\tilde \chi}^+_j \nonumber
\\ &-& g \sum_{\ell=i,j} \overline{{\tilde \chi}_\ell^+} \biggl\{
\biggl[V_{1\ell} \, \gamma_L - \left(  {m_{\nu_s}
U^\ast_{3\ell}\over \sqrt{2} m_W \sin\theta_k} + {m_{e_s}
U^\ast_{4\ell}\over \sqrt{2} m_W \cos\theta_k}\right) \gamma_R
\biggr] e_s \, {\tilde \nu}^\ast_{L s} \nonumber \\ &+&
\biggl[U^\ast_{2\ell} \,  \gamma_R - \left(  {m_{\nu_s}
V_{3\ell}\over \sqrt{2} m_W \sin\theta_k} + {m_{e_s} V_{4\ell}\over
\sqrt{2} m_W \cos\theta_k}\right) \gamma_L \biggr] e_s \, {\tilde
\nu}^\ast_{R s}\biggr\}, \label{eq:lagrangian-vertices} \eeqa where
$\gamma_{L,R}= (1 \mp \gamma_5) / 2$ and
\beqa \label{eq:OLij} O^L_{ij} &=&  V^\ast_{2 i } \,  V_{2 j} + V^\ast_{3 i } \,  V_{3 j} + V^\ast_{4 i } \,  V_{4 j}, \\
\label{eq:ORij} O^R_{ij} &=& U^\ast_{2 i } \, U_{2 j} + U^\ast_{3 i } \, U_{3 j} + U^\ast_{4 i } \, U_{4 j} \\
 \label{eq:OLpij} O^{'L}_{ij} &=& -  V^\ast_{1 i } V_{1 j} - {1 \over 2}( V^\ast_{3 i }
 V_{3 j} +  V^\ast_{4 i } V_{4 j}) + \delta_{ij} \sin^2\theta_W, \\
\label{eq:ORpij}  O^{'R}_{ij} &=& -  U_{1 i } U^\ast_{1 j} - {1
\over 2}( U_{3 i }
 U^\ast_{3 j} +  U_{4 i } U^\ast_{4 j}) + \delta_{ij}
 \sin^2\theta_W.
\eeqa

Thus, using Eqs.  \eqref{eq:Vaj} and  \eqref{eq:Uaj} the basic
coupling vertices of the interaction Lagrangian
\eqref{eq:lagrangian-vertices} and the so called bilinear and
quartic charges determining the cross-section-type physical
observables, can be entirely written in terms of reduced projectors
and eigenphases.

For instance, using Eqs.  \eqref{eq:Vaj} and  \eqref{eq:Uaj} we can
write the coupling (\ref{eq:OLij}-\ref{eq:ORpij}) in terms of the
eigenphases and reduced projectors, namely \beqa
\label{eq:coupling1} O^L_{ij} &=& \zeta_i \zeta^\ast_j \,
\sqrt{P^V_{i11} P^V_{j11}}
(p_{i 2} p^\ast_{j2} + \delta_{i3} \delta_{j3} + p_{i 4} p^\ast_{j4}), \\
\label{eq:coupling2} O^R_{ij} &=& \sqrt{P^{U^\ast}_{i11}
P^{U^\ast}_{j11}}
({\tilde p}^\ast_{i2} {\tilde p}_{j2} + {\tilde p}^\ast_{i3} {\tilde p}_{j3} + \delta_{i3} \delta_{j3}), \\
\label{eq:coupling3} O^{'L}_{ij} &=& \delta_{ij} \sin^2\theta_W -
\zeta_i \zeta^\ast_j \sqrt{P^V_{i11} P^V_{j11}} \left[ (1 -
\delta_{i3}) (1 - \delta_{j3}) + {1\over 2} \left( \delta_{i3}
\delta_{j3} +  p_{i4} p^\ast_{j4} \right) \right] ,
\\\label{eq:coupling4}
O^{'R}_{ij} &=& \delta_{ij} \sin^2\theta_W - \sqrt{P^{U^\ast}_{i11}
P^{U^\ast}_{j11}}\left[ (1 - \delta_{i3}) (1 - \delta_{j3}) +
{1\over 2} \left( \delta_{i3} \delta_{j3} + {\tilde p}_{i3} {\tilde
p}^\ast_{j3} \right) \right] , \eeqa where we have used the fact
that $p_{i3}={\tilde p}_{i4}=\delta_{i3},$ $p_{i1}= {\tilde
p}_{i1}=1 - \delta_{i3},$ and the choice ${\tilde \eta}_j=1,
j=1,\ldots,4.$

Note that when $i=j,$ the coupling
(\ref{eq:coupling1}-\ref{eq:coupling4}) are independent on the
eigenphases, they only depends  on the norm of the reduced
projectors. On the other hand, when $i \ne j,$ using Eqs.
\eqref{eq:sysV} and \eqref{eq:tsysU}, we can show that these
couplings in addition to the explicit dependence on the eigenphases,
only depends on  the norm of four $V$-type and four $U^\ast$-type
reduced projectors.

Let us write $p_{ij}$ in the form \be p_{ij}= |p_{ij}| e^{i
\beta_{ij}}, \ee where $\beta_{ij}$ is a real phase. Inserting this
result into \eqref{eq:sysV}, and  splitting the real and imaginary
part, we get \be 1 + a^{(2)}_{ij}  \cos(\beta_{i2} -\beta_{j2}) +
a^{(4)}_{ij} \cos(\beta_{i4} -\beta_{j4}) =0 \ee and \be
a^{(2)}_{ij} \sin(\beta_{i2} -\beta_{j2}) + a^{(4)}_{ij}
\sin(\beta_{i4} -\beta_{j4})=0,\ee where $a^{(k)}_{ij}= |p_{ik}|
|p_{jk}|, \,  i,j=1,\ldots,4, \, ( i > j \perp  i,j \ne 3).$ Solving
these equations for the unknown $\beta_{ij}$ variables, we obtain
\be \label{eq:betaX} \beta_{i2} - \beta_{j2}= \pm \, \arccos(X_{ij})
\ee and \be \label{eq:betaY}\beta_{i4} - \beta_{j4}= \mp \,
\arccos(Y_{ij}),\ee where \be X_{ij}= {- 1 +  \left[
(a^{(4)}_{ij})^2  - (a^{(2)}_{ij})^2 \right] \over 2 a^{(2)}_{ij}}
\ee and \be Y_{ij}={- 1 - \left[(a^{(4)}_{ij})^2 -
(a^{(2)}_{ij})^2\right]  \over 2 a^{(4)}_{ij}}, \ee when $ i > j
\perp  i,j \ne 3.$

 The same relations are valid for the norm
and phases of the  $U^\ast$-type reduced projectors. Thus, according
to Eqs. \eqref{eq:betaX} and \eqref{eq:betaY} (and the corresponding
ones for the $U^\ast$-type reduced projectors)
 all phase differences in (\ref{eq:coupling1}-\ref{eq:coupling4}) only depends on the norm
 of the
reduced projectors.

Moreover, Eq. \eqref{eq:betaX} allow us  eliminate two phases and
obtain a real algebraic equations relating the norm of the reduced
projectors, namely \be \label{eq:beta22} \beta_{22} = \beta_{12} \pm
\arccos X_{21}, \ee \be \beta_{42}= \beta_{12} \pm  \arccos
X_{41}\ee
 and \be \arccos X_{21} - \arccos X_{41} + \arccos
X_{42}=0. \label{eq:x32} \ee  In the same way, from Eq.
\eqref{eq:betaY}, we get \be \beta_{24} = \beta_{14} \mp  \arccos
Y_{21}, \ee \be \beta_{44}= \beta_{14} \mp \arccos Y_{41}\ee  and
\be \arccos Y_{21} - \arccos Y_{41} + \arccos Y_{42}=0.
\label{eq:y32}\ee

Note that from Eq. \eqref{eq:tsysU},  analogous equations for the
phases and norms of the reduced projectors of the $U^\ast$-type can
be obtained  by defining ${\tilde p}_{ij}=|{\tilde p}_{ij}| e^{i
{\tilde \beta}_{ij}},$ ${\tilde a}^{(k)}_{ij}= |{\tilde p}_{ik}|
|{\tilde p}_{jk}|, \, k=2,3, \,   i,j=1,\ldots,4, \, ( i
> j \perp i,j \ne 3).$

Thus, from Eqs. (\ref{eq:beta22}-\ref{eq:y32}), and the analogous
ones for the $U^\ast$-type reduced projectors, we deduce that the
number of independent parameters determining  the complete set of
reduced projectors is  twelve.  The independent parameters can be
chosen to be $|p_{12}|, |p_{22}|, |p_{14}|, |p_{24}|,
 |{\tilde p}_{12}|, |{\tilde
p}_{22}|, |{\tilde p}_{13}|,|{\tilde p}_{23}|,
\beta_{12},\beta_{14}, {\tilde \beta}_{12},{\tilde \beta}_{13}.$

 Moreover, using Eqs. (\ref{eq:cdelta1j-V}-\ref{eq:cdelta4j-U}),
the independent phases can be expressed in terms of the fundamental
L-R SUSY parameters and physical masses as \be \tan \beta_{1j} =
{{\mathrm{Im\,}} (p_{1j}) \over {\mathrm{Re\,}}( p_{1j})} =
{{\mathrm{Im\,}} (\Delta^\ast_{j1}) \over
{\mathrm{Re\,}}(\Delta^\ast_{j1})} \ee when $j=2,4,$ and \be \tan
{\tilde \beta}_{1j} = {{\mathrm{Im\,}} ({\tilde p}_{1j}) \over
{\mathrm{Re\,}}({\tilde p}_{1j})} = {{\mathrm{Im\,}} ({\tilde
\Delta}^\ast_{j1}) \over {\mathrm{Re\,}}({\tilde \Delta}^\ast_{j1})}
\ee when $j=2,3.$ Note that the independent phases are connected
with the eigenphases through Eqs. \eqref{eq:zetaj-ReZ-non-zero} and
\eqref{eq:zetaj3}, i.e., eventually we could also choose the
eigenphases as independent parameters in place of them.

On the other hand, if two of the chargino physical masses are known,
for instance $m_{{\tilde \chi}^\pm_{1,2}},$  combining  Eqs.
\eqref{eq:MR-pj-etaj-thetaj} and \eqref{eq:zetaj-ReZ-non-zero} we
get that both the $\zeta_j$ eigenphases, $ j=1,2,4,$ and
$\tan\theta_k$ parameter can be expressed in terms of the above
mentioned set of twelve independent parameters and two given
chargino physical masses. Moreover, as $\zeta_3= - \mu / |\mu|,$
with the help of  Eq. \eqref{eq:mu-pj-etaj-thetaj} when $j=1$ or
$j=2,$ we can also express the $\zeta_3$ eigenphase in terms of this
set of parameters and physical masses. In this way, we conclude that
all the L-R SUSY fundamental parameters can be expressed in terms of
above mentioned set of independent parameters and physical masses.

In sum, when two chargino physical masses are known, the quartic
charges determining the cross-section-type observables, in addition
to the known chargino masses, depend on a subset of reduced
projector norms and phases. Due to the large number of independent
parameters we need to determine (twelve, if two physical chargino
masses are known, or ten, if four physical chargino masses are
known), the problem for extracting some information on the
fundamental L-R SUSY parameters by measuring this class of
observables is complex. Comparing with the corresponding case in the
MSSM, in similar circumstances, the quartic charges only depends on
two independent parameters\cite{key22,key23}, in our approach they
can be taken to be the norm of a $U^\ast$-type reduced projector and
the one of a $V$-type  reduced projector.

Returning to the counting of independent variables, in an idealized
situation where all the physical chargino masses are known we need
to determine ten parameters corresponding exactly to the number of
independent equations obtained by measuring polarized or unpolarized
chargino pair production total cross section $\sigma_T : e^+  e^- \,
\rightarrow {\tilde \chi}^+_{i} {\tilde \chi}^+_{j}, i,j=1,\ldots,4;
i \ge j.$ The same occurs in the context of the MSSM where the
number of independent parameters is two (when the two lightest
chargino masses are known). Further independent relations as
forward-backward asymmetries in the case of unpolarized beams or
left-right asymmetries in the case of polarized beams could
constitute a fundamental complement  to fix all unknown reduced
projectors, i.e., the unknown L-R SUSY parameters, up to a discrete
ambiguity. Moreover, as in the MSSM \cite{key22}, this discrete
ambiguity could be resolved by analyzing manifestly CP-non invariant
observables such as those related to normal polarization of the
charginos or T-odd asymmetries related to the chargino pair
production with longitudinally polarized beams and subsequent decays
of one of the charginos into a sneutrino and anti-lepton. Evidently,
from both the analytical and numerical point of view the situation
is not evident.

In the next section, we propose a suitable parametrization for the
independent reduced projectors, i.e., basic couplings and
cross-section-type physical observables that could help us to
visualize the solution to our problem in a more excellent manner.

\subsection{Parametrization in spherical coordinates}
The particular choice of the set of independent reduced projectors
determining the structure of the $V$ and $U^\ast$ matrices allows us
to use spherical coordinates to parameterize the physical cross
section observables. Indeed, from Eq. \eqref{eq:Vaj} and the unitary
character of the $V$ matrix, we have \be P^V_{j11} (1 + |p_{j2}|^2 +
|p_{j4}|^2 ) = 1, \qquad j=1,2. \ee Then, we can define ($j=1,2$)
\beqa
\sin\theta^{(j)} \cos\phi^{(j)} &=& \sqrt{P^V_{j11}} \; |p_{j2}|,\\
\sin\theta^{(j)} \sin\phi^{(j)} &=& \sqrt{P^V_{j11}} \; |p_{j4}|, \\
 \cos\theta^{(j)} &=& \sqrt{P^V_{j11}},
\eeqa where $ 0 \le \theta^{(j)} \le \pi /2  $ and  $ 0 \le
\phi^{(j)} \le \pi .$

In the same way, from Eq. \eqref{eq:Uaj} and the unitary character
of the $U^\ast$ matrix, we have \be P^{U^\ast}_{j11} (1 + |{\tilde
p}_{j2}|^2 + |{\tilde p}_{j3}|^2 ) = 1, \qquad j=1,2. \ee Then,
again we can define ($j=1,2$) \beqa
\sin{\tilde \theta}^{(j)} \cos{\tilde \phi}^{(j)} &=& \sqrt{P^{U^\ast}_{j11}} \; |{\tilde p}_{j2}|,\\
\sin{\tilde \theta}^{(j)} \sin{\tilde \phi}^{(j)} &=& \sqrt{P^{U^\ast}_{j11}} \; |{\tilde p}_{j3}|, \\
 \cos{\tilde \theta}^{(j)} &=& \sqrt{P^{U^\ast}_{j11}},
\eeqa where $ 0 \le {\tilde \theta}^{(j)} \le \pi /2  $ and  $ 0 \le
{\tilde \phi}^{(j)} \le \pi .$

Thus, all independent reduced projector are expressed in terms of
the spherical coordinated in the form $(j=1,2)$ \be |p_{j2}|=
\tan{\theta}^{(j)} \cos{\phi}^{(j)}, \qquad |p_{j4}|=
\tan{\theta}^{(j)} \sin{\phi^{(j)}} \ee  \be |{\tilde p}_{j2}| =
\tan{\tilde \theta}^{(j)}  \sin{\tilde \phi}^{(j)}, \qquad |{\tilde
p}_{j3}| = \tan{\tilde \theta}^{(j)} \cos{\tilde \phi}^{(j)}.\ee

The norms  $|p_{42}|\equiv \sqrt{X}$ and $ |p_{44}|\equiv \sqrt{Y},$
are obtained by solving the following system of equations deduced
from Eqs. \eqref{eq:x32} and \eqref{eq:y32}: \beqa \label{eq:XYa}
a_{20} X^2 +
a_{11} X Y + a_{02} Y^2 + a_{10} X  + a_{01} Y + a_{00}=0, \\
 b_{20} X^2 + b_{11} X Y + b_{02} Y^2 + b_{10} X  + b_{01} Y +
b_{00}=0, \label{eq:XYb} \eeqa where the $a_{\alpha \beta}$
coefficients, $\alpha,\beta=1,2$ are given by  \beqa a_{20}&=&
{1\over 4}\biggl\{ 1 + \cos^2 \phi^{(1)} \, \tan^2
\theta^{(1)}  +  {1 \over 2}  \, \tan^2 \theta^{(2)} \nonumber \\
&\times& \left[ {\left( \cos (2\,\phi^{(1)}) + \cos (2\,\phi^{(2)})
\right) \, \sec^2 \theta^{(1)}}
    +  2  \sin^2 \phi^{(1)} \right]
       \biggr\},\eeqa
\beqa a_{11}&=& {1 \over 4} \biggl\{ -   \sin^2 \phi^{(2)} \, \tan^2
\theta^{(2)} \, \left( 2 + \cos^2 \phi^{(1)} \, \tan^2 \theta^{(1)}
\right)
       \nonumber \\ &-& \tan^2 \phi^{(1)} -  \tan^2 \phi^{(2)} +
    \sin^2 \phi^{(1)} \, \tan^2 \theta^{(1)}  \nonumber \\ &\times&  \,\biggl[ -2 +
       \tan^2 \theta^{(2)} \, \left( -  \cos^2 \phi^{(2)} +
\          \sin^2 \phi^{(2)} \,\left( \tan^2 \phi^{(1)} + \tan^2
\phi^{(2)} \right)  \right)  \biggr]  \biggr\}, \eeqa
\beqa
a_{02}&=& {1\over 8} \, \biggl\{2\,  \sin^2 \phi^{(1)}\, \tan^2
\theta^{(1)} \, \tan^2 \phi^{(1)} +
    \biggl[ 2\, \sin^2 \phi^{(2)} \, \tan^2 \theta^{(2)} \nonumber \\ &+&
       \biggl( 2 + \left( \cos (2\,\phi^{(1)}) + \cos (2\,\phi^{(2)}) \right) \,
       \tan^2 \theta^{(1)}\, \tan^2 \theta^{(2)} \biggr)
          \, \tan^2 \phi^{(1)} \biggr] \, \tan^2 \phi^{(2)} \biggr\}, \eeqa
\beqa a_{10} &=& {1 \over 16} \,  \biggl\{ \cot^2 \theta^{(1)} \,
\cot^2 \theta^{(2)} \,
 \sec^2 \phi^{(1)} \, \sec^2 \phi^{(2)} \,
 \biggl[ 2 + 2\,  \cos^2 \phi^{(1)} \, \tan^2 \theta^{(1)} \nonumber \\
 &+&
      \left( \left( \cos (2\,\phi (1)) + \cos (2\,\phi (2)) \right) \,
      \sec^2 \theta^{(1)} + 2\, \sin^2 \phi^{(1)} \right) \,
       \tan^2 \theta^{(2)} \biggr] \, \nonumber \\
       & \times & \biggl[ 2 + \left( \cos (2\,\phi^{(1)}) + \cos (2\,\phi^{(2)}) \right)
\, \tan^2 \theta^{(1)} \, \tan^2 \theta^{(2)} \biggr] \biggr\},\eeqa
\beqa a_{01} &=&  {1\over 8} \, \biggl\{ \biggl[ -4 - \sec^2
\phi^{(2)}\,\biggl( 2\, \cot^2 \theta^{(2)} +
          \left( \cos (2\,\phi^{(1)}) + \cos (2\,\phi^{(2)}) \right) \, \tan^2 \theta^{(1)} \biggr)  \biggr]
           \, \tan^2 \phi^{(1)}
    \nonumber  \\ &-& \biggl[ 4 + \sec^2 \phi^{(1)} \,\left( 2\, \cot^2 \theta^{(1)} +
          \left( \cos (2\,\phi^{(1)}) + \cos (2\,\phi^{(2)}) \right) \, \tan^2 \theta^{(2)} \right)  \biggr]
           \, \tan^2 \phi^{(2)} \biggr\}, \eeqa
and \be a_{00}= {1\over 4} \, \biggl\{ \left( 1 + \cot^2 \theta^{
(1)} \, \sec^2 \phi^{(1)} \right) \,
   \left( 1 + \cot^2 \theta^{(2)} \, \sec^2 \phi^{(2)} \right)  - \tan^2 \phi^{(1)} \, \tan^2
   \phi^{
   (2)} \biggr\},
   \ee
and the $b_{\alpha \beta}$ coefficients are obtained from the
corresponding  $a_{\beta \alpha}$ by replacing $\sin\phi^{(j)}$ by
$\cos\phi^{(j)},$ $j=1,2$ or vice versa,  in all the terms
containing these functions in either the explicit or implicit form.

Similarly, the norms  $|{\tilde p}_{42}|\equiv \sqrt{{\tilde X}}$
and $ |{\tilde p}_{43}|\equiv \sqrt{{\tilde Y}},$ are obtained by
solving the following equation system: \beqa \label{eq:XYta} {\tilde
a}_{20} {\tilde X}^2 +
{\tilde a}_{11} {\tilde X} {\tilde Y} + {\tilde a}_{02} {\tilde Y}^2 + {\tilde a}_{10} {\tilde X}  + {\tilde a}_{01} {\tilde Y} + {\tilde a}_{00}=0, \\
 {\tilde b}_{20} {\tilde X}^2 + {\tilde b}_{11} {\tilde X} {\tilde Y} + {\tilde b}_{02} {\tilde Y}^2 + {\tilde b}_{10} {\tilde X}  + b_{01} {\tilde Y} +
{\tilde b}_{00}=0, \label{eq:XYtb}\eeqa where the ${\tilde
a}_{\alpha \beta}$ and ${\tilde b}_{\alpha \beta}$ coefficients,
$\alpha,\beta=1,2$ are the same as the  $a_{\alpha \beta}$ and
$b_{\alpha \beta}$ coefficients, respectively, with the obvious
changes.

Note that from the geometric point of view, the independent reduced
projector-type parameters lie on four disconnected spherical
surfaces and the dependent ones are determined by intersecting two
conical sections.

Thus, for instance,  from Eqs. (\ref{eq:OLij}-\ref{eq:ORpij}), we
deduce that the coupling constants in spherical coordinates are
given by \beqa O^L_{ij} &=& \delta_{ij} - \zeta_i \zeta_j^\ast
\cos\theta^{(i)} \cos\theta^{(j)}, \\
O^R_{ij} &=& \delta_{ij} -
\cos{\tilde \theta}^{(i)} \cos{\tilde \theta}^{(j)}, \\
O^{'L}_{ij} &=& \delta_{ij} \sin^2 \theta_W  - {\zeta}_i
{\zeta}_j^\ast \cos\theta^{(i)} \cos\theta^{(j)} \nonumber \\
&\times&  \biggl\{ 1 + {1\over 2} \tan\theta^{(i)} \tan\theta^{(j)}
\sin\phi^{(i)} \sin\phi^{(j)} \left[ Y_{ij} \mp {\bf i} \sqrt{1-
Y^2_{ij}}\right] \biggr\}, \\
O^{'L}_{ij} &=& \delta_{ij} \sin^2 \theta_W  - \cos{\tilde \theta}^{(i)} \cos{\tilde \theta}^{(j)} \nonumber \\
&\times&  \biggl\{ 1 + {1\over 2} \tan{\tilde \theta}^{(i)} {\tilde
\tan\theta}^{(j)} \sin{\tilde \phi}^{(i)} \sin{\tilde \phi}^{(j)}
\left[ {\tilde Y}_{ij} \mp {\bf i} \sqrt{1- {\tilde
Y}^2_{ij}}\right] \biggr\}, \eeqa when $i,j=1,2,$ and \be O^L_{3j}
=O^L_{j3} =O^R_{3j} =O^R_{j3} =\delta_{3j}, \qquad O^{'L}_{3j}
=O^{'L}_{j3} =O^{'R}_{3j} =O^{'R}_{j3}= \delta_{3j} \,
\left(\sin^2\theta_W- {1\over 2}\right), \ee when $j=1,2,3.$ Here,
\be Y_{ij}=  {- 2\, +
      \left[ \cos (2\,\phi^{(i)}) + \cos (2\,\phi^{(j)}) \right] \,\tan^2 \theta^{(i)} \,\tan^2 \theta^{(j)}
      \over 4 \,
 \tan \theta^{(i)}\,
      \tan
\theta^{(j)} \sin \phi^{(i)}\, \sin \phi^{(j)} \,  } \ee and
$Y_{ii}\equiv1,$  with analogous expressions for ${\tilde Y}_{ij}.$
As we have pointed out, when the two lightest chargino masses are
known, from  Eq. \eqref{eq:MR-pj-etaj-thetaj} we can express the
fundamental parameter $\tan\theta_k$ in terms of these masses and
the independent reduced projector-type parameters, i.e., in terms of
the two lightest chargino masses, the spherical angles defined in
this section and the four independent reduced projector phases, see
Appendix \ref{sec-tanthetak}. Moreover, using  Eq.
\eqref{eq:zetaj-ReZ-non-zero},  $\zeta_j, \, j=1,2,$  can also be
expressed in terms of this set of parameters. Hence, in a first
stage, we should analyze those observables  that depend directly on
these parameters.

On the other hand, when either $i=4$ or $j=4,$ the couplings in Eqs.
(\ref{eq:OLij}-\ref{eq:ORpij}) in addition to the basic independent
parameters depend on the norms $|p_{42}|,|p_{44}|,|{\tilde
p}_{42}|,|{\tilde p}_{43}|, $ i.e., we must solve Eqs.
(\ref{eq:XYa}-\ref{eq:XYb}) and Eqs. (\ref{eq:XYta}-\ref{eq:XYtb})
to obtain the  explicit dependence of these coupling in terms of the
physical chargino masses and the complete set of angular independent
parameters. From both, the analytical and numerical point of view,
the determination of the independent parameters through the
experimental measurements of observables containing these dependent
terms  is more complex.

\section{Conclusions}
In this paper we have studied the consequences produced by the
introduction of CP-phases into the chargino mass matrix in the
context of the L-R SUSY model. We have analyzed the chargino mass
spectrum and  treated  the inverse parameter problem.  Thus, the
chargino mass matrix was described by eight real fundamental
parameters, i.e., the four usual parameters $|M_L|,$ $|\mu|,$ $M_R,$
and $\tan\theta_k $ in addition to the four real phases $\Phi_L,
\Phi_\mu, {\tilde \Phi}_L$ and ${\tilde \Phi}_R.$ To find analytical
expressions for the chargino physical masses $m_{{\tilde
\chi}^{\pm}_{j}}, \, j=1,\ldots, 4, $ and some connecting relations
among the parameters, at the tree level, we have diagonalized the
non-symmetric chargino mass matrix by constructing  two
diagonalizing unitary matrices. The masses, obtained by solving the
associated characteristic polynomial to this problem, have been
ordered by sizes and plotted as a function of the Higgsino parameter
$\mu,$  and also as a function of $\tan\theta_k$ by considering some
possible CP-conserving and CP-violating scenarios. We have observed
by comparing different plots that the effects on the mass spectrum
are more significative when the mixing phases ${\tilde \Phi}_L$ or
$\Phi_L$ varies. Some comparisons with the MSSM have been given.

The inverse problem consisting to determine the fundamental
parameters in terms of the chargino physical masses, the reduced
projectors and the eigenphases have been solved using the projector
formalism based on the construction of the two mentioned
diagonalizing matrices of this problem. Also, combining  these
results with the corresponding  Jarlskog's formulas we have obtained
an alternative way  to disentangle the unknown L-R SUSY parameters.
In particular, under some conditions, we have found analytic
expressions to disentangle the parameters $|M_L|,$ $\Phi_L,$
${\tilde \Phi_L},$ from the rest of the parameters. Thus, we have
considered three types of CP-violating scenarios, all of them
characterized by a big rate between $k_u$ and $k_d.$ In one of these
scenarios we have supposed that the quantities that could be first
measured are  the two lightest chargino masses and the two lightest
neutralino masses. In this case, we have observed that analytical
and numerical expressions can be given which allow us disentangle
the parameters $|M_L|,$ $\Phi_L$ and ${\tilde \Phi}_L$ up to a
twofold discrete ambiguity. Also, an additional equation should
allows us to extract, at least numerically, an additional parameter.

In a more general schema, we have demonstrated that the fundamental
L-R SUSY parameters can be expressed in terms of twelve independent
parameters associated to the reduced projectors, which can be
represented by four pair of spherical angles an four independent
reduced projector phases. The analytical or numerical determination
of these parameters through the measurement of cross-section-type
observables could allow us, in principle, to known all the
fundamental L-R SUSY parameters. However, the treatment of this
problem is not simple since the large number of involved parameters
difficult its resolution. A possible issue to the problem could be
to give some appropriate physical inputs in the chargino and
neutralino sectors.

The formulas deduced  in this article allow us to considerer other
possibilities of input data. For instance, some scenarios where the
lightest chargino and  neutralino masses are known. Also, as the
basic system of equations given in Eqs. \eqref{eq:gen-inversionU}
and \eqref{eq:gen-inversionV} only involve the matrix elements of
the original chargino mass matrix, the projectors, the reduced
projectors, the chargino physical masses and the eigenphases, and
these equations  remains uncoupled with respect to the index $j,$
the present formalism can be directly generalized to any chargino
number and any complex non-symmetric mass matrix. For instance, the
inverse problem for determining the fundamental parameters including
all terms of the Lagrangian \eqref{eq:lagrangiano-chargino} can be
treated in the same way, i.e., we could determine the contribution
of the charged right-handed higgsino fields ${\tilde \Delta}_R$ and
${\tilde \delta}_R$ to the determination of the fundamental
parameters. However, in this case, to determine the unknown physical
chargino masses in terms of the fundamental parameters we must solve
a quintic equation which requires some additional work. These last
aspects could be treated in a separate communication.

\label{sec-detparameters}
\section*{Acknowledgments}
The author dedicates this article to the memory of his friend Jeff
Delbecque, {\it Celui qui savait paler aux femmes}. He thank
Patricia Aguilar, member of the organization {\it Baobab Familial}
of Montreal, for valuable support, Artorix de la Cruz de O\~na and
Mariana Frank for enlightening discussions, and the referees for
valuable lecture and suggestions in order to  enhance the
presentation and contents of this paper.

\appendix

\section{Chargino  mass spectrum}
\label{sec-chargino-MASS} \setcounter{equation}{0} According to Eq.
\eqref{eq:MD2},  the chargino masses predicted by the present model
are given by the positive roots of the eigenvalues associated to
either  the Hermitian matrix $ H\equiv M^{\dag} \,M$ or the
Hermitian matrix $\tilde H \equiv M \, M^\dag.$ These eigenvalues
can be obtained by solving the common characteristic equation
associated to these matrices. Indeed, starting from Eq.
\eqref{eq:MD2}, we get \be \label{eq:EVP} (M^{\dag}\,M) \, V  - V \,
M^{2}_{D} =0, \ee and \be \label{eq:EUP} (M\, M^{\dag}) \, U^\ast -
U^\ast \, M^{2}_{D} =0, \ee which expressed by  components writes
\be \label{eq:FE-EVP}(H_{ii} - m_{{\tilde \chi}^{\pm}_{j}}^2)
V_{ij}+ \sum_{k \ne i}^{4} H_{ik} V_{kj}, \quad i,j=1,\ldots,4, \ee
and \be \label{eq:FE-EUP}({\tilde H}_{ii} - m_{{\tilde
\chi}^{\pm}_{j}}^2) U^\ast_{ij} + \sum_{k \ne i}^{4} {\tilde H}_{ik}
U^\ast_{kj}, \quad i,j=1,\ldots,4, \ee respectively, where $H_{ij}=
\sum_{k=1}^4 M^{\ast}_{ki}
M_{kj}:$ \beqa \nonumber H_{11}&=&  |M_L|^2 + 2 |{\tilde M}_L|^2 \sin^2\theta_k, \\
\nonumber H_{22} &=&  M_{R}^2 + 2  |{\tilde M}_R|^2 \sin^2\theta_k,
\\ \nonumber H_{33}&=& |\mu|^2,
\\ \nonumber H_{44}&=& |\mu|^2 + 2 (|{\tilde M}_L|^2 + |{\tilde
M}_R|^2) \cos^2\theta_k,
\\\nonumber H_{12}&=& H_{21}^\ast =  2 |{\tilde M}_L|  |{\tilde
M}_R| e^{i ({\tilde \Phi}_R - {\tilde \Phi}_L)} \sin^2\theta_k,
\\
\nonumber H_{13}&=& H_{31}^\ast =  0,
\\
\nonumber H_{14}&=& H_{41}^\ast = \sqrt{2}|{\tilde M}_L| \bigl[e^{i
({\tilde \Phi}_L - \Phi_L)} |M_L| \cos\theta_k \\ \nonumber &-& e^{i
( \Phi_\mu - {\tilde \Phi}_L)} |\mu| \sin\theta_k\bigr],
\\
\nonumber H_{23}&=& H_{32}^\ast = 0,
\\
\nonumber H_{24}&=& H_{42}^\ast = \sqrt{2}|{\tilde M}_R| \bigl[e^{i
{\tilde \Phi}_R } M_R  \cos\theta_k \\ \nonumber &-&  e^{i (
\Phi_\mu - {\tilde \Phi}_R)} |\mu| \sin\theta_k\bigr],
\\ \nonumber H_{34}&=& H_{43}^\ast = 0. \\
\label{eq:Hij} \eeqa and  ${\tilde H}_{ij}= \sum_{k=1}^4 M_{ik}
M^{\ast}_{jk}
:$ \beqa \nonumber {\tilde H}_{11}&=&  |M_L|^2 + 2 |{\tilde M}_L|^2 \cos^2\theta_k, \\
\nonumber {\tilde H}_{22} &=&  M_{R}^2 + 2  |{\tilde M}_R|^2
\cos^2\theta_k,
\\ \nonumber {\tilde H}_{33}&=& |\mu|^2 + 2 (|{\tilde M}_L|^2 + |{\tilde
M}_R|^2) \sin^2\theta_k,
\\ \nonumber {\tilde H}_{44}&=& |\mu|^2,
\\\nonumber {\tilde H}_{12}&=& {\tilde H}_{21}^\ast =  2 |{\tilde M}_L|  |{\tilde
M}_R| e^{i ({\tilde \Phi}_L - {\tilde \Phi}_R)} \cos^2\theta_k,
\\
\nonumber {\tilde H}_{13}&=& {\tilde H}_{31}^\ast = \sqrt{2}|{\tilde
M}_L|
\bigl[e^{i (\Phi_L - {\tilde \Phi}_L )} |M_L| \sin\theta_k \\
\nonumber &-& e^{i ({\tilde \Phi}_L - \Phi_\mu)} |\mu|
\cos\theta_k\bigr],\\
\nonumber {\tilde H}_{14}&=& {\tilde H}_{41}^\ast =  0,
\\
\nonumber {\tilde H}_{23}&=& {\tilde H}_{32}^\ast = \sqrt{2}|{\tilde
M}_R| \bigl[e^{- i {\tilde \Phi}_R } M_R  \sin\theta_k \\ \nonumber
&-& e^{i ({\tilde \Phi}_R - \Phi_\mu  )} |\mu| \cos\theta_k\bigr],
\\
\nonumber {\tilde H}_{24}&=& {\tilde H}_{42}^\ast = 0,
\\ \nonumber {\tilde H}_{34}&=& {\tilde H}_{43}^\ast = 0. \\
\label{eq:THij} \eeqa

For fixed  $j,$  each of the Eqs. \eqref{eq:FE-EVP} and
\eqref{eq:FE-EUP} represents  a system of homogeneous linear
equations depending on only one of the chargino masses. Thus, the
chargino masses can be determined by solving the characteristic
equation associated to these systems, that is \be
 X^{4} - a\,X^{3}
+  b\,X^{2} - c\, X + d=0\,, \label{eq:CEQ} \ee where \beqa
\nonumber  a &=& |M_L|^2 +  M_{R}^2 \\\label{eq:a-term} &+& 2 \,
(|{\tilde M}_L|^2 + |{\tilde M}_R|^2 + |\mu|^2 ), \eeqa \beqa
\nonumber b &=&  |\mu|^4 + 2 |\mu|^2 ( |M_L|^2 + M_R^2 + |{\tilde
M}_L|^2 +  |{\tilde M}_R|^2) \\ \nonumber &+& 2 |\mu| \sin(2
\theta_k)  \bigl[ |M_L| |{\tilde M}_L|^2 \cos(\Phi_L - 2 {\tilde
\Phi}_L + \Phi_\mu ) \\ \nonumber &+& M_R |{\tilde M}_R|^2 \cos(2
{\tilde \Phi}_R -  \Phi_\mu )\bigr] + 2 M_R^2 |{\tilde M}_L|^2
\\ \nonumber &+& |M_L|^2 (M_R^2 + 2 |{\tilde M}_R|^2 ) \\
\label{eq:b-term} &+& (|{\tilde M}_L|^2 + |{\tilde M}_R|^2)^2 \sin^2
(2 \theta_k), \eeqa \beqa \nonumber c &=& |\mu|^4 (M_R^2 + |M_L|^2)
+ 2 |\mu|^3 \sin(2\theta_k) \\ \nonumber &\times& \bigl[|M_L|
|{\tilde M}_L|^2 \cos(\Phi_L - 2 {\tilde \Phi}_L + \Phi_\mu) \\
\nonumber &+& M_R |{\tilde M}_R|^2 \cos(2 {\tilde \Phi}_R -
\Phi_\mu)\bigr] + |\mu|^2
\\ \nonumber &\times& \bigl[ (|{\tilde M}_L|^2 + |{\tilde M}_R|^2)^2
\sin^2(2 \theta_k)  + 2 |M_L|^2  |{\tilde M}_R|^2 \\
\nonumber &+&  2 M_R^2 (|M_L|^2 + |{\tilde M}_L|^2)\bigr]+ 2 |\mu|
|M_L| M_R \sin(2 \theta_k) \\ \nonumber &\times& \bigl[M_R |{\tilde
M}_L|^2 \cos(\Phi_L - 2 {\tilde \Phi}_L + \Phi_\mu)
\\ \nonumber &+& |M_L| |{\tilde M}_R|^2 \cos(2 {\tilde \Phi}_R -
\Phi_\mu)\bigr] + \sin^2 ( 2 \theta_k) \\ \nonumber &\times& \bigl[
 2 M_R |M_L| |{\tilde M}_L|^2 |{\tilde M}_R|^2  \cos(\Phi_L - 2
{\tilde \Phi}_L + 2 {\tilde \Phi}_R) \\ \label{eq:c-term} &+&
M_R^2|{\tilde M}_L|^4 + |M_L|^2 |{\tilde M}_R|^4 \bigr]\eeqa and
\beqa \nonumber d &=& |\mu|^2  \bigl\{\sin^2(2\theta_k) \bigl[M_R^2
|{\tilde M}_L|^4 + |M_L|^2 |{\tilde M}_R|^4 \\ \nonumber &+& 2 |M_L|
M_R |{\tilde M}_L|^2 |{\tilde M}_R|^2  \cos(\Phi_L - 2 {\tilde
\Phi}_L + 2 {\tilde \Phi}_R) \bigr]
\\ \nonumber &+& 2 M_R |\mu| |M_L| \sin(2 \theta_k)
\\ \nonumber &\times& \bigl[ M_R |{\tilde M}_L|^2  \cos(\Phi_L - 2
{\tilde \Phi}_L + \Phi_\mu) \\ \nonumber &+&  |M_L||{\tilde M}_R|^2
\cos(2 {\tilde \Phi}_R - \Phi_\mu)] + |M_L|^2 M_R^2 |\mu|^2
\bigr\}.\\\label{eq:d-term}
 \eeqa
Solving Eq. \eqref{eq:CEQ}, we get the  analytic formulas for the
chargino  masses \beqa \label{eq:EIGU} m_{{\tilde \chi}^{\pm}_{1}}^2
,
m_{{\tilde \chi}^{\pm}_{2}}^2 &=& \frac{a}{4}-\frac{\alpha}{2}\mp\frac{1}{2}\,\sqrt{\beta - \varpi-\frac{\lambda}{4\alpha}},\\
m_{{\tilde \chi}^{\pm}_{3}}^2,m_{{\tilde \chi}^{\pm}_{4}}^2 &=&
\frac{a}{4}+\frac{\alpha}{2}\mp\frac{1}{2}\,\sqrt{\beta
-\varpi+\frac{\lambda}{4\alpha}},\label{eq:EIGV} \eeqa
where \beqa \alpha
&=& \nonumber \sqrt{{\beta \over 2} + \varpi},\\
\nonumber \varpi &=&\frac{\epsilon}{3 \ 2^{\frac{1}{3}}} +
\frac{(2^{\frac{1}{3}}\,\gamma)}{3\,\epsilon},\\ \nonumber \epsilon
&=& (\delta+ \sqrt{\delta^2 - 4 \gamma^3})^{\frac{1}{3}},
\\ \nonumber \beta &=& \frac{a^{2}}{2}-\frac{4b}{3},\\\nonumber
 \lambda &=& \nonumber a^{3}- 4\,a\,b + 8\,c\\ \nonumber
\gamma &=& \nonumber b^{2}- 3\,a\,c+12\,d, \\ \delta &=& \ 2\,b^{3}-
9\,a\,b\,c+27\,c^{2}+ 27\,a^{2}\,d-72\,b\,d.
\label{eq:relation-parameters}\eeqa We note that the physical masses
have been ordered according to their increasing magnitude. Also, as
$|\mu|^2$ is an exact root of Eq. \eqref{eq:CEQ}, then it is always
possible to find a neighborhood in the fundamental parameter space
where one of the chargino physical masses take the value $|\mu|.$

\section{Disentangling \boldmath $\tan\theta_k$}
\label{sec-tanthetak}
 When the two lightest physical chargino masses are
known, from Eq. \eqref{eq:MR-pj-etaj-thetaj}, we get \be f_4 \tan^4
\theta_k + f_3 \tan^3 \theta_k + f_2 \tan^2 \theta_k  + f_1 \tan
\theta_k + f_0 =0, \label{eq:thetak-disentangled} \ee where \beqa
f_{0}&=& \sin^2 \theta^{(1)} \, \sin^2 \theta^{(2)}\, \sin^2
\phi^{(1)} \, \sin^2 \phi^{(2)}
  \biggl\{  m^2_{{\tilde \chi}^\pm_1} \, \cos^2 {\tilde \phi}^{(2)} \, \cos^2 \phi^{(1)} \,
  \sin^2 {\tilde \theta}^{(2)} \,
     \sin^2 \theta^{(1)} \nonumber \\ &-&   m^2_{{\tilde \chi}^\pm_2} \cos^2 {\tilde \phi}^{(1)} \, \cos^2 \phi^{(2)} \,
     \sin^2 {\tilde \theta}^{(1)} \, \sin^2 \theta^{(2)} \biggr\}, \eeqa
\beqa f_{1} &=&  2 \, \sin^2 \theta^{(1)} \, \sin^2 \theta^{(2)} \,
\sin \phi^{(1)}\, \sin \phi^{ (2)}\, \nonumber \\ & \times&
  \biggl\{ \cos [\beta_{22} - \beta_{24} +  {\tilde \beta}_{22} - {\tilde
  \beta}_{23}]\,  \cos {\tilde \phi}^{(2)}\,
  \cos \phi^{(2)} \,   \sin^2 {\tilde \theta}^{(2)}\, \sin {\tilde \phi}^{(2)}\, \sin \phi^{(1)} \nonumber \\ &\times &
     \biggl(  m^2_{{\tilde \chi}^\pm_2} \, \cos^2 {\tilde \phi}^{(1)} \, \sin^2 {\tilde \theta}^{(1)}
      - m^2_{{\tilde \chi}^\pm_1} \,
       \cos^2 \phi^{(1)} \, \sin^2 \theta^{(1)}\biggr)  + \cos [\beta_{12} - \beta_{14} + {\tilde \beta}_{12} - {\tilde \beta}_{13}]
       \nonumber \\ &\times&
    \cos {\tilde \phi}^{(1)} \,
    \cos \phi^{(1)} \,
     \sin^2 {\tilde \theta}^{(1)} \, \sin {\tilde \phi}^{(1)} \, \sin \phi^{(2)} \nonumber \\ &\times &  \biggl( m^2_{{\tilde \chi}^\pm_2}
            \, \cos^2 \phi^{(2)} \, \sin^2 \theta^{(2)}  -  m^2_{{\tilde \chi}^\pm_1} \, \cos^2 {\tilde \phi}^{(2)} \,
     \sin^2 {\tilde \theta}^{(2)}\biggr)  \biggr\}, \eeqa
\beqa f_{2} &=& m^2_{{\tilde \chi}^\pm_1} \, \sin^2 {\tilde
\theta}^{(2)} \, \sin^2 \theta^{ (2)}
 \nonumber \\ &\times&
   \biggl[ \cos^2 {\tilde \phi}^{(1)} \, \cos^2 {\tilde \phi}^{(2)} \, \sin^4 {\tilde \theta}^{(1)}\,
      \sin^2 {\tilde \phi}^{(1)} \, \sin^2 \phi^{(2)} \nonumber \\ &+&
     \cos^2 \phi^{(1)} \, \cos^2 \phi^{(2)} \, \sin^2 {\tilde \phi}^{(2)} \, \sin^4 \theta^{(1)}\, \sin^2 \phi^{(1)}
      \biggr] \nonumber \\
      &+&  4\, \left( m^2_{{\tilde \chi}^\pm_1} -  m^2_{{\tilde \chi}^\pm_2}
   \right) \,
      \cos [\beta_{12} - \beta_{14} + {\tilde \beta}_{12} - {\tilde \beta}_{13}]\,
      \cos [\beta_{22} - \beta_{24} + {\tilde \beta}_{22} - {\tilde \beta}_{23}]
 \nonumber \\
&\times&
        \cos {\tilde \phi}^{(1)}\, \cos {\tilde \phi}^{(2)}  \, \cos \phi^{(1)}\, \cos \phi^{(2)}\,
        \sin^2 {\tilde \theta}^{(1)} \,
       \sin^2 {\tilde \theta}^{(2)}  \nonumber \\ &\times&  \sin {\tilde \phi}^{(1)}\,
      \sin {\tilde \phi}^{(2)}\,  \sin \phi^{(1)}\, \sin \phi^{(2)}\,
       \sin^2 \theta^{(1)} \, \sin^2 \theta^{(2)} \nonumber \\
      &-& m^2_{{\tilde \chi}^\pm_2} \,  \sin^2 {\tilde \theta}^{(1)} \, \sin^2 \theta^{(1)} \
     \nonumber \\ &\times& \biggl[ \cos^2 {\tilde \phi}^{(1)} \, \cos^2 {\tilde \phi}^{(2)}\,
         \sin^4 {\tilde \theta}^{(2)} \, \sin^2 {\tilde \phi}^{(2)} \, \sin^2 \phi^{(1)} \nonumber \\
         &+&
        \cos^2 \phi^{(1)} \, \cos^2 \phi^{(2)} \, \sin^2 {\tilde \phi}^{(1)}\, \sin^4 \theta^{(2)}\,
         \sin^2 \phi^{(2)}
         \biggr],     \eeqa  \beqa
f_3 &=&  2 \sin^2 {\tilde \theta}^{(1)}\,
  \sin^2 {\tilde \theta}^{(2)} \, \sin {\tilde \phi}^{(1)}\, \sin {\tilde \phi}^{(2)}
\biggl\{  \, \cos[\beta_{22} - \beta_{24} + {\tilde \beta}_{22} -
{\tilde \beta}_{23}] \nonumber \\ & \times & \cos {\tilde
\phi}^{(2)} \,   \cos \phi^{(2)}\,
     \sin {\tilde \phi}^{(1)}\, \sin^2 \theta^{(2)} \, \sin \phi^{(2)} \nonumber \\ &\times&
\biggl( m^2_{{\tilde \chi}^\pm_2}  \,  \cos^2 \phi^{(1)} \, \sin^2
\theta^{(1)} -  m^2_{{\tilde \chi}^\pm_1}  \,
       \cos^2 {\tilde \phi}^{(1)} \,  \sin^2 {\tilde \theta}^{(1)} \biggr)
      + \cos [ \beta_{12} - \beta_{14} + {\tilde \beta}_{12} - {\tilde \beta}_{13}] \nonumber \\ &\times& \cos {\tilde \phi}^{(1)} \,
     \cos \phi^{(1)} \, \sin {\tilde \phi}^{(2)}\,
     \sin^2 \theta^{(1)} \, \sin \phi^{(1)} \nonumber \\ &\times&
     \biggl( m^2_{{\tilde \chi}^\pm_2} \,
       \cos^2 {\tilde \phi}^{(2)} \,
       \, \sin^2 {\tilde \theta}^{(2)} -  m^2_{{\tilde \chi}^\pm_1} \,  \cos^2 \phi^{(2)} \, \sin^2 \theta^{(2)}
        \biggr)  \biggr\} \eeqa and
\beqa f_{4} &=& \sin^2 {\tilde \theta}^{(1)} \, \sin^2 {\tilde
\theta }^{(2)} \, \sin^2 {\tilde \phi}^{(1)} \,
  \sin^2 {\tilde \phi}^{(2)} \,  \biggl\{  m^2_{{\tilde \chi}^\pm_1} \, \cos^2 {\tilde \phi}^{(1)} \, \cos^2 \phi^{(2)} \,
\sin^2 \theta^{(2)} \, \sin^2 {\tilde \theta}^{(1)} \nonumber \\ &-&
m^2_{{\tilde \chi}^\pm_2} \,
    \cos^2 {\tilde \phi}^{(2)} \, \cos^2 \phi^{(1)} \,
    \sin^2 \theta^{(1)} \,
     \sin^2 {\tilde \theta}^{(2)} \biggr\}.
  \eeqa

The four solutions of Eq. \eqref{eq:thetak-disentangled} are
obtained from  Eqs. (\ref{eq:EIGU}-\ref{eq:relation-parameters}) by
putting \be a= - {f_3 \over f_4}, \qquad b= {f_2  \over f_4}, \qquad
\, c=- {f_1  \over f_4}, \qquad   d= {f_0 \over f_4} . \ee

Note that $\tan\theta_k$ depend on the phases $\beta_{12},
\beta_{14}, {\tilde \beta}_{12}$ and ${\tilde \beta}_{13}$  through
the phase differences  $\beta_{12} - \beta_{14}$ and ${\tilde
\beta}_{12} - {\tilde \beta}_{13}.$

\section{Summary of results}
\label{sec-tablas} In this section we summarize  the principal
relations among the parameters deduced from the chargino sector of
the L-R SUSY model. Here, we don't display  neither the results for
the disentangled quantities obtained by considering the two lightest
neutralino physical masses as input parameters or the disentangled
expressions for $M_R$ and redefined eigenphases obtained as a
consequence of introducing a novel parametrization. With respect to
this last point, using this novel parametrization it is possible to
think other scenarios starting with different sets of input
parameters.

\end{document}